\PassOptionsToPackage{dvipsnames,table}{xcolor}
\documentclass[sigconf]{acmart}

\AtBeginDocument{%
  \providecommand\BibTeX{{%
    \normalfont B\kern-0.5em{\scshape i\kern-0.25em b}\kern-0.8em\TeX}}}

\copyrightyear{2023}
\acmYear{2023}
\setcopyright{rightsretained}
\acmConference[CHI '23]{Proceedings of the 2023 CHI Conference on Human Factors in Computing Systems}{April 23--28, 2023}{Hamburg, Germany}
\acmBooktitle{Proceedings of the 2023 CHI Conference on Human Factors in Computing Systems (CHI '23), April 23--28, 2023, Hamburg, Germany}\acmDOI{10.1145/3544548.3581337}
\acmISBN{978-1-4503-9421-5/23/04}

\definecolor{lightgray}{gray}{0.9}
\usepackage{balance}       
\usepackage{graphics}      
\usepackage[T1]{fontenc}   
\usepackage{color}
\usepackage{booktabs}
\usepackage{textcomp}
\usepackage{makecell}
\usepackage{xspace}
\usepackage{fancyvrb}
\usepackage{amsthm}
\usepackage{hyperref}
\usepackage{subcaption}
\usepackage{todonotes}
\usepackage{capt-of}
\usepackage{graphicx}
\usepackage{dcolumn} 
\usepackage{tipa}
\usepackage{varwidth}
\usepackage{tabularx}
\usepackage{caption}
\usepackage{color,soul}

\usepackage{microtype}        
\usepackage{ccicons}          

\newcommand{\eg}{\textit{e.g.}\@\xspace}
\newcommand{\ie}{\textit{i.e.}\@\xspace}
\newcommand{\etal}{\textit{et al. }}

\usepackage{multirow}
\usepackage{enumitem}
\setlist[description]{leftmargin=\parindent,labelindent=\parindent}

\def\plainkeywords{datasets; privacy; human-centered AI; disability; data ownership}

\begin{document}


\title{Contributing to Accessibility Datasets: Reflections on Sharing Study Data by Blind People}

\author{Rie Kamikubo}
\affiliation{%
    \institution{College of Information Studies}
    \institution{University of Maryland, College Park}
    \streetaddress{4130 Campus Dr}
    \country{United States}
}
\email{rkamikub@umd.edu}
\author{Kyungjun Lee}
\affiliation{%
    \institution{Department of Computer Science}
    \institution{University of Maryland, College Park}
    \streetaddress{4130 Campus Dr}
    \country{United States}
}
\email{kyungjun@umd.edu}

\author{Hernisa Kacorri}
\affiliation{%
    \institution{College of Information Studies}
    \institution{University of Maryland, College Park}
    \streetaddress{4130 Campus Dr}
    \country{United States}
}
\email{hernisa@umd.edu}

\renewcommand{\shortauthors}{Kamikubo et al.}

\begin{abstract}
To ensure that AI-infused systems work for disabled people, we need to bring accessibility datasets sourced from this community in the development lifecycle. However, there are many ethical and privacy concerns limiting greater data inclusion, making such datasets not readily available. We present a pair of studies where 13 blind participants engage in data capturing activities and reflect with and without probing on various factors that influence their decision to share their data via an AI dataset. We see how different factors influence blind participants' willingness to share study data as they assess risk-benefit tradeoffs. The majority support sharing of their data to improve technology but also express concerns over commercial use, associated metadata, and the lack of transparency about the impact of their data.  These insights have implications for the development of responsible practices for stewarding accessibility datasets, and can contribute to broader discussions in this area.

\end{abstract}



\maketitle
\section{Introduction}

With advances in artificial intelligence (AI), there is a potential for emerging technologies to improve the lives of people who experience barriers to inclusion such as the disability community. We have seen many efforts in this direction. For example, many AI-infused\footnote{We follow the term used by Amershi \etal~\cite{amershi2019guidelines} referring to having ``features harnessing AI capabilities that are directly exposed to the end user.'' } assistive applications for supporting blind\footnote{In this paper, we adopt identity-first language (\eg, ``blind people'' and ``disabled people'') instead of person-first (\eg, ``people with visual impairments'' and ``people with disabilities''). We acknowledge that there are active and ongoing discussions around the two with some of them captured in Sharif \etal~\cite{sharif2022should}.} people, the community of focus in this work, employ computer vision for better access to physical and virtual spaces~\cite{microsoft2017seeing,google2019lookout,bigham2010vizwiz,kacorri2017people}. However, the potential benefits may not be realized if the data used to build these systems do not represent the end users and the contexts within which they operate~\cite{morris2020ai,guo2019toward,whittaker2019disability}. On the contrary, they may harm.
Yet, the majority of large computer vision models are trained on photos taken by sighted people~\cite{ren2010figure,fathi2011learning,russakovsky2015imagenet}, performing poorly on photos taken by blind users~\cite{brady2013visual, kacorri2017people,tseng2022vizwiz}, a gap that is only increasing~\cite{cao2022whats}.

Despite their critical role, researchers have identified myriads of challenges in collecting and sharing accessibility datasets~\cite{blaser2020why,morris2020ai,abbott2019local,trewin2018ai}. Primary barriers are privacy and ethical considerations to protect those represented in the data~\cite{treviranus2019value,hamidi2018should,abbott2019local,morris2020ai}. Collecting
data from small populations increases the risk of re-identification, which can amplify concerns for further discrimination pertaining to sensitive disability status~\cite{abbott2019local,morris2020ai}. Sharing accessibility datasets also poses risks of data abuse and misuse without proper laws and regulation enforcement (\eg, building a hiring algorithm that can make biased decisions based on disability~\cite{engler2019some}). 

With this paper, we contribute to discussions around increasing the availability of accessibility datasets by surfacing the motivating and challenging factors involved in data sharing decisions of disabled people. We focus on the blind community and image data, a challenging scenario where those contributing the data may not be able to inspect them when deciding to share. 
We designed a pair of studies that aim to surface blind participants’ perspectives on data sharing in a situated, rather than simulated (\eg,~\cite{park2021designing,mcnaney2022exploring,hamidi2018should}), context. To achieve this, we teamed up with researchers who were interested in evaluating an AI-infused application in the homes of blind participants. The application, a teachable object recognizer~\cite{kacorri2017people}, was deployed on smartglasses. Blind participants took photos and used them to finetune a computer vision model and test its performance. The team was interested in sharing study data with the broader research community and was looking at best practices. Situated in this context, we designed a semi-structured interview as a follow up. Typically within the span of a few days, 13 blind participants both (i) performed data capturing activities, and (ii) were interviewed on their perspectives towards sharing their study data via a public AI dataset.

We found various factors that could play into blind participants' willingness to share their study data (\ie, photos and labels of objects), revealing the need for better assessment of benefits and risks. Many perceived potential risks such as re-identification as minimal and supported sharing practices to improve AI-infused technology for greater benefit to both disabled and non-disabled people. Yet, they were hesitant to contribute their data for commercial purposes and companies handling the use of their data, due to notions of ``distrust'' even though almost all frequently used AI-infused applications (\eg SeeingAI, SuperSense, and Lookout) to read text and identify objects or shared their camera view with sighted helpers (\eg via Aira and BeMyEyes).
They also expressed concerns for sharing demographic metadata (\eg, age, gender, race) along their study data, due to not only privacy and safety but also the ambiguity of its value for building AI-infused technology. This suggests that the process of collecting and stewarding accessibility datasets requires greater transparency of data use, especially to challenge inclusivity issues in AI fairness~\cite{kamikubo2022data,buolamwini2018gender}. Some participants showed further interest in learning about the potential impact of their data, an option that is neither supported in current informed consent processes nor is there a way to practically implement it yet.

Our intention is to bring potential data contributors from the disability community to the forefront of data sharing discussions. We acknowledge that our focus on a specific population and context limits the generalizability of our insights. To overcome some of these limitations, we carefully connect and contrast our observations with existing literature. More so, we incorporate prior questionnaires (\eg, Park \etal~\cite{park2021designing}) with the disability community on related topics. To facilitate replicability, we share our questionnaire with expanded scenarios and interview questions (more than 80\% new content). We see the main contribution of this work being empirical. By investigating data sharing from the perspectives and experiences of blind people, we contribute to the larger call-to-action for the research community, industry, and policy makers in shaping future data practices that are inclusive of disability. We also see how our approach of eliciting participants' perspectives before asking them to decide on whether they want their study data to be shared and how, could be leveraged by future researchers who want to engage participants in decisions around sharing of their study data. 
\section{Related Work}
In this section, we cover prior work on creating and sharing accessibility datasets to provide a clear picture of the challenges and significance. We also extend to current efforts in informing data practices across different disciplines, with few studies exploring how disabled people view their data being sourced and used.

\subsection{Accessibility Datasets}
    The need for data is growing in the field of accessibility, especially for accelerating innovation around assistive technology~\cite{kacorri2017teachable,morris2020ai,bragg2019sign}. Notable examples include photos taken by blind users to build object recognition applications~\cite{lee2019hands,zhong2013real,sosa2017hands} and sign language videos from Deaf signers to train machine translation applications~\cite{hassan2020isolated,huenerfauth2014release}. To facilitate the discovery and re-use of currently available data in this space, Kacorri \etal~\cite{kacorri2020incluset, kacorri2020data} put together a collection of accessibility datasets sourced from disabled people over the last decade that could be leveraged for training and evaluating machine learning models. However, dataset availability is found to be sparse across different communities of focus~\cite{kamikubo2021sharing}, with challenges in data diversity persisting for greater inclusion of marginalized communities~\cite{kamikubo2022data}. More so, discussions around unique challenges for data collection involving disability communities are ongoing~\cite{blaser2020why,morris2020ai,sears2011representing,trewin2018ai,bragg2020exploring}. For example, Blaser and Ladner~\cite{blaser2020why} raised issues with inconsistent measures of how disability is elicited, making it difficult to aggregate and combine different data sources to facilitate large-scale datasets. 
    
   We have seen efforts to address the lack of larger, more diverse datasets in the field. Some leverage crowdsourcing or telemetry data collection methods~\cite{bragg2022exploring,bot2016mpower}, while others deploy assistive applications (\eg, VizWiz~\cite{bigham2010vizwiz}) in the real-world~\cite{gurari2018vizwiz,kacorri2016supporting,massiceti2021orbit}. Indeed, these strategies have complemented data contributions from certain communities of focus; those sourced from the blind community typically include larger numbers of contributors compared to other disability communities~\cite{kamikubo2021sharing}. Blind people have been early adopters of technologies, often in the context of taking photos of objects/scenes to access visual information~\cite{bigham2010vizwiz,adams2013qualitative}. However, given that blind people cannot inspect their data such as the photos they took, it is left to data stewards to protect the privacy and safety of those represented in the datasets~\cite{gurari2018vizwiz,theodorou2021disability}. Thus, we see the opportune involvement of this community to discuss how to ethically contribute data.
        
    \subsection{Tensions in Data Sharing}
    Kop~\cite{kop2021machine} cited data sharing as an essential practice for a successful AI ecosystem in analyzing and processing high-quality training datasets. In general, many academic disciplines and industries have seized the opportunity to promote open datasets~\cite{lyle2014openicpsr, calzolari2010lrec, microsoft2018microsoft,awssearch} to drive innovation or create new knowledge and shared resources. The health community is not an exception (\eg, using patient data to improve care and outcomes~\cite{luo2020interrelationships, walport2011sharing, verhulst2014open,mishra2018supporting}). In some venues, researchers are required to submit data sharing statements for clinical trials along with their manuscripts~\cite{taichman2017data}. Even so, such data sharing schemes have raised issues as data subjects' preferences and control are rarely addressed, with their participation limited in governance structures for sharing medical data~\cite{krutzinna2019ethical}. Advances in pervasive and wearable technologies also bring attention to enabling users to track and share self-collected data via health apps~\cite{shirazi2013already, kim2017omnitrack, luo2019co, mentis2017crafting}. At the same time, scholars have warned of the risks including privacy and security surrounding the use of data by third parties which is not always transparent~\cite{tangari2021mobile,vitak2020trust,shilton2009four}. 
    
    Similar conversations are seen in accessibility, calling for a careful balancing act in how data are shared~\cite{abbott2019local,whittaker2019disability,kacorri2016data,theodorou2021disability}. There are many ongoing issues with disability-inclusive data; they are highly sensitive and can raise concerns for privacy and data protection~\cite{abbott2019local,morris2020ai,trewin2018ai}. While such concerns are prevalent across disciplines~\cite{wacharamanotham2020transparency,fiesler2019ethical}, the consequences are severe in accessibility, with the risk of re-identification along the potential for discrimination~\cite{abbott2019local,whittaker2019disability,nakamura2019my}. Data sharing also raises ethical considerations for data re-use, when it could lead to abuse and misuse outside of the original intention~\cite{whittaker2019disability}. Considering the possibility of algorithms to detect disability status~\cite{white2018detecting}, accessibility datasets could exacerbate further bias and marginalization through systems built~\cite{morris2020ai}.

\subsection{Data Contributors in Future Data Practices}
All the concerns related to potential risks of data sharing practices make discussions and frameworks around data ethics more pressing. Recently, there is a body of work within and beyond accessibility to involve potential data contributors in the data collection or sharing contexts to inform future practices~\cite{park2021designing,mozersky2020research,shah2019motivations,mcnaney2022exploring,gilbert2021measuring,nicholas2019role,hamidi2018should}. In addition, these prior efforts surfaced a variety of situational and contextual factors that could influence the contributors' judgments of concern and risk. For instance, while many disabled people were open to contributing data to an AI dataset with the prospect of helping the disability community, they were hesitant depending on the data type that could be more or less personal~\cite{park2021designing}. Meanwhile, Shah \etal~\cite{shah2019motivations}, in a biomedical domain, found that with whom data are shared played a role in their judgments than what data types are shared. Mozersky \etal~\cite{mozersky2020research} also observed a sense of trust towards researchers, receiving broad support for data contributions in qualitative research. In contrast, McNaney \etal~\cite{mcnaney2020future} identified fears and concerns around how commercial companies might use health data (\eg, targeted advertising). Privacy and security concerns were common themes across these research areas, yet still depended on a number of elements including represented populations (\eg, visible vs hidden disability groups~\cite{park2021designing}) or awareness through consent~\cite{gilbert2021measuring}.

Our research complements previous work that has highlighted multifaceted motivations and concerns relating to data sharing. It also prompts the unaddressed questions concerning how accessibility datasets should be sourced and used to drive an AI ecosystem, when data contributors' broader views on data sharing are often contextual and situationally dependent, such as impacted by data types, research or application domains, or data use purposes. In this work, we add new dimensions to these discussions by enabling potential data contributors to reflect on the contexts of sharing data sourced from settings where the technology is being deployed (\ie, home). Furthermore, prior efforts investigating concerns of disabled people to inform better technical and legal frameworks for data stewardship (\eg,~\cite{park2021designing,mcnaney2022exploring}) are conducted with regards to simulated datasets and environments. Prior literature has warned about the impact of direct vs. indirect experiences on the development of knowledge, attitudes, and behavior~\cite{duerden2010impact}, which can engender differing opinions on issues such as privacy~\cite{lee2020pedestrian} or risk beliefs~\cite{viscusi2015relative}. Conversations around data sharing practices related to accessibility need to be further attuned to the communities of focus, with the blind community being centered in this work, capture the extent to which implications can be drawn about how they would want data sharing to occur in the real world \eg, while participating in a user study or engage with technology in their homes.

\section{Methods}

To have direct conversations with the blind community on how data that they may contribute should be shared via an AI dataset, we consider a cross-sectional study design with blind people who have been exposed to a novel AI-infused assistive application and asked to evaluate it. Specifically, we teamed up with researchers who were developing an object recognition application on smartglasses and deploying it in the homes of blind participants. The team was interested in best ways to share the study data used in their analysis (\eg, photos of objects) both for the purpose of replicability but also for motivating future work in this area \eg use the data to train or test future machine learning models.

\begin{figure*}[h]
    \centering
    \includegraphics[width=0.8\textwidth]{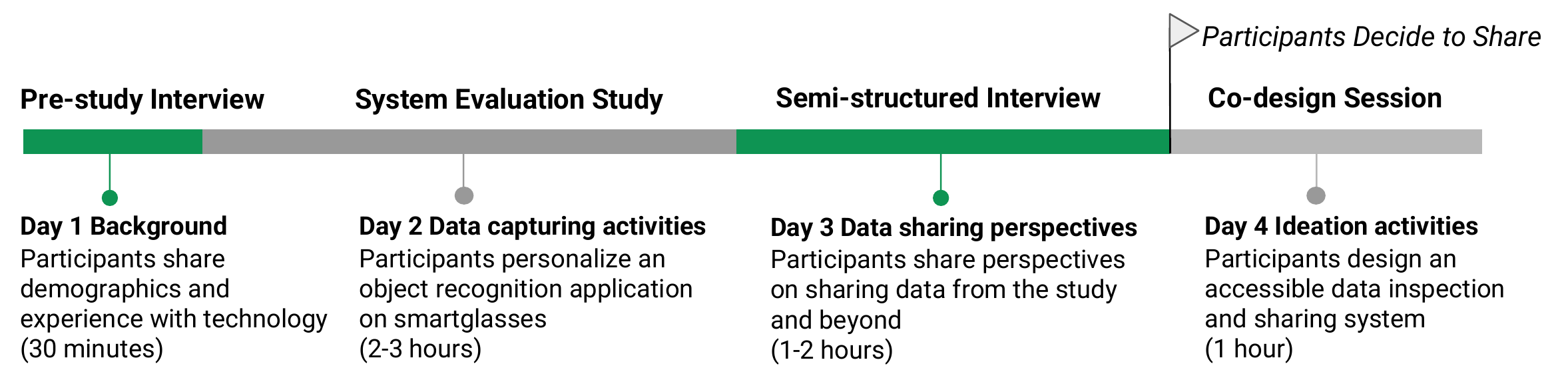}
    \caption{We present findings from a semi-structured interview contextualized within a larger study that includes a short interview on demographics and experiences with technology, a system evaluation study, and a follow up co-design session.}
    \label{fig:study_design}
    \Description[This figure shows four different sessions of the entire study.]{From left to right, pre-study interview, system evaluation study, semi-structured interview, and co-design session are shown. During pre-study interview, participants share demographics and experience with technology, and it took on around 30 minutes. In system evaluation study, which took 2 to 3 hours, participants personalize an object recognition application on smartglasses. In semi-structured interview that took 1 to 2 hours, participants share perspectives on sharing data from the study and beyond. In the last one, co-design session, participants design an accessible data inspection and sharing system, and it took on average an hour.}\vspace*{-2pt}
\end{figure*}

As shown in Figure~\ref{fig:study_design}, this pairing of studies allows us to surface blind participants' perspectives in a situated, rather than simulated, context---enabling their decisions about sharing study data and preferences for data control to be reflected in a real-world sharing context. Unlike previous research where the findings are synthesized as implications for broader research practices~\cite{shah2019motivations, park2021designing}, we shift to active participation of potential data contributors to guide the data sharing process that the development team of the AI-infused application will go through to release the AI dataset from the study. Also, their user study provided great context to gain empirical insights that would not be as generalizable from an in-lab study; typically, datasets have value when collected in naturalistic settings (\ie, where the technology is meant to be deployed such as in people's home~\cite{theodorou2021disability}). More so, privacy risks are more heightened in these settings~\cite{heumann2016privacy}. The camera in the smartglasses may capture the home environment in the background potentially revealing more about the person and their life.

The larger study spanned multiple days within the May-September 2022 period. It started with a 30-minute long Zoom call to capture participant demographics, attitudes, and experience with technology. Some of this information is presented in Section 3.1 to provide context for our analysis.  A day or two later, participants joined a longer study where they performed remotely from their homes a series of data capturing activities presented in Section 3.2. Usually within a week, they participated via Zoom in a semi-structured interview, the focus of this paper. Any time after this, participants could indicate to the researcher, the data stewards, their decision around sharing of their study data as a response to an email they received. The need to make a decision was communicated with participants early on in the study and was included in their consent forms. 
Participants could also opt to join a follow up co-design session, typically conducted a few weeks later. This last session focused on the design of an accessible data inspection interface that allowed participants to go over the photos they collected. Some opted to confirm their decision on data sharing after this session.

We briefly describe the system evaluation study as it provides the critical context of the data capturing activities that the participants engaged in. However, the specifics of the technological contributions and findings from that study are beyond the scope of this paper. Data from the co-design session and participants' final decisions to share remain yet to be analyzed. 
Our semi-structured interview is described in detail in Section 3.3 followed by our analysis approach (Section 3.4),  which aims to reveal (i) the factors related to one's decision to share study data as well as  (ii) potential ethical, legal, and technical implications for mitigating risks and concerns related to data sharing.
To facilitate future research on exploring perspectives from other communities or in other AI data sharing contexts we make our scenarios, context probes, and questions available at \textbf{\textit{\url{https://go.umd.edu/datasharing_questionnaire}}}.

\subsection{Recruitment and Participants}
Our pre-study interview (day 1) captured demographic information including age, gender, education, and occupation as well attitudes and experiences with technology that are relevant to the AI-infused application being evaluated. At the end of the semi-structured interview (day 3), participants were given an option to choose the information that they do not wish to be made available on publication. Reflecting their consent, Table~\ref{tab:participant} shows the demographics for our 13 blind participants. Nine were totally blind and four were legally blind. On average, participants were 53.46 years old (STD=14.94). Out of the responses we received, six self-identified as women and five as men. Participants were compensated \$15 per hour, with an average of 2.75 hours (STD=0.5) spent for the experimental study (including the opening demographic and experience questionnaires) and 1.72 hours (STD=0.33) for the interview on data sharing.

\renewcommand{\tabcolsep}{1pt}
\begin{table}[t]
\caption{Self-reported participant information including participant ID, vision level, age, gender, education, occupation, and AI familiarity on a 4-point scale: 1 = not familiar at all (have never heard of it), 2 = slightly familiar (have heard of it but don’t know what it does), 3 = somewhat familiar (have a broad understanding of what it is and what it does, 4 = extremely familiar (have extensive knowledge). A dash (-) indicates that the participant did not consent to disclose.}
\label{tab:participant}
\small\renewcommand{\tabcolsep}{1pt}
\centering
\begin{tabular}{ | >{\centering\arraybackslash}m{0.08\linewidth} 
|  >{\centering\arraybackslash}m{0.12\linewidth}
| >{\centering\arraybackslash}m{0.08\linewidth} 
| >{\centering\arraybackslash}m{0.13\linewidth} 
| >{\centering\arraybackslash}m{0.16\linewidth} 
| >{\centering\arraybackslash}m{0.19\linewidth} 
| >{\centering\arraybackslash}m{0.19\linewidth} |}

	\hline
PID & Vision & Age & Gender & Education & Occupation & AI Familiarity  \\
\hline
P1 & legally blind & 59 & man & bachelor & budget analyst & somewhat  \\
\rowcolor{lightgray}
P2 & totally blind & - & - & -& - & somewhat  \\
P3 & totally blind & 72 & woman & bachelor & retired & extremely  \\
\rowcolor{lightgray}
P4 & totally blind & 67 & man & master & retired & slightly  \\
P5 & totally blind & 48 & man & master & social worker & somewhat  \\
\rowcolor{lightgray}
P6 & legally blind & 32 & woman & bachelor & operations & somewhat  \\
P7 & totally blind & 53 & woman & master & retired & somewhat  \\
\rowcolor{lightgray}
P8 & totally blind & 67 & woman & master & retired & somewhat \\
P9 & totally blind & - & - & - & IT solution architect & somewhat  \\
\rowcolor{lightgray}
P10 & totally blind & 37 & woman & law school & attorney & somewhat  \\
P11 & legally blind & 52 & woman & master & application developer & somewhat  \\
\rowcolor{lightgray}P12 & legally blind & 71 & man & post-bachelor & retired & somewhat \\ 
P13 & totally blind & 67 & man & some college & computer technician & extremely  \\
\hline
\end{tabular}
\end{table}

\begin{figure}[t]
    \centering
    \includegraphics[width=1.05\linewidth]{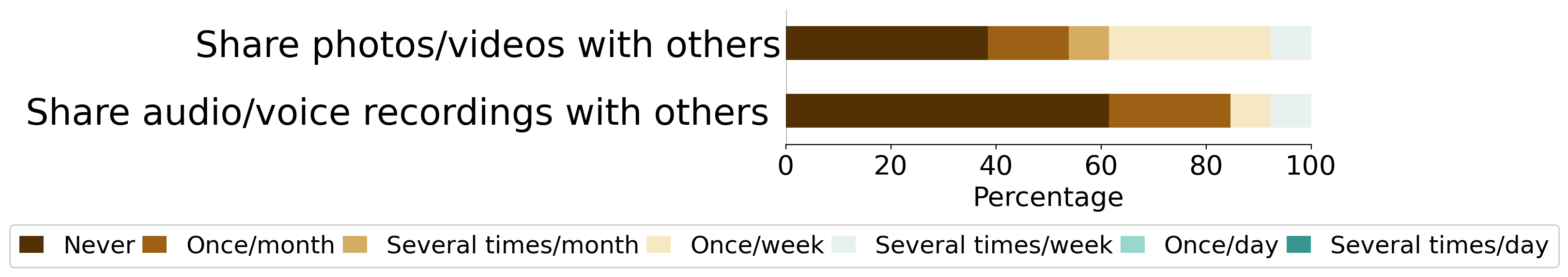}
    \caption{Responses to how frequently participants shared data such as photos, videos, and audio recordings with others. }
    \label{fig:sharing_exp}
    \Description[This figure shows stacked bars for participants' responses to how frequently they shared photos/videos or audios/voices with others.]{38.5\% of the participants never shared photos/videos. 15.4\% of them shared photos/videos once a month, and 7.7\% answered sharing such data several times a month. 30.8\% of them reported sharing photos/videos once a week while 7.7\% mentioned sharing such data several times a week. Regarding how frequently they shared audio/voices with others, 61.5\% of the participants never shared such data, while 23.1\% said they shared such data once a month. 7.7\% of them reported sharing audio/voices once a week while the rest (7.7\%) answered sharing such data several times a week.}
\end{figure}

To better contextualize our findings, we report participants' technology use and attitudes responses. All but two (P1, P11) participants reported using assistive applications for accessing visual information in their surroundings such as Seeing AI (n=8), Aira (n=4), BeMyEyes (n=4), VoiceDreamReader (n=3), BlindSquare (n=2), BlindShell (n=1), BeSpecular (n=1), CurrencyReader (n=1), ColorIdentifier (n=1), Google Lookout (n=1), KNFBReader (n=1), and Supersense (n=1). P1 and P11, who are legally blind, typically relied on built-in camera features such as magnification \textit{``to read bus signs''} (P1) or \textit{``to check their [own] appearance''} (P11). As shown in Figure~\ref{fig:sharing_exp}, more than half of the participants (n=$8$) reported sharing photos or videos with others at least once a month. Often this was done to get sighted help from family and friends for recognition tasks. Some (n=5) never did. Sharing of voice and audio recordings was often less frequent.

\begin{figure}[h]
    \centering
    \includegraphics[width=\linewidth]{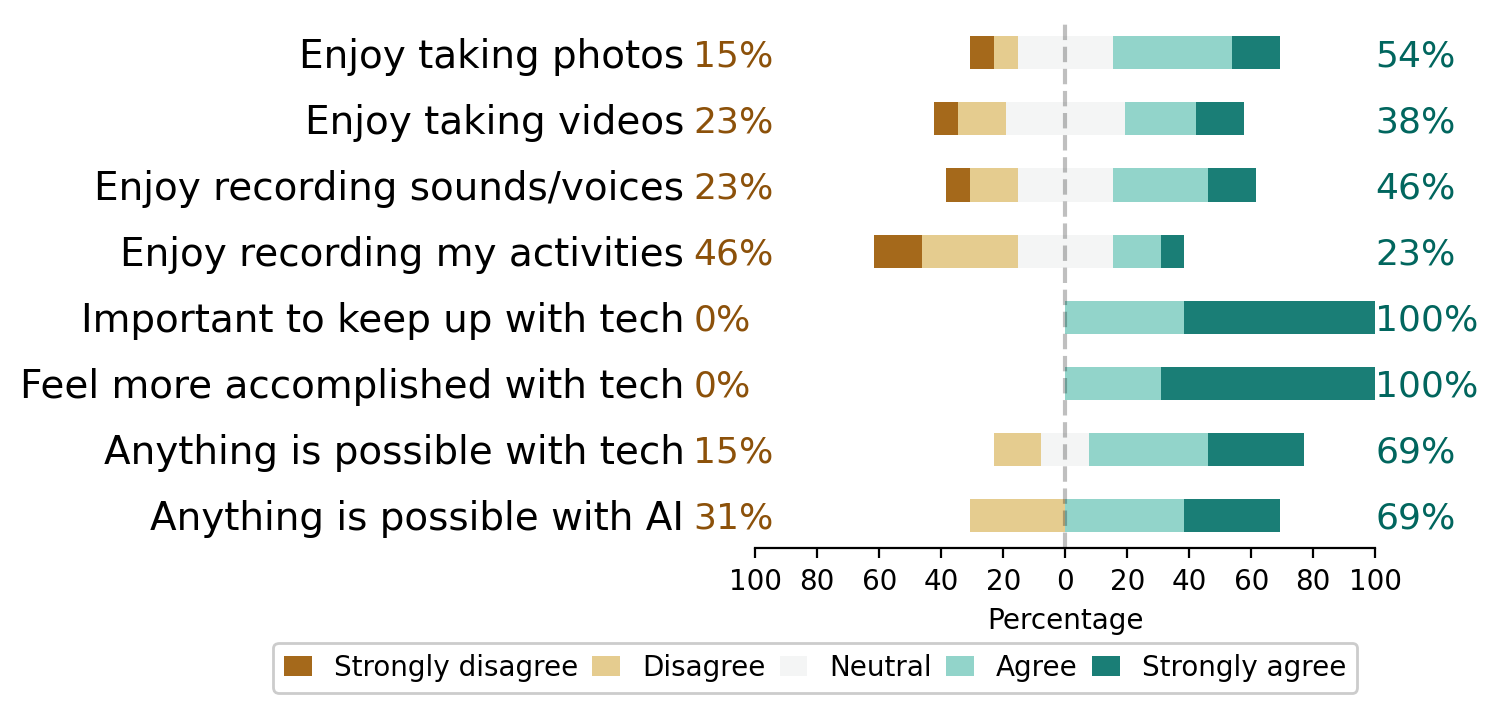}
    \caption{Participants' attitudes towards AI, technology, and data capturing such as tracking their activities, recording sound or voices, and taking videos and photos.}
    \label{fig:tech_attitude}
    \Description[This figure shows stacked bars for participants' attitudes towards AI, technology, and keeping up with it.]{Regarding the statement, "I enjoy taking photos", strongly disagree=7.7\%; disagree=7.7\%; neutral=30.8\%; agree=38.5\%, strongly agree=15.4\%. Regarding "I enjoy taking videos", strongly disagree=7.7\%; disagree=15.4\%; neutral=38.5\%; agree=23.1\%; strongly agree=15.4\%. Regarding "I enjoy recording sounds/voices", strongly disagree=7.7\%; disagree=15.4\%; neutral=30.8\%; agree=30.8\%; strongly agree=15.4\%. Regarding "I enjoy recording my activities (e.g., steps, sitting, running)", strongly disagree=15.4\%; disagree=30.8\%; neutral=30.8\%; agree=15.4\%; strongly agree=7.7\%. Regarding "I think it is important to keep up with the latest trends in technology", strongly disagree=0\%; disagree=0\%; neutral=0\%; agree=38.5\%; strongly agree=61.5\%. Regarding "I feel that I get more accomplished because of technology", strongly disagree=0\%; disagree=0\%; neutral=0\%; agree=30.8\%; strongly agree=69.2\%. Regarding "With technology anything is possible", strongly disagree=0\%; disagree=15.4\%; neutral=15.4\%; agree=38.5\%; strongly agree=30.8\%. Regarding "With advances in artificial intelligence anything is possible", strongly disagree=0\%; disagree=30.8\%; neutral=0\%; agree=38.5\%; strongly agree=30.8\%.}
\end{figure}

Most participants were positive about the potential of AI and technology, as shown in Figure~\ref{fig:tech_attitude}. Indeed, all agreed or strongly agreed on statements such as \textit{``It is important to keep up with tech''} and \textit{``Feel more accomplished due to tech.''}  Almost half of the participants (n=$6$) disagreed or strongly disagreed that they enjoy recording their activities. When it came to videos, photos, and sounds or voices, some disagreed (n=$2$, n=$3$, and n=$3$, respectively).

\subsection{Evaluation Study: Data Capturing Context}

During this session, participants evaluated a working prototype of a teachable object recognition application deployed on smartglasses. The term \textit{teachable} refers to the fact that participants could \textit{teach} the underlying machine learning model to recognize objects of their choice by providing a few photo examples of those objects as well as labels (\ie object names that are spoken upon recognition). The application is meant to facilitate personalization as it promises a better fit for real-world scenarios by significantly constraining the machine learning task to a specific user and their environment. It does not require any machine learning expertise from the user. More so, the interactive nature of teachable applications could help users uncover basic machine learning concepts and gain familiarity with AI (\eg,~\cite{hitron2019can, queiroz2020ai, carney2020teachable, hong2020crowdsourcing, dwivedi2021exploring}). We see similar evidence for studies with blind participants \cite{kacorri2017people, hong2022blind} where participants reflect on the value of diversity in training data. Thus, the data capturing tasks in this study seemed a great fit providing a realistic data contribution scenario while exposing participants to the value of AI data for training and testing.

\begin{figure}[t]
\includegraphics[width=.5\linewidth]{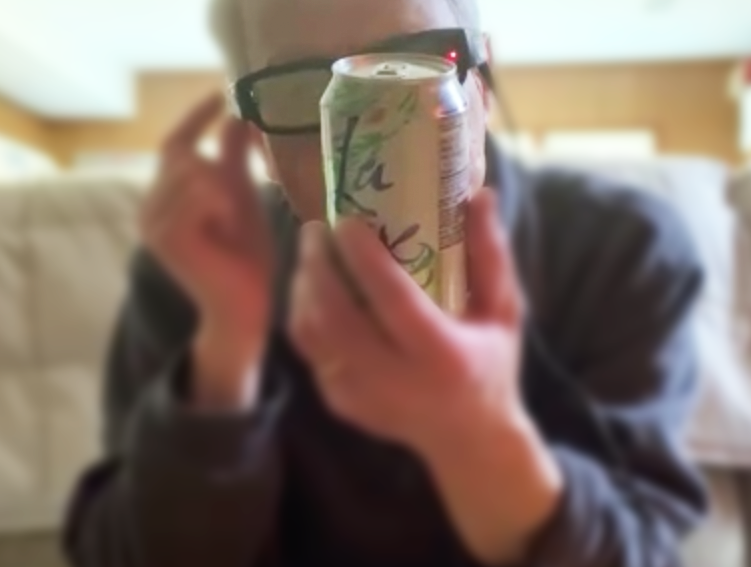}
    \caption{A blind participant using an object recognition application installed on smart glasses, which they use to take multiple photos of a soda can. The object recognition model is fine-tuned on these photos to personalize the application.}
    \label{fig:glasses}
    \Description[A participant wearing smartglasses.]{A person wearing smartglasses is holding a soda can and putting it in front of the smartglasses. Tapping the smartglasses, the person is capturing a soda can with smartglasses camera.}
\end{figure}

\begin{figure}
    \includegraphics[width=\linewidth]{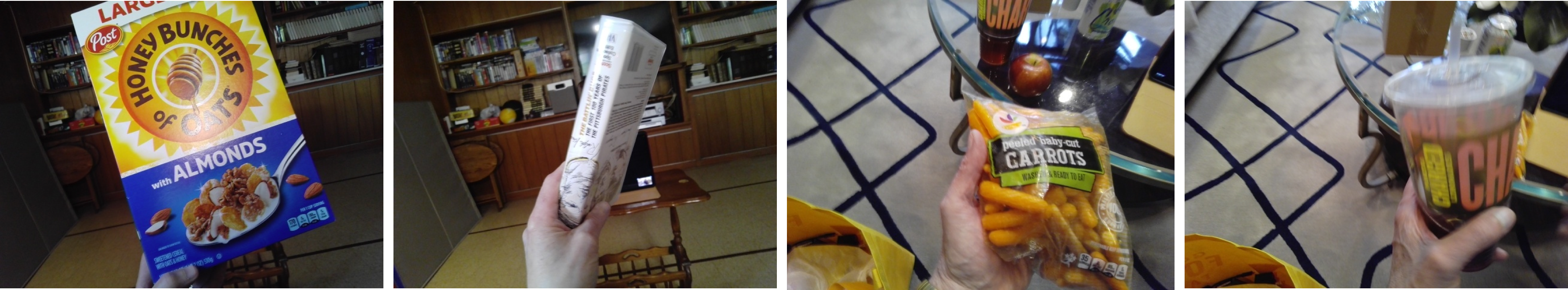}
    \caption{Examples of photos of objects captured and labeled by our blind participants (from left to right) as: ``Cereal,'' ``Videotape,'' ``Carrots,'' and ``Panera Cup.'' }
    \label{fig:objects}
    \Description[This figure shows four different objects in four different photos, respectively.]{From left to right, a photo of cereal, a photo of videotape, a photo of carrots, and a photo of plastic cup are shown. All these photos show that blind participants are holding those objects with one hand when capturing those objects with smartglasses camera.}
\end{figure}

In detail, participants were instructed to find a sitting area in their homes where they feel comfortable setting up the laptop with the Zoom call and interacting with the stimuli objects while wearing a pair of smartglasses. To familiarize themselves with the smartglasses and the application, they first practice taking photos and providing labels for 2 objects. The research team provided practice objects. 
As shown in Figure~\ref{fig:glasses}, participants used the touchpad on the smartglasses located along the temple to navigate the menu and trigger the photo taking and labeling functions, which are communicated through text-to-speech. Voice commands were also implemented for entering, correcting, and confirming the object label. Once familiar with the system, participants are asked to complete data capturing activities that involved taking multiple photos per object for a total of 6 objects and providing associated labels for training and evaluating a classifier. Half of these objects were stimuli engineered to be visually distinct but nearly identical by touch (\ie different bags of snacks). They were fixed across all participants and were provided by the research team. The rest of the objects were up to the participants; they could choose anything in their home. Typically, they opted for somewhat similar objects to the stimuli including everyday products, as shown in Figure~\ref{fig:objects}.
Participants answered questions related to their experience in between the data capturing tasks and at the end.

By the end of this session, each participant generated on average 222 photos (STD=59.9) across the 6 object labels. Both the photos and the labels generated from these activities were referred to the participants as \textit{``your study data''} throughout the communication with the research team.  This wording aimed to situate participants in the context of sharing via a public AI dataset. 
 
\begin{table*}[t]
    \renewcommand{\arraystretch}{1.1}
    \small
    \caption{Scenarios and context probes to guide our semi-structured interviews.}
    \label{tab:scenarios}
    \begin{tabular}{ | @{\hspace{5pt}}p{12.5em}@{\hspace{5pt}} | @{\hspace{5pt}}p{16em}@{\hspace{5pt}}| @{\hspace{5pt}}p{16em}@{\hspace{5pt}} | @{\hspace{5pt}}p{12.5em}@{\hspace{5pt}} | } 
    \hline
    \textbf{Part 1 Benefits} & \textbf{Part 2 Risks} & \textbf{Part 3 Contexts} & \textbf{Part 4 Mitigating Risks} \\ 
    \hline
        \textit{\textbf{Broader impact}}\newline
        \textit{``Datasets may not just benefit blind people via assistive tech but really anyone who may interact with a smart app or appliance...Imagine a robot that one can ask to fetch things for them. Your data could be used to make such robots function better for everyone.''}
    &
        \textit{\textbf{Re-identification}}\newline
        \textit{``People can do some guesswork. For example, they could see in the publication that Participant 3, who is someone around 40 years old, identifies as male and uses a guide dog, took a bunch of photos of t-shirts associated with specific events...it turns out they happened to know someone who fits the description.''}\newline
        
        Non-consenting disclosure:
        \textit{``Would you be concerned if others (your friends or employers) might find out that your data is included in the AI dataset?''}\newline\newline
        \textit{\textbf{Data abuse/misuse}}\newline
        \textit{``Imagine someone building an algorithm that given an image it can figure out whether a blind person took it. They may not be able to guess who but they may be able to guess if one has a disability or not without their consent for disclosure.''}
    &
        \textit{\textbf{Type of modality, object,}}\newline
        \textit{\textbf{environment, demographics}}\newline\newline
        \textit{\textbf{Data access methods}}\newline
        Open access:\newline
        \textit{Anyone on the Internet can download study data.}\newline\newline
        Authenticated access:\newline
        \textit{Anyone who registers their information such as name, email address, organization can download study data.}\newline\newline
        Consented access:\newline
        \textit{Anyone who registers their user profile and agrees to the terms of use can download study data.}\newline\newline
        Authorized access:\newline
        \textit{Anyone who registers their use profile, agrees to the terms of use, as well as submit the purpose of data use for approval can download study data.}
    &
        \textit{\textbf{Stakeholders}}\newline
        Data stewards:\newline
        \textit{Those putting together datasets}\newline\newline
        Policy-/law-makers:\newline
        \textit{Those making policies or regulations.}\newline\newline
        Data sharing entities:\newline
        \textit{Those operating methods of access and sharing}\newline\newline
        Data contributors:\newline
        \textit{Those contributing data to an AI dataset} \\
    \hline
    \end{tabular}
\end{table*}

\subsection{Semi-structured Interview: Reflections} 

The interview was conducted via Zoom and was audio-recorded for analysis. We used scenarios and context probes (Table~\ref{tab:scenarios}) to guide the interviews, with 15\% of the questions either re-used or expanded from the questionnaire shared by Park \etal~\cite{park2021designing}. The interviews were structured as follows:\\
\textbf{Part 1 Benefits.} We first asked our participants about their understanding of any benefits in sharing their study data via a public AI dataset. We then presented a scenario describing potential benefits (Table~\ref{tab:scenarios} Part 1) to probe their willingness for data sharing. We followed up with a few questions to gauge their motivations. \\
\textbf{Part 2 Risks.} We asked participants about their understanding of any risks in sharing their study data via a public AI dataset. We then presented two scenarios describing potential risk cases raised in the field: (i) re-identification of individuals from anonymized datasets~\cite{morris2020ai} and (ii) data abuse/misuse given that ``it’s difficult to \nobreak ensure that data won’t be reused in ways that could cause harm''~\cite{whittaker2019disability}. We followed up with a few questions to gauge their concerns including non-consenting disability disclosure. \\
\textbf{Part 3 Contexts.} We then explored the contexts that may impact their decision to share data (within and beyond this study). We asked about their level of comfort depending on the type of modality, object, environment, and demographic information being shared. Similar to the design of validated questionnaires for measuring attitudes (\eg,~\cite{findler2007multidimensional}), we employed a 7-point Likert scale and asked for rationale where it was appropriate. For example, we asked ``On a scale of 1 to 7, where 1 is not comfortable at all and 7 is very comfortable, please rate your level of comfort with sharing photos of \textit{medication}?'' We explored different types of objects accroding to object instances chosen by blind participants to personalize an object recognizer~\cite{kacorri2017people}, and the types of demographic information were informed by the metadata of accessibility datasets~\cite{kamikubo2022data}. For each question, we probed conditions on how others can access the data, ranging from being openly accessible to anyone to accessible only by those who are authorized (Table~\ref{tab:scenarios} Part 3), to gain a broader understanding of the factors that our participants would consider when reasoning about sharing their data. \\
\textbf{Part 4 Mitigating Risks.} Separately, we explored possible approaches to reduce their concerns surrounding the potential risks and challenges discussed in Part 2 and 3. We first asked a set of open-ended questions on actions that our participants want to see from different stakeholders (Table~\ref{tab:scenarios} Part 4)---\eg, ``What actions would you like to see from \textit{data stewards}, those collecting the data like for example our team, against such risk scenarios that might influence your decision about sharing your study data like the photos and labels of objects?'' To concretize the discussion on risk mitigating strategies, we later followed up with existing data sharing purposes, methods, or regulations and asked the participants to rate these approaches by their level of comfort or acceptance with sharing their study data on a 7-point Likert scale.

\subsection{Analysis}
We transcribed the audio recordings of the interviews which included both open-ended questions (Part 1, 2, and 4) and Likert scale questions with shorter qualitative responses (Part 3 and 4). For qualitative data, we applied a reflexive thematic analysis~\cite{braun2006using,braun2019reflecting} to explore our interpretations on data. One member of the research team went through the process for data familiarization, inductive coding, and development of initial themes~\cite{braun2021can, braun2021one}. The research team then reviewed the themes, followed by discussions to conceptualize them as unifying concepts~\cite{braun2019reflecting}. For example, we explored motivation and risk factors that were brought from other science fields to conceptualize the themes capturing patterns in how participants perceived data sharing. While the controversial discussions on ``quantitizing'' qualitative data exist~\cite{maxwell2010using, sandelowski2001real,sandelowski2009quantitizing}, we report the number of participants whose responses included such themes. We adopt ``quasi-statistics''~\cite{becker1970field} only to support statements such as ``many'', ``some'', or ``a few'' in the description of the qualitative data; percentages are not used given the small sample size which may lead to misinterpretation of the analysis.

For the quantitative responses, we used descriptive statistics to caption emerging patterns and tendencies. There are tensions in the field regarding how Likert scales should be analyzed~\cite{jamieson2004likert}. As Likert scale ordinal data do not follow a normal distribution, using the mean can be of limited value as a measure of central tendency~\cite{barry2017not}. Instead, we adopt frequencies (count of responses for each point of the scale) and median as recommended by~\cite{sullivan2013analyzing}. Anticipating a small sample size and having a large number of questions, we did not pre-register any hypothesis for inferential statistics.

\label{sec:limit}
\subsection{Limitations} 
Our methods come with limitations. We highlight them here to help one better interpret the findings that follow. 

\paragraph{Recruiting participants} Our study involves a small sample, though it is reflective of local standards at CHI~\cite{caine2016local}. We employ non-probability sampling, a combination of convenience, voluntary response, and snowball sampling. Some participants might have previously joined studies by our research institution. This could bias perspectives; they can be trusting of the team or institution and more willing to contribute their data with fewer concerns. The degree of concerns can be also dependent on the awareness of existing social and political issues around data --- \eg, a few participants who self-reported working in the IT or security field expressed more negative perspectives regarding data sharing.

\paragraph{Eliciting Responses.} Prompting scenarios are considered effective in capturing participants’ opinions and perspectives~\cite{schoenberg2000using,jackson2015using}, but they can also impact responses. Within a category, we typically ask questions before and after a scenario is given. But scenarios can have an effect on the next categories of questions. For example, more potential risk cases to reflect on (Part 2 Risks) might trigger more concerns, affecting participants' level of comfort with sharing (Part 3 Contexts of Data Sharing). There could be also an order effect as our interview questions proceeded from benefits to risks to underlying elements behind the decision to share; responses could differ were the conversations in a different order.

\paragraph{Generalizing findings.} Our focus on a specific disability community, country, and context makes it challenging to obtain generalizable implications.  For example, this study might not tell us much about the concerns and motivations for the Deaf community to contribute to sign language video data capturing their face, body, and background. Such limitations are not unique to our study. Adopting Nissenbaum’s notion of \textit{contextual integrity}~\cite{nissenbaum2004privacy}, Barkhuus questions altogether ``\textit{the viability of obtaining universal answers in terms of people’s ‘general’ privacy practices}''~\cite{barkhuus2012mismeasurement}. \\
\indent We also recognize an inevitable limitation of qualitative analysis. Our own positionality and reflexivity may bias the interpretations of the findings~\cite{berger2015now}. Therefore, the analysis is exploratory, and it would be meaningful to facilitate the transferability of findings to different communities and contexts in future research. Nonetheless, given that we are interested in contributing to conversations and initiatives on responsible practices for stewarding accessibility datasets, we make a conscious effort to connect our findings with the larger theory around privacy and data sharing as well as with prior work including other disability communities (\eg,~\cite{park2021designing,mozersky2020research,shah2019motivations}).

\section{Findings}
We summarize the findings related to the perceptions of 13 blind participants towards benefits, risks, as well as contexts of sharing their data (\ie photos and labels of objects) via an AI dataset.

\subsection{Willingness to Share Given Benefits \& Risks}
We explored whether and how benefits and risks could contribute to participants' views on data sharing. Overall, we found that many focused on the greater benefits and perceived potential risks as minimal. However, their willingness to share is related to a number of elements, which we go deeper into below. How the participants assessed benefits and risks also reflected their attitudes towards data sharing. Some considered that the benefits would outweigh the risks, and those who foresaw the risks as severe remained hesitant to share their study data despite the considerations for benefits.

\subsubsection{Benefits of Data Sharing}

When initially asked about any benefits of data sharing \textbf{without prompting}, a majority (n=10) of the participants identified instances in which sharing their study data could lead to beneficial outcomes. Often the benefits were related to the improvement and evaluation of object recognition technology (6 out of 10), to help \textit{``developers figure out what worked out and what could be improved''} (P2) and \textit{``people or companies who are into this work build from this [data] and advance it''} (P6). We saw a similar trend in~\cite{park2021designing} where disabled participants from different communities were more willing to contribute their data for future AI applications if their contribution would be a dependent factor for the success of the development. This may correspond with the motivation factors laid out by Batson \etal~\cite{batson2002four} suggesting that acting for the common good is not only driven by self-benefits (\textit{egoism}). Acting for the community (\textit{collectivism}) or for specific others (\textit{altruism})---\eg, to help scientists advance their research---are also noted as potential sources of motivations~\cite{lotfian2020framework}. 

Looking at other \textbf{unprompted} benefits reported, we further observed the interplay of factors motivating willingness to share study data. Some (4 out of 10) perceived benefits that were directed towards the user community, as referred by P5: \textit{``If you've got more people sharing the data, everybody doesn't have to build their own independent library...Libraries can kind of benefit from each other's because they have different data and that could help AI to learn something more.''} This can be seen as a \textit{collectivist} motivation and support previous anecdotal evidence where social factors come into play behind \eg, community involvement~\cite{rotman2012dynamic} or information sharing~\cite{jean2012information}. We received a few (n=2) additional comments, which in some ways resembling \textit{altruistic motivation} often driven by \textit{empathy} to help others who are ``perceived to be in need''~\cite{batson2002four}. The perceived benefits were geared to the interest of specific other blind users: \textit{``people who are much younger than me and just starting out''} (P12) and \textit{``people who were born blind''} (P13).

Participants saw even more benefits when \textbf{prompted} with a positive scenario for others (\ie, building a robot that can serve wider audiences), with a majority (n=11) of them expressing that they would be more open to sharing their study data. Some (n=4) were motivated by the next future technology, while others (n=4) considered the potential to help other disabled people \ie those experiencing mobility challenges: \textit{``They might need assistance, when you talked about fetching things, I mean, folks who are paralyzed or whatever. I could see that would be helpful to them, and I would be thrilled to be part of helping that''} (P8). Furthermore, seeking broader impacts also factored into our participants' willingness to share their data (n=4), articulating \textit{``Best type of help would benefit everyone''} (P2) and \textit{``I would be even more anxious to share. More people can benefit the better''} (P4). Following these motivations, some (n=4) expanded to other application domains where sharing their study data would benefit, for identifying objects in a shipping inventory (P3), describing photos on social media (P10), or helping with translation and second language learning (P5, P7). 

\subsubsection{Risks of Data Sharing} We subsequently explored our participants' awareness and reactions towards potential risks pertaining to data sharing.
With \textbf{no probing} at the start and within the context of their study data, only a few (n=4) participants identified potential risks in sharing through a public AI dataset.
The majority (3 out of 4) of their concerns revolved around the secondary use of data. Concerns were broad as \textit{``How do I know that if you shared it with somebody else, they would ethically treat the data?''} (P9). Some included a specific secondary use scenario such as ``targeted advertising'' that might lead to potentially negative consequences: \textit{``I don't want companies to get my information, and then they just start targeting me with advertisement, like hey, we know that you are probably blind because you sent this data...Who knows how things can be used if it's in the wrong hands''} (P6). This can become a critical factor for contributing to an AI dataset, particularly when considerations for re-use cases are lacking for existing datasets sourced from blind people~\cite{massiceti2021orbit,lee2019hands,gurari2018vizwiz}---\eg, permitting commercial and private use of the data~\cite{massiceti2021orbit}. The concerns of the fourth participant (P7) focused on privacy and location identification. They considered where the photos were taken (\ie, home) and whether the photos would be geo-tagged (though no geo-tags are embedded in their study data). 

Further, we \textbf{probed} perception towards potential risks by prompting scenarios that our participants might foresee negative consequences: re-identification (with a follow up question related to non-consenting disclosure) and data abuse and misuse. We explored how these risk elements would impact willingness to share.

\paragraph{Re-Identification.} We first provided a scenario that, even in anonymous or pseudonymous datasets, people could do some guesswork about an individual who contributed data based on released demographic information (\eg, age and gender along disability status on publication). Most (n=10) participants expressed no or minimal concerns in terms of risk of re-identification. Some (4 out of 10) responded to the scenario that it is hard to imagine what could go wrong with the information disclosed, as they described the data captured as \textit{``not really personal data''} (P1). Perhaps, they would still be willing to share photos of objects despite the risk factor prompted, as referred by P12: \textit{``I think the benefits far outweigh any minor drawbacks that could occur...If someone says, oh, there was a 71 year old guy that took pictures of Lay's potato chips, you know, if that's all that ever happens to me, I'm okay, as long as they don't take my credit cards or anything.''} This brings attention to the interaction between perceived benefits and risks. According to privacy calculus (risk–benefit analysis), privacy concerns are measured based on the perceived value of disclosing personal information relative to the perceived costs~\cite{xu2011personalization}. For example, Verheggen \etal~\cite{verheggen1998determinants} found that patients who agreed to participate in a clinical trial were likely to weigh the benefits more than the risks, whereas it was the reverse for those who declined to participate. Indeed, when looking at those who showed concerns when prompted (n=3), 2 of them did not report previously any unprompted benefits.

In response to the given scenarios, participants with concerns were strongly against sharing demographics along with their data. P9 mentioned, \textit{``Why would you even be collecting that information? It's for that reason that I don't answer demographic questions generally...You don't need to be collecting that demographic information in the first place. It's not relevant to you to the task of the AI.''}  Similarly, P2 responded, ``People being able to figure things out...that is actually part of my reserve [in sharing data].'' Thus, it is not a surprise to see these two participants opting not to include their age, gender, and education in Table~\ref{tab:participant}. While P7 shared her demographics for the context of data for this study, she also raised the importance of privacy: \textit{``Is it necessary to say that this person has a guide dog? Because you don't see a whole bunch of people walking around with a guide dog so it's easy to kind of pinpoint who that person is. So protect people's privacy.''} 

\paragraph{Non-Consenting Disclosure} When asked whether they would be concerned about others (\eg, family, friends, current or future employers) finding out that their data is included in AI datasets, all except one said no. The rationale for their response could be partially related to this specific study data. For example, P11 said, \textit{``I'm not revealing any confidential information of my workplace. And I don't care whether they see something related to me like a picture taken by me. Is that a problem? I don't think so.''} A few (2 out of 12) justified this lack of concerns by contrasting it to everyday risks: \textit{``We deal with a lot of online information that we share, or we interact with. We are not really dealing with any more risk than what we have already''} (P1). This high response agreement could also relate to the specific community. Kamikubo \etal~\cite{kamikubo2021sharing} observed that among all accessibility datasets, those sourced from the blind community, such as our participants, tend to be shared publicly and typically include larger numbers of contributors. This is in contrast to datasets sourced from communities that encompass so called ``invisible disabilities'' which are less apparent to others and perhaps more sensitive to disclosure. When asking a similar question, Park \etal~\cite{park2021designing} observed heightened concerns from people experiencing ``non-apparent forms of disabilities'' (\eg, ADHD).

The one participant who said yes to the question, added, ``especially, [I] wouldn't want potential employers seeing that, because I just didn't want them to infer anything or assume anything. So I wouldn't care about my friends and family knowing but I wouldn't want anyone else to know'' (P10). As they are a legal professional, perhaps this may be reflective of Judge Richard Posner's view towards privacy as \textit{``power to conceal information about themselves that others might use to [the individuals’] disadvantage''}~\cite{solove2008understanding}.

\paragraph{Data Abuse and Misuse} When \textbf{prompted} with a negative scenario of repurposing data (\ie, building an algorithm that can detect disability and be used against disabled people), many (n=11) participants were not overly concerned about the potential consequences. Perhaps they saw the potential risks as unimaginable and minimal. Some (4 out of 11) even responded that they would be open to sharing, as referred by P4: ``I think it's very, very unlikely, though I do think it's possible as you describe it. But, it would not change my decision [to share].'' Similar to the previous conversations about privacy, some of their rationale pertained to the specific study data and the benefit-risk tradeoff: \textit{``I can't imagine given what I took pictures of, it's that critical. You know, it doesn't bother me. I mean, it would be unfortunate if somebody kind of concluded something negative about people with disabilities or found it funny or like, laughable that they couldn't take pictures or something, but it doesn't bother me. You know, I think the good outweighs the bad''} (P8). Interestingly, the one participant who had been raising stronger concerns expressed a similar argument: \textit{``I'm not as worried about that. I do understand that you can't control what other people might use the dataset for. However, that's why it's important to be careful [about] what you collect. At that point, I'm still more concerned about the collection than what other people might use it for''} (P9).

While raising minimal concerns for the given scenario, our participants followed up with some degree of concerns around the ethics of data use. A few (n=3) of them posed mixed feelings, articulating that such data abuse and misuse risks are unavoidable in the digital world: \textit{``I can't control, you know, and nobody else can really control. If people use this information for something else, honestly, that happens everywhere. That's always a concern''} (P6). P1 reflected on the lack of concerns as \textit{``we either got numb or gave up.''} Such attitudes could be explained by their nuanced understanding of data policies. For example, camera-based assistive technologies like Aira or SeeingAI which our participants reported to use, provide no clear indication of whether and for what purpose personal visual data are shared with third parties~\cite{stangl2022privacy}. This perhaps could lead to something of a paradox widely discussed in privacy literature~\cite{kokolakis2017privacy}. Brown~\cite{brown2001studying} found that, despite the general worries people seemed to have about privacy, they would still give out information for perceived benefits. In fact, the two participants who reacted to both scenarios (\ie, potential risk cases of re-identification, data abuse and misuse) with stronger concerns and hesitancy to data sharing, reported to use Be My Eyes---one of the few services that explicitly indicates dissemination of video streams to third parties~\cite{stangl2022privacy}.

\subsection{Data Sharing Given Data Access Methods}

\begin{figure*}[!th]

\begin{subfigure}{1\textwidth}
  \centering
  \includegraphics[width=1\linewidth]{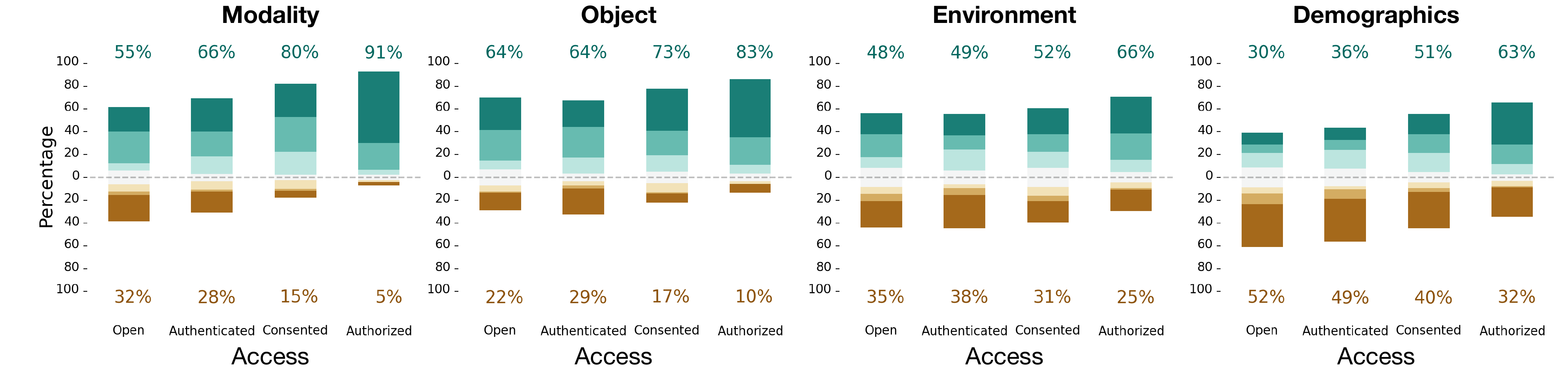} 
  \caption{}
  \label{fig:combinedlikert}
  \Description[Likert data vertical bar charts indicating concatenated responses on comfort of sharing by categories: modality, object, environment, and demographic information. Each row begins with a variable relating to data access conditions: open, authenticated, consented, and authorized, followed by percentages of respondents selecting lower comfort, neutral, and higher comfort options to the variable.]{
Modality
Open           (32\%, 13\%, 55\%)
Authenticated  (28\%, 6\%, 66\%)
Consented      (15\%, 5\%, 80\%)
Authorized     (5\%, 4\%, 91\%)

Object
Open           (22\%, 14\%, 64\%)
Authenticated  (29\%, 7\%, 64\%)
Consented      (17\%, 10\%, 73\%)
Authorized     (10\%, 7\%, 83\%)

Environment
Open           (35\%, 17\%, 48\%)
Authenticated  (38\%, 13\%, 49\%)
Consented      (31\%, 17\%, 52\%)
Authorized     (25\%, 9\%, 66\%)

Demographics
Open            (52\%, 18\%, 30\%)
Authenticated   (49\%, 15\%, 36\%)
Consented       (40\%, 9\%, 51\%)
Authorized      (32\%, 5\%, 63\%)
  }
  
\end{subfigure}

\begin{subfigure}{1\textwidth}
  \centering
   \includegraphics[width=1\linewidth]{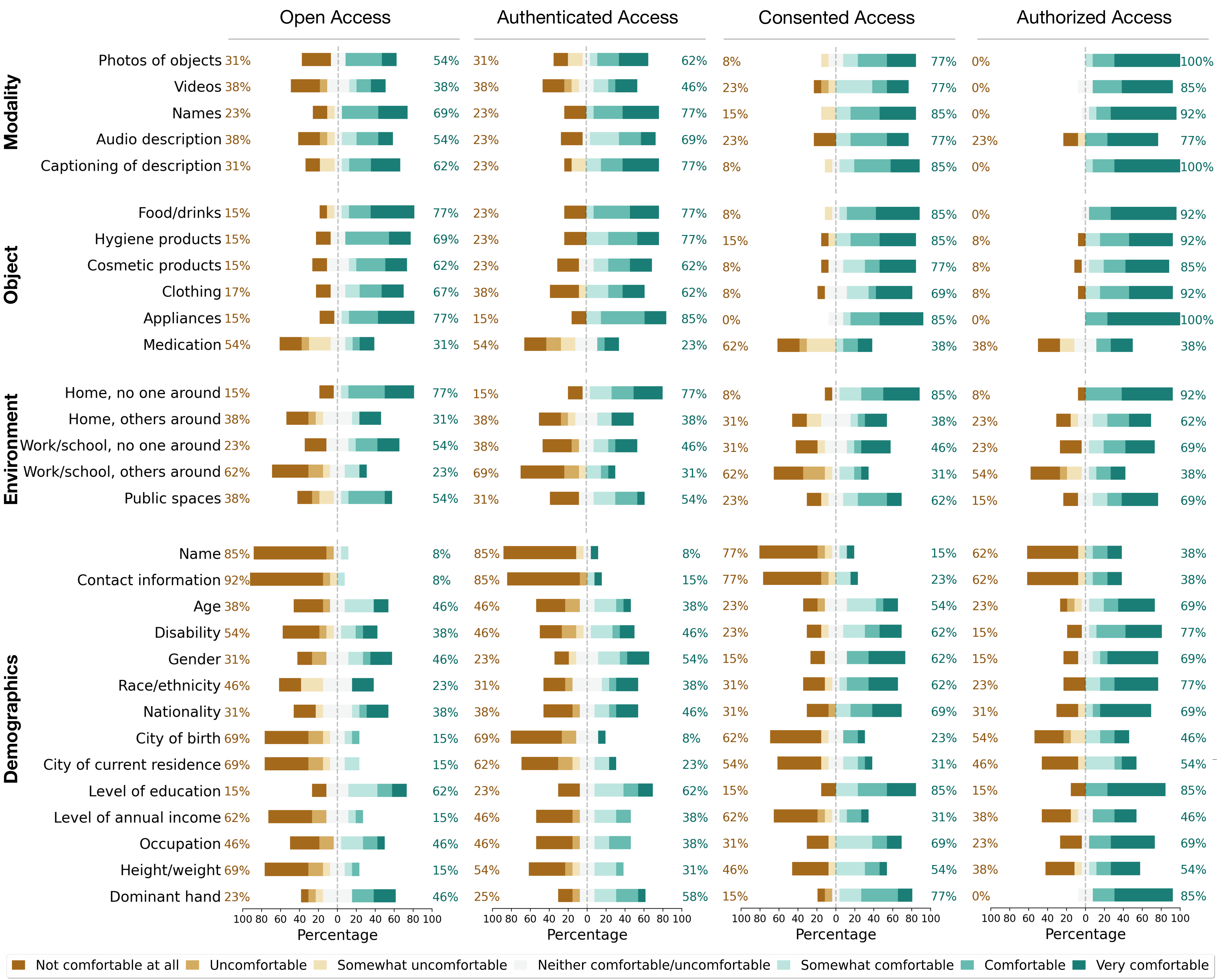}
  \caption{}
      \Description[Likert data horizontal bar charts indicating comfort of sharing by categories: modality, object, environment, and demographic information. Each row begins with a variable relating to the category, followed by percentages of respondents selecting lower comfort, neutral, and higher comfort options to the variable across data access methods: open, authenticated, consented, and authorized.]{
Modality                   Open                         Authenticated                Consented                   Authorized
Photos of objects          (31\%, 15\%, 54\%)  (31\%, 7\%, 62\%)   (8\%, 15\%, 77\%)  (0\%, 0\%, 100\%)
Videos                     (38\%, 24\%, 38\%)  (38\%, 16\%, 46\%)  (23\%, 0\%, 77\%)  (0\%, 15\%, 85\%)
Names                      (23\%, 8\%, 69\%)   (23\%, 0\%, 77\%)   (15\%, 0\%, 85\%)  (0\%, 8\%, 92\%)
Audio description          (38\%, 8\%, 54\%)   (23\%, 8\%, 69\%)   (23\%, 0\%, 77\%)  (23\%, 0\%, 77\%)
Captioning of description  (31\%, 7\%, 62\%)   (23\%, 0\%, 77\%)   (8\%, 7\%, 85\%)   (0\%, 0\%, 100\%)

Object             Open                         Authenticated                Consented                   Authorized
Food/drinks        (15\%, 8\%, 77\%)   (23\%, 0\%, 77\%)   (8\%, 7\%, 85\%)   (0\%, 8\%, 92\%)
Hygiene products   (15\%, 16\%, 69\%)  (23\%, 0\%, 77\%)   (15\%, 0\%, 85\%)  (8\%, 0\%, 92\%)
Cosmetic products  (15\%, 23\%, 62\%)  (23\%, 15\%, 62\%)  (8\%, 15\%, 77\%)  (8\%, 7\%, 85\%)
Clothing           (17\%, 16\%, 67\%)  (38\%, 0\%, 62\%)   (8\%, 23\%, 69\%)  (8\%, 0\%, 92\%)
Appliances         (15\%, 8\%, 77\%)   (15\%, 0\%, 85\%)   (0\%, 15\%, 85\%)  (0\%, 0\%, 100\%)
Medication         (54\%, 15\%, 31\%)  (54\%, 23\%, 23\%)  (62\%, 0\%, 38\%)  (38\%, 24\%, 38\%)

Environment                 Open                         Authenticated                Consented                    Authorized
Home, no one around         (15\%, 8\%, 77\%)   (15\%, 8\%, 77\%)   (8\%, 7\%, 85\%)    (8\%, 0\%, 92\%)
Home, others around         (38\%, 31\%, 31\%)  (38\%, 24\%, 38\%)  (31\%, 31\%, 38\%)  (23\%, 15\%, 62\%)
Work/school, no one around  (23\%, 23\%, 54\%)  (38\%, 16\%, 46\%)  (31\%, 23\%, 46\%)  (23\%, 8\%, 69\%)
Work/school, others around  (62\%, 15\%, 23\%)  (69\%, 0\%, 31\%)   (62\%, 7\%, 31\%)   (54\%, 8\%, 38\%)
Public spaces               (38\%, 8\%, 54\%)   (31\%, 15\%, 54\%)  (23\%, 15\%, 62\%)  (15\%, 16\%, 69\%)

Demographic Information    Open                         Authenticated                Consented                    Authorized
Name                       (85\%, 7\%, 8\%)    (85\%, 7\%, 8\%)    (77\%, 8\%, 15\%)   (62\%, 0\%, 38\%)
Contact information        (92\%, 0\%, 8\%)    (85\%, 0\%, 15\%)   (77\%, 0\%, 23\%)   (62\%, 0\%, 38\%)
Age                        (38\%, 16\%, 46\%)  (46\%, 16\%, 38\%)  (23\%, 23\%, 54\%)  (23\%, 8\%, 69\%)
Disability                 (54\%, 8\%, 38\%)   (46\%, 8\%, 46\%)   (23\%, 15\%, 62\%)  (15\%, 8\%, 77\%)
Gender                     (31\%, 23\%, 46\%)  (23\%, 23\%, 54\%)  (15\%, 23\%, 62\%)  (15\%, 16\%, 69\%)
Race/ethnicity             (46\%, 31\%, 23\%)  (31\%, 31\%, 38\%)  (31\%, 7\%, 62\%)   (23\%, 0\%, 77\%)
Nationality                (31\%, 31\%, 38\%)  (38\%, 16\%, 46\%)  (31\%, 0\%, 69\%)   (31\%, 0\%, 69\%)
City of birth              (69\%, 16\%, 15\%)  (69\%, 23\%, 8\%)   (62\%, 15\%, 23\%)  (54\%, 0\%, 46\%)
City of current residence  (69\%, 16\%, 15\%)  (62\%, 15\%, 23\%)  (54\%, 15\%, 31\%)  (46\%, 0\%, 54\%)
Level of education         (15\%, 23\%, 62\%)  (23\%, 15\%, 62\%)  (15\%, 0\%, 85\%)   (15\%, 0\%, 85\%)
Level of annual income     (62\%, 23\%, 15\%)  (46\%, 16\%, 38\%)  (62\%, 7\%, 31\%)   (38\%, 16\%, 46\%)
Occupation                 (46\%, 8\%, 46\%)   (46\%, 16\%, 38\%)  (31\%, 0\%, 69\%)   (23\%, 8\%, 69\%)
Height/weight              (69\%, 16\%, 15\%)  (54\%, 15\%, 31\%)  (46\%, 0\%, 54\%)   (38\%, 8\%, 54\%)
Dominant hand              (23\%, 31\%, 46\%)  (25\%, 17\%, 58\%)  (15\%, 8\%, 77\%)   (0\%, 15\%, 85\%)
}
  \label{fig:singlelikert}
\end{subfigure}
\vspace{-1.5em}
\caption{Participants' level of comfort with data sharing using a Likert scale (1: not comfortable at all, 7: very comfortable), with percentages of responses for lower comfort and higher comfort levels varies widely across contexts. At the top in \textbf{(a)}, we show overall trends of comfort by four data access conditions: Open, Authenticated, Consented, and Authorized aggregated across different modalities, objects, environments, and demographics. Below that in \textbf{(b)}, we provide a breakdown of responses for each type of modality, object, environment, and demographic information, also stratified by the data access conditions.}
\label{fig:access_responses}
\end{figure*}

To discuss beyond specific data and environment (\ie, photos and labels of objects generated in the home environment), we investigated various factors that could influence participants' decision to share their data. Inspired by prior work measuring people's comfort and acceptance of their data being collected and used~\cite{park2021designing,gilbert2021measuring}, we questioned how different data modalities, objects, environments, and demographic metadata would affect such measurements. As shown in Figure~\ref{fig:access_responses}, we explored their responses to these data sharing contexts conditioned by data access methods. Though we observed the unsurprising trends for increased comfort as more restrictions are applied on the level of access (Figure~\ref{fig:combinedlikert}), there were specific contexts that the participants raised concerns consistently across different data access methods. For certain information or settings (\eg, audio description, photos of medication, photos taken around bystanders), some participants rated comfort on the negative end of the scale even under authorized access (Figure~\ref{fig:singlelikert}). In the following, we go deeper into these topics of concern.

\subsubsection{Type of Modality}
We found that some data modalities relate to the level of comfort for sharing. In particular, we gathered concerns for videos of objects and audio description of images. Under open access, five participants rated lower on the comfort scale for these two modalities (Video: median=4, Audio Description: median=5) compared to other types of modality like photos and names of objects (median=6). When comparing videos with photos, audio partially factored into their concern: ``Videos got a little more concern because of whatever might be heard in the background'' (P5). In fact, videos of objects from blind people in the ORBIT dataset respected this aspect; audio was never collected~\cite{theodorou2021disability}. Even so, the degree of concerns for videos degraded as we moved to the authorized access condition (median=7). Whether photos or videos, what information these data captured was at the root of their concerns, as P13 briefly mentioned: \textit{``It's totally dependent on what you're sharing.''}

In terms of sharing audio description, four participants were especially worried about being identified from their voice. P9 elaborated the rationale: ``That (voice) leads to potentially identifiable information. Especially with the current type of voice fingerprinting software that is starting to be developed now. You actually can, with a fairly consistent degree of accuracy, match someone's voice print.'' Following these worries, three participants kept their comfort level at the lower end across all access methods for audio.  

\subsubsection{Type of Object} Our participants identified certain objects as `private.' One noticeable trend was that they did not feel comfortable sharing photos of medication, which was rated low on the comfort scale from open (median=3) to authorized (median=4) access conditions. Five participants expressed stronger concerns, as briefly explained by P1: \textit{``Medication goes to private status or private characteristics of a person''} (P1). In comparison, the ratings were relatively higher for other types of objects even under open access, including hygiene and cosmetic products, food/drinks, and appliances (median=6). These objects could be perceived as more general items. Thus, it was not a surprise to see differences in their perception towards prescription and general medications: \textit{``If it's like a general like Tylenol medicine or something over the counter stuff, then yes, I would give it a seven. If it's like prescription then no, because my information is on there''} (P6). Perhaps this can be a double edge sword given the potential of object recognition technology supporting identification of medication that is often a challenge for blind people~\cite{brady2013visual}. These concerns can further limit the availability of images for such 'private' objects. This indicates the need for privacy-preserving discussions in data collection, as explored by Gurari \etal~\cite{gurari2019vizwiz} to enable building image recognition algorithms while safeguarding private information.

Participants also expressed hesitancy in sharing objects of `personal' nature. Though the comfort scores for clothing were relatively at the higher end from open (median=5.5) to authorized (median=6) access conditions, concerns still remained regarding privacy: \textit{``It's just like, you wouldn't invite your friends to see your closet right? They come to your house. You invite them to a party but they will never go into your bedroom or your walk-in closet, that kind of thing...I still want the privacy no matter who they are''} (P11). We can expect similar attitudes depending on the types of hygiene or cosmetic products, that may be more or less personal. Additionally, our participants listed other items that they do not want to capture in the photos. These were mostly objects of `sensitive' nature including personal documents (\eg, driver's license, passport, insurance cards, credit cards, bills). They further indicated safety concerns for objects that can reveal or trace their identity (\eg, vehicles, diploma, friends/family photos). The characteristics of these objects with `personal' and `sensitive' nature could shed light on the taxonomy of \textit{what is private in images}~\cite{gurari2019vizwiz}.

\subsubsection{Type of Environment} When asked to rate their level of comfort by the environment, the presence of bystanders factored into their perception. In general, our participants were comfortable with sharing photos captured in the home space especially where bystanders are not present; 10 participants rated their comfort at the higher end even under the open access condition (median=6). In the same access condition but in the presence of bystanders such as family members, their perceived comfort was lower (median=4). Concerns often revolved around the background of photos, as P5 explained: \textit{``Something I could do very diligent is making sure there's not identifying things in the photos. That would be my only concern.''} We saw that these concerns were more associated with work or school spaces; in the presence of bystanders, participants rated low on the comfort scale from open (median=2) to authorized (median=3) access conditions. This might be due to more information being available to identify them (\eg, company logos) or others around (\eg, co-workers) who didn't give consent to share. P6 remarked, \textit{``I took a photo of an object and somebody was in the background. I guess I wouldn't want that to be displayed. Again, just because that person may not feel comfortable with.''} 

While these concerns for bystanders resonate the privacy perspectives that are being reported with technology use in public (\eg, for blind people to detect pedestrians through wearable glasses~\cite{lee2020pedestrian}), our participants showed a bimodal reaction for sharing photos generated in public spaces such as streets or plazas. Interestingly, these places were characterized differently; even under open access, participants indicated higher levels of comfort (median=5). Though some expressed much stronger concerns than others, such as \textit{``I am more concerned if people can identify the space that is attached to you''} (P8), other participants noted less privacy comparing to home or work environments: \textit{``I think when people are in public, they don't have expectation of privacy that they might have in my house or even at work''} (P4). P11 also justified the lack of concerns as \textit{``everyone is taking photos, who cares what these photos are?''}, yet expressed concerns if they were neighborhoods where family, friends, or neighbors could be identified. 

\subsubsection{Type of Demographic Information} Unsurprisingly, participants were hesitant to sharing identifiable information, such as their name or contact information receiving low comfort ratings across all data access conditions (median=1). Even so, 5 participants rated higher on their comfort level for sharing such information under authorized access, as referred by P13: \textit``{It depends on what I'm sharing with...Sometimes you get on Facebook and you think you're gonna find some person that has a name and you see this 25 people with the same name. So I don't have a problem with as long as it's not attached to anything that could be damaging.''} They did not see anything harmful from \textit{``a bag of potato chips that we took photo of and we labeled''} (P13). While the ratings for comfort started low for sharing city of birth or current residence under open access (median=2), more participants became comfortable with sharing city of current residence under authorized access (median=5). It was not the case for sharing city of birth (median=3), which P1 commented \textit{``That goes back probably to my nationality''} considering the potential linkage. P4 expressed hesitancy in sharing such information as \textit{``I don't see the value of that except for negative reasons. Someone trying to build a profile of me.''}

While privacy or security concerns often reflected the lower comfort ratings for sharing certain demographic information (\eg, annual income), participants raised ethical concerns as well. They found sensitivity of disclosing race/ethnicity or nationality; P12 (whose comfort level was 1 across data access methods) described it as \textit{``invasive because there's still a lot of discrimination with various nationalities''} and suggested that participants should have the option to not answer such demographic information. P2 specially warned the risk of sharing demographic information including disability in general: \textit{``I think people could make false assumptions or incorrect judgments about you.''} Our participants also raised other demographic information that they would not want to share, including marital status, number of children or siblings, employment history, and religious affiliation.

We observed other concerns that could factor into consideration, including the ambiguity of the data collection purpose. For example, our participants were unsure of the usefulness of height/weight or dominant hand information as part of the AI dataset. P7 noted, \textit{``What's the purpose? Why is it important? Why do you want to know that?''}, rating the comfort level as 1 across different data access methods except for authorized access users whom the participant could expect a clear purpose for its use. We found a similar trend with other demographic information including gender and race/ethnicity, as articulated by P5 \textit{``I don't really know how important the gender pieces to this as far as AI developers doing what they need to do''} and P3 \textit{``I don't care about whether they know I'm black or white or whatever, but I really don't think it's that important.''} This indicates the importance of increasing the understanding of how such information could contribute to AI development, especially raising awareness around issues of fairness for underrepresented groups~\cite{buolamwini2018gender, tatman2017gender, bolukbasi2016man}.

\subsection{Mitigating Risks Given Stakeholders \& Regulations} 
\begin{figure*}[h]
    \includegraphics[width=\textwidth]{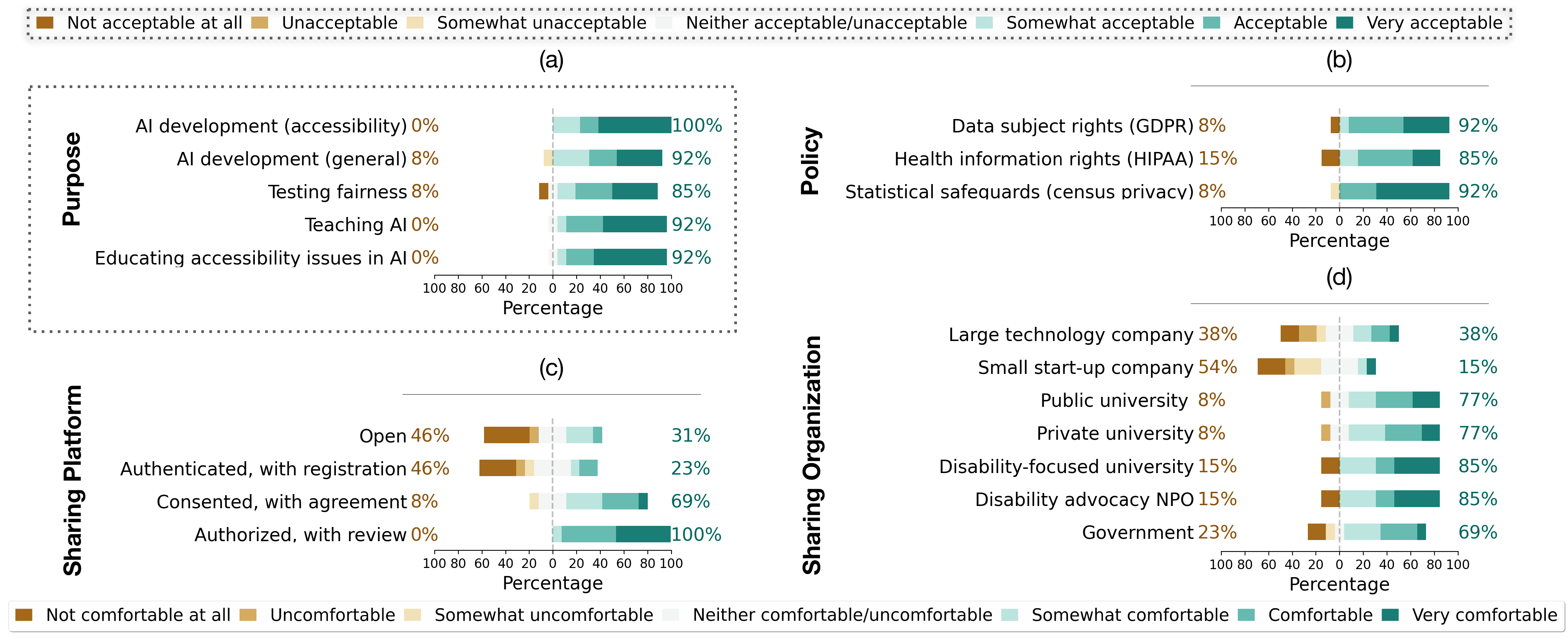}
    \vspace*{-5mm}
    \caption{Participants’ level of acceptance (1: not acceptable at all, 7: very acceptable) or comfort (1: not comfortable at all, 7: very comfortable) with data sharing varies across data use purposes (a), policies (b), sharing platforms (c), and organizations (d).}
    \label{fig:other_responses}
    \Description[Likert data horizontal bar charts indicating acceptance of sharing across data purposes and comfort of sharing via data policies, data sharing platforms, and data sharing organizations. Each row lists percentages of respondents selecting lower, neutral, and higher ratings to the measured variable.]{

Purpose
AI development (accessibility)        (0\%, 0\%, 100\%)
AI development (general)              (8\%, 0\%, 92\%)
Testing fairness                      (8\%, 7\%, 85\%)
Teaching AI                           (0\%, 8\%, 92\%)
Educating accessibility issues in AI  (0\%, 8\%, 92\%) 

Policy
Data subject rights (GDPR)               (8\%, 0\%, 92\%)
Health information rights (HIPAA)        (15\%, 0\%, 85\%)
Statistical safeguards (census privacy)  (8\%, 0\%, 92\%)

Platform
Open                              (46\%, 23\%, 31\%)
Authenticated, with registration  (46\%, 31\%, 23\%)
Consented, with agreement         (8\%, 23\%, 69\%)
Authorized, with review           (0\%, 0\%, 100\%)

Organization
Large technology company       (38\%, 24\%, 38\%)
Small start-up company         (54\%, 31\%, 15\%)
Public university              (8\%, 15\%, 77\%)
Private university             (8\%, 15\%, 77\%)
Disability-focused university  (15\%, 0\%, 85\%)
Disability advocacy NPO        (15\%, 0\%, 85\%)
Government                     (23\%, 8\%, 69\%)

}
\end{figure*}

We elicited responses from our participants about risk mitigating strategies for data sharing. We first asked what actions by different stakeholders or regulations could possibly address their concerns. We then expanded the conversations by listing existing practices and explored their reactions via comfort of sharing or acceptance ratings (see Figure~\ref{fig:other_responses}). Their responses revealed different perspectives on research or legal practices as well potential strategies that can help minimize the harms and impose safety measures.

\subsubsection{Data Stewards} When asked how data stewards, those putting together a dataset (\ie, researchers of this study), could mitigate the risks, all our participants expected some forms of actions --- \eg, screening the photos and removing any personal or sensitive information if caught by mistake (n=6), collecting data in a privacy-respecting way such as codifying participants' names and cropping out the background from photos (n=2), ensuring that the data are protected from security breaches when storing and sharing (n=3), or fully informing the use of their data to the contributors including secondary data use cases (n=2). 

Our participants expanded on how they would like data stewards to restrict the use cases of their data being shared. Many (n=9) favored to restrict them by specific domains or usage types --- \eg, for research purposes only, for object recognition technology development, for purposes defined originally, informed, and permitted by data contributors. A participant, however, noted the importance of supporting broader purposes: \textit{``It's easy to say, it's only should be for research purposes. And that's fine. But at some point, let's say you guys have a final ready to go, market ready [object recognition technology], then at that point, you guys are starting your library all over, which has its pros and cons''} (P5). Broadening the use of data (\eg, commercialization) yet raises more questions and challenges such as ownership, intellectual property, or data agreements~\cite{thomas2016big}. In fact, this has been a `wicked problem' across fields~\cite{zuiderwijk2016wicked, wilczek2021archival,kostkova2016owns}, especially when no criteria exist for determining the correctness or value for the use of open data~\cite{zuiderwijk2016wicked}.

To expand on the use of their study data, we prompted a list of data purposes (see Figure~\ref{fig:other_responses}a). Participants reacted with somewhat high degree of acceptance towards data use for accessibility (median=7). While ratings are also on the higher end for teaching AI and educating accessibility issues in AI (median=7), two participants left some remarks, as referred by P9: \textit{``The goal is good but it all depends on how it's implemented.''} P2 asked \textit{``What's the purpose? What application?''} factoring into their neutral rating. Lacking specifics of the purpose could explain why the acceptance was slightly lower for other prompted purposes, developing AI for general purposes and testing fairness (median=6). Echoing P9's comment on fairness as \textit{``a word that is so vague these days,''} efforts to increase the knowledge about contextual factors influencing fairness (\eg, issues of bias in AI) might be necessary to support their conceptualizations~\cite{bogina2022educating}. P10 gave a neutral rating to the purpose of testing fairness although they seemed to value it: \textit{``It (data) is not being used for what I thought it was used for, which is helping with the object identification and recognition.''}

\subsubsection{Policy- and Law-Makers} When asked what actions policy- and law-makers could take to prevent and minimize the risks, some (n=5) stressed legal actions to penalize individuals who misbehaved, such as those sharing data without consent (P2), using data for a wrong purpose (P1, P8), or not keeping the promise to handle data safely and responsibility (P7). A participant (P4) articulated to \textit{``establish accountability or illegality''} of misbehaved individuals. A few others (n=3) suggested ways to reduce the chances of misbehaviors, such as keeping records of who accessed the data \textit{``so that if that information does get used inappropriately, for whatever reason, you can at least have a narrower field of suspects''} (P5), or what people claimed to use the data \textit{``to get that assurance so you have some protection, in case you have to bring up a case, you know, you misused my data, but you said you weren't going to misuse it''} (P13). Even so, P13 raised a limitation point: \textit{``You really don't know if they're misusing your data or not until something happens.''} Similarly, one participant expressed further limitations given that policies are often not sufficient to protect data contributors: \textit{``I'm less optimistic about lawmakers...They've instituted the GDPR, the general data protection regulation, but even that, it outlines a whole bunch of scenarios, and legal requirements, a lot of which I'm happy with. But all you have to do to get out of the scope of the GDPR is simply moving data outside of Europe. That's not difficult. Lawmakers, I don't really think have a lot of power in this scenario.''}

We prompted existing privacy policies (see Figure~\ref{fig:other_responses}b) to further probe participant perception. Overall, participants reacted positively to these policies for protecting their personal or identifiable information. For example, all except two participants rated data subject rights (\eg, GDPR~\cite{linden2018privacy} giving control over data) high on the scale as they would feel comfortable sharing their data (median=6). However, they added some remarks regarding the lack of guarantee, as referred by P5: \textit{``Policies are not always followed at different places.''} They reacted similarity to privacy protection for health information such as HIPPA~\cite{kulynych2003new}; though higher in ratings (median=6), some concerns also remained in the lack of guarantee as \textit{``data can always be breached''} (P6). Additionally, P7 wondered whether such data policies like GDPR would actually apply to study data and context: \textit{``Not fully trusting the existing policy, so trying to see if the existing policy concept can be transferable to like study data.''} Indeed, there are still open challenges for the uncertainty about how such existing policies would apply to scientific research practices, such as in informed consent and anonymization~\cite{ienca2019general}.

\subsubsection{Data Sharing Entities}

We explored data sharing practices by prompting different conditions for platforms and organizations. Our participants rated these by their level of comfort for sharing their data. As shown in Figure~\ref{fig:other_responses}c, 9 participants reacted positively to conditions where a Terms of Use agreement is applied (median=5). With further authorization to screen people's access by the purpose of data use, all participants indicated higher ratings of comfort (median=6) due to more \textit{accountability} (P2, P9), \textit{control} (P5), and \textit{security} (P6, P12). However, the lack of trust remained depending on how platforms operate: \textit{``There's still a slight chance that [data] could get in the wrong hands. You don't know who is operating''} (P7). 

Concerns for operation were mainly reflected within the context of technology companies (see Figure~\ref{fig:other_responses}d), especially with smaller start-ups which only 2 participants rated high on the comfort scale for sharing their data (median=3). Often the reason was behind security measures, with a participant articulating \textit{``doubt that they would have controls in place''} (P9). Their concerns resonate with public perceptions reported in anecdotal evidence --- \eg, small companies were seen to be less stable~\cite{mcnaney2022exploring}. Large technology companies were also perceived as profit driven~\cite{hill2013let}, which could explain their lower comfort ratings (median=4) towards them compared to organizations supporting education/research or disability communities. Regarding their comfort with sharing, 10 participants reacted positively towards public universities (median=6) and private universities (median=5). The participants saw that these organizations have \textit{``a fewer reasons to misuse the data''} (P4) and \textit{``to be a little bit more discreet''} (P7). Similar responses were found for disability-focused universities and non-profit organizations, where 11 participants felt comfortable sharing through them (median=6). When asking similar questions, Park \etal~\cite{park2021designing} also observed that organizations oriented towards disability communities are seen as reliable, yet comfort ratings for large technology companies closely aligned with ratings for universities. As the work came from company-based research~\cite{park2021designing}, we might be seeing the effects of sampling as P2 noted: \textit{``I wouldn't be concerned if I do a study with someone. I trust them. It's just that anyone else that might get a hold of stuff that might have been shared or supposedly shared by another.''}

Following the lack of trust, 3 participants raised the importance of transparency, ensuring that they are informed about purposes and restrictions and kept in the loop for any actions made regarding their data. A participant strongly called for regulations: \textit{``If researchers are going to share data with any third parties, that third party also needs to have a person's consent and offer full and complete disclosure, and honor their promise. The third parties need to honor it to the researcher as well as to the individual [contributing data]''} (P2). \textit{Datasheets for Datasets}~\cite{gebru2018datasheets}, while being targeted for dataset creators and dataset consumers to promote transparency and accountability, includes questions regarding data collection and distribution---\eg, whether individuals contributing their data consented to the use of their data, or whether it will be distributed to third parties. We see this as a meaningful resource to be reviewed with potential data contributors and also as a way to discuss preferences to keep them in the loop in the process. Furthermore, given the individual differences that we found in terms of sharing demographic information, especially in the context of sharing entities (\eg, open access vs. authorized access; data-sharing organizations), these data sharing factors need to be considered together when adopting inclusive practices for datasheets.

\subsubsection{Data Contributors}
Finally, we asked what actions our participants as potential data contributors could take to protect themselves from the risks. One recurring theme among the participants was to carefully consider what data they would be sharing (n=5), ensuring that objects and information contained in the photos are not sensitive and they are disclosing as little demographic information as possible. We also observed two participants selecting the lobby of the apartment and one participant selecting the patio area to generate the photos. This might be reflective of individual concerns that were induced by what could be captured in their data inadvertently, in addition to other potential concerns (\eg, safety related to COVID-19 or letting an experimenter in their home). P6 reflected on the data capturing activities for ways to minimize capturing unintended information in the future: \textit{``I think I would definitely be more cautious about making sure that there's no identifiable photos. So probably, maybe I would just take all my photos in front of a white wall or something, you know. So that's really up to me, to make sure that I don't take photos that I'm uncomfortable with.''} Such workarounds, while efficient, might not capture the real-world contexts for image recognition tasks. To facilitate the collection of high-variation conditions (\eg, in a wide variation of backgrounds~\cite{massiceti2021orbit}), Theodorou \etal~\cite{theodorou2021disability} incorporated a manual validation process to check and remove data containing personally identifying information (PII) in their dataset creation. This brings more attention to better approaches that allow automatic detection of PII to blur or erase from images~\cite{gurari2019vizwiz}.

Our results also indicate the importance of supportive materials and communication to help potential data contributors assess the benefit-risk tradeoffs. A few (n=3) participants showed further interest in learning more about how their data would be used, as articulated by P13: \textit{``I would love to understand better how AI use the data. But again, it's just my curiosity. Yeah, just wanted to understand a little bit better''}. This could help bridge the knowledge gap and establish trust; for example, we saw people's reluctance to share their demographic information as it was deemed `not relevant' for AI development. Participants also asked for a better sense of the impacts of the data they share: \textit{`` I personally just want to see what other people are using it for. Because, you know, I don't have all of the answers. I don't know how people could be using this technology to improve their quality of life. I mean, we're all here to learn from one another''} (P6). This is an option that is not supported in current informed consent processes and needs to be reflected in future research practice discussions~\cite{ahern2012informed, nusbaum2017communicating}.

\section{Discussion}

When data are at the core of innovation, creating and releasing annotated AI datasets becomes critical. This has been illustrated in the progress of shared data resources in broader computing fields~\cite{kaggle,zenodo,asuncion2007uci,calzolari2010lrec,microsoft2018microsoft,brickley2019google}. However, many application domains still lack sufficient data for accessibility~\cite{bragg2019sign,kacorri2017teachable,park2021designing}, due to sensitivity surrounding smaller population groups and potential harms that may arise along privacy and ethics. In the face of these challenges, we reflect on the empirical insights gained from the blind community to consider how we can better configure data practices and study methods to guide future directions for promoting more transparency, trust, and engagement in these practices. These insights come with several limitations that threaten their validity and generalizability, detailed in our Methods (under Section~\ref{sec:limit}).  

\subsection{Assessing Benefit-Risk Tradeoffs in Data Sharing}
Participants were aware of many benefits of contributing their study data via an AI dataset. Even without probing, they described how their study data would be important for technological innovations for accessibility and beyond. Additionally, with probing, they further expressed motivations behind data sharing, which were often associated with broader impacts of their data contribution. Some wished to learn more about the role of data in AI to assess the positive impacts better. This strengthens prior anecdotal evidence about data sharing ``for the greater good''~\cite{mcnaney2022exploring} indicating the willingness of disabled people to share their data for the benefits of the broader communities~\cite{park2021designing,mozersky2020research}.

While recognizing the benefits, many participants were not overly concerned about risk factors behind data sharing as long as these data are anonymized and not disclosing sensitive information. Such assessment is not unique to the blind community; data harms are often overlooked and underestimated, surfacing limitations for how well people can anticipate negative impact~\cite{kroger2021data,solove2014reasons}. Indeed, Hamidi \etal, building on the privacy threat framework~\cite{deng2011privacy}, referred to \textit{unawareness} of end users not realizing the consequences of sharing their data as a serious concern~\cite{hamidi2018should}. In return, we have seen attempts to raise awareness about potential harms, such as releasing real-life data misuse examples (\eg, Data Harm Record~\cite{redden2020data}, Inventory of Risks and Harms~\cite{centre2016risk}). However, an important caveat revealed from our study is that assessment of risks can be too nuanced to be directly addressed, as the participants reacted to our risk scenarios as \textit{``hard to imagine if they are critical''} (P8). This extends to similar conversations on how documents aimed to increase awareness by informing data sharing practices (\eg, privacy policy) are not always effective; they are often difficult to read and understand the implications~\cite{mcdonald2008cost, earp2005examining}.
 
One way we could better assess benefits and risks is to incorporate the participatory nature to analyze the impacts and implications for data sharing. While intended for dataset creators and users, \textit{Datasheets for Datasets} elicited a similar point: ``Has an analysis of the potential impact of the dataset and its use on data subjects (e.g., a data protection impact analysis) been conducted?''~\cite{gebru2018datasheets} 
This needs to be addressed and communicated with individuals or communities who could be most affected, adopting many other guided questions related to the collection and distribution of datasets. Our study methods seem to reveal partial effects by providing an opportunity for participants to engage more deeply in these topics. As we probed more contexts to our participants to reflect on how their data should be shared, they began to analyze a set of rules that need to be considered in implementing data sharing --- \eg, restriction of data use for research purposes only (especially for improving object recognition technology) or control of information they are disclosing with their study data such as what's being captured in the background of photos. Also, with probing of benefit and risk scenarios, we started to see participants making a tradeoff assessment as ``I think the benefits far outweigh any minor drawbacks that could occur...If someone says, oh, there was a 71 year old guy that took pictures of Lay's potato chips, you know, if that's all that ever happens to me, I'm okay'' (P12). In this vein, to support critical reflections among stakeholders on navigating ``wicked problems''~\cite{berger2017wicked}, we encourage practices such as Value Sensitive Design (VSD)~\cite{friedman2002value} or Speculative Design~\cite{dunne2013speculative}, along with design fictions and participatory workshops~\cite{ballard2019judgment,soden2019chi} as important directions. We further suggest extending the line of work around co-designing accessible impact assessment tools or checklists, highlighting Madaio \etal~\cite{madaio2020co} as a methodological inspiration.

\subsection{Bringing Trustworthiness in Data Ecosystems: Transparency and Engagement}

Our findings indicate the importance of complementing trust across stakeholders in data practices. As highlighted throughout this work, many were hesitant to contribute their study data for ambiguous purposes or purposes not aligning with their perceived benefits, revealing notions of ``distrust.'' For example, we similarly observed in our study public perceptions about small companies which they were seen to lack regulations~\cite{mcnaney2022exploring} and larger tech companies seen as profit driven~\cite{hill2013let}. They also expressed concerns for sharing demographic metadata (\eg, gender, race) along their data, not only due to privacy but also uncertainty of its importance and preconceived notions about what kind of data should be used to train AI models. To fill the disconnect between the disability and AI community, the process of stewarding accessibility datasets requires greater transparency of data use as well as awareness~\cite{bogina2022educating}, especially to challenge inclusivity issues that are pressing for marginalized communities~\cite{kamikubo2022data,buolamwini2018gender}. 

To further support transparency, our participants wanted to be kept in the loop for any actions made regarding their data, always obtaining a person's consent and providing complete disclosure of the way their data are being used. While this is important over the data lifecylce, we see practical challenges in sustaining such long term relationships between data contributors and data stewards. Though GDPR is designed to support such connections, there are still open challenges in applying to scientific research practices~\cite{ienca2019general}, further inhibited by lack of resources and expertise~\cite{sivan2022varieties}. Perhaps proper implementation and maintenance at the institutional level may be necessary. Reflecting on trustworthy certification in human-centered AI~\cite{shneiderman2020human,shneiderman2020bridging, chan2021supporting, garibay2023six}, we echo oversight structures, with institutional interventions and reviews, to support an ongoing communication to align data practices with individual concerns and values. More so, we see benefits in implementing systems that enable community efforts and coordination, such as ``data consortia'' which have been developed as institutional frameworks in archives to navigate the challenges behind ethical data collection and sharing~\cite{jo2020lessons}.

We also recognize oversight at an individual level. Incorporating a ``framework for participatory data stewardship''~\cite{ai2021participatory}, we suggest researchers to consider data sharing from the beginning of their projects and further consider mechanisms beyond transparency to engage participants in decision making about the data they consider sharing for AI development. As Rake \etal~\cite{rake2017personalized} explored personalized consent flows for data sharing with stakeholders in medical research, we see future directions in favor of practices that empower potential data contributors to help shape and govern their own data. We perhaps see benefits in data cooperativeness to turn ``distrust'' into shared understanding of how data sharing should be carried out~\cite{machirori2021turning}.

\section{Conclusion}

There are important normative questions around the use of accessibility datasets sourced from disabled people --- they can be used against them by uncovering their identity and (mis)detecting disability status without consent. However, AI-infused systems trained on data lacking in terms of inclusion and diversity \nobreak further increase the risks of unfair or discriminatory outcomes for underrepresented groups. To drive AI efforts that are inclusive of disability, this \nobreak research aims to shape data practices that align with the concerns and values of disabled data contributors. We conducted a case study engaging blind participants in `in-situ' data capture activities to inquire about how their data should be used and shared via an AI dataset. Our findings have highlighted the opportunities for making their data accessible for the research community in AI development, through proper actions and restrictions around data sharing and re-use. We hope this research helps discussions that aim to improve the norms for collecting and sharing data by understanding the facilitators and barriers that are more attuned to the communities of focus in accessibility.

\begin{acks}
We thank Hal Daum\'e III for providing valuable feedback on our preliminary work. We also thank Katie Shilton, Beth St. Jean, \nobreak Jessica Vitak, and our anonymous reviewers for their expertise and comments to further strengthen this paper. Special thanks go to Farnaz Zamiri Zeraati for her help in running the study. Finally, we want to acknowledge our participants for their time and the research advisory council from The National Federation of the Blind for assisting us with participant recruitment. This work is supported by National Institute on Disability, Independent Living, and Rehabilitation Research (NIDILRR), ACL, HHS (\#90REGE0008).
\end{acks}

\bibliographystyle{ACM-Reference-Format}
\bibliography{ref}


\begin{thebibliography}{151}


\ifx \showCODEN    \undefined \def \showCODEN     #1{\unskip}     \fi
\ifx \showDOI      \undefined \def \showDOI       #1{#1}\fi
\ifx \showISBNx    \undefined \def \showISBNx     #1{\unskip}     \fi
\ifx \showISBNxiii \undefined \def \showISBNxiii  #1{\unskip}     \fi
\ifx \showISSN     \undefined \def \showISSN      #1{\unskip}     \fi
\ifx \showLCCN     \undefined \def \showLCCN      #1{\unskip}     \fi
\ifx \shownote     \undefined \def \shownote      #1{#1}          \fi
\ifx \showarticletitle \undefined \def \showarticletitle #1{#1}   \fi
\ifx \showURL      \undefined \def \showURL       {\relax}        \fi
\providecommand\bibfield[2]{#2}
\providecommand\bibinfo[2]{#2}
\providecommand\natexlab[1]{#1}
\providecommand\showeprint[2][]{arXiv:#2}

\bibitem[\protect\citeauthoryear{Abbott, MacLeod, Nurain, Ekobe, and
  Patil}{Abbott et~al\mbox{.}}{2019}]%
        {abbott2019local}
\bibfield{author}{\bibinfo{person}{Jacob Abbott}, \bibinfo{person}{Haley
  MacLeod}, \bibinfo{person}{Novia Nurain}, \bibinfo{person}{Gustave Ekobe},
  {and} \bibinfo{person}{Sameer Patil}.} \bibinfo{year}{2019}\natexlab{}.
\newblock \showarticletitle{Local Standards for Anonymization Practices in
  Health, Wellness, Accessibility, and Aging Research at CHI}. In
  \bibinfo{booktitle}{\emph{Proceedings of the 2019 CHI Conference on Human
  Factors in Computing Systems}} (Glasgow, Scotland Uk)
  \emph{(\bibinfo{series}{CHI '19})}. \bibinfo{publisher}{Association for
  Computing Machinery}, \bibinfo{address}{New York, NY, USA},
  \bibinfo{pages}{1–14}.
\newblock
\showISBNx{9781450359702}
\urldef\tempurl%
\url{https://doi.org/10.1145/3290605.3300692}
\showDOI{\tempurl}


\bibitem[\protect\citeauthoryear{Adams, Morales, and Kurniawan}{Adams
  et~al\mbox{.}}{2013}]%
        {adams2013qualitative}
\bibfield{author}{\bibinfo{person}{Dustin Adams}, \bibinfo{person}{Lourdes
  Morales}, {and} \bibinfo{person}{Sri Kurniawan}.}
  \bibinfo{year}{2013}\natexlab{}.
\newblock \showarticletitle{A qualitative study to support a blind photography
  mobile application}. In \bibinfo{booktitle}{\emph{Proceedings of the 6th
  International Conference on PErvasive Technologies Related to Assistive
  Environments}}. \bibinfo{pages}{1--8}.
\newblock


\bibitem[\protect\citeauthoryear{Ahern}{Ahern}{2012}]%
        {ahern2012informed}
\bibfield{author}{\bibinfo{person}{Kathy Ahern}.}
  \bibinfo{year}{2012}\natexlab{}.
\newblock \showarticletitle{Informed consent: are researchers accurately
  representing risks and benefits?}
\newblock \bibinfo{journal}{\emph{Scandinavian Journal of Caring Sciences}}
  \bibinfo{volume}{26}, \bibinfo{number}{4} (\bibinfo{year}{2012}),
  \bibinfo{pages}{671--678}.
\newblock


\bibitem[\protect\citeauthoryear{AI}{AI}{2021}]%
        {ai2021participatory}
\bibfield{author}{\bibinfo{person}{JUST AI}.} \bibinfo{year}{2021}\natexlab{}.
\newblock \showarticletitle{Participatory data stewardship}.
\newblock \bibinfo{journal}{\emph{Reading time}} (\bibinfo{year}{2021}).
\newblock


\bibitem[\protect\citeauthoryear{Amazon}{Amazon}{2021}]%
        {awssearch}
\bibfield{author}{\bibinfo{person}{Amazon}.} \bibinfo{year}{2021}\natexlab{}.
\newblock \bibinfo{title}{Registry of Open Data on AWS}.
\newblock \bibinfo{howpublished}{\url{https://registry.opendata.aws/}}.
\newblock


\bibitem[\protect\citeauthoryear{Amershi, Weld, Vorvoreanu, Fourney, Nushi,
  Collisson, Suh, Iqbal, Bennett, Inkpen, Teevan, Kikin-Gil, and
  Horvitz}{Amershi et~al\mbox{.}}{2019}]%
        {amershi2019guidelines}
\bibfield{author}{\bibinfo{person}{Saleema Amershi}, \bibinfo{person}{Dan
  Weld}, \bibinfo{person}{Mihaela Vorvoreanu}, \bibinfo{person}{Adam Fourney},
  \bibinfo{person}{Besmira Nushi}, \bibinfo{person}{Penny Collisson},
  \bibinfo{person}{Jina Suh}, \bibinfo{person}{Shamsi Iqbal},
  \bibinfo{person}{Paul~N. Bennett}, \bibinfo{person}{Kori Inkpen},
  \bibinfo{person}{Jaime Teevan}, \bibinfo{person}{Ruth Kikin-Gil}, {and}
  \bibinfo{person}{Eric Horvitz}.} \bibinfo{year}{2019}\natexlab{}.
\newblock \showarticletitle{Guidelines for Human-AI Interaction}. In
  \bibinfo{booktitle}{\emph{Proceedings of the 2019 CHI Conference on Human
  Factors in Computing Systems}} (Glasgow, Scotland Uk)
  \emph{(\bibinfo{series}{CHI '19})}. \bibinfo{publisher}{Association for
  Computing Machinery}, \bibinfo{address}{New York, NY, USA},
  \bibinfo{pages}{1–13}.
\newblock
\showISBNx{9781450359702}
\urldef\tempurl%
\url{https://doi.org/10.1145/3290605.3300233}
\showDOI{\tempurl}


\bibitem[\protect\citeauthoryear{Asuncion and Newman}{Asuncion and
  Newman}{2007}]%
        {asuncion2007uci}
\bibfield{author}{\bibinfo{person}{Arthur Asuncion} {and}
  \bibinfo{person}{David Newman}.} \bibinfo{year}{2007}\natexlab{}.
\newblock \bibinfo{title}{UCI machine learning repository}.
\newblock
\newblock


\bibitem[\protect\citeauthoryear{Ballard, Chappell, and Kennedy}{Ballard
  et~al\mbox{.}}{2019}]%
        {ballard2019judgment}
\bibfield{author}{\bibinfo{person}{Stephanie Ballard},
  \bibinfo{person}{Karen~M. Chappell}, {and} \bibinfo{person}{Kristen
  Kennedy}.} \bibinfo{year}{2019}\natexlab{}.
\newblock \showarticletitle{Judgment Call the Game: Using Value Sensitive
  Design and Design Fiction to Surface Ethical Concerns Related to Technology}.
  In \bibinfo{booktitle}{\emph{Proceedings of the 2019 on Designing Interactive
  Systems Conference}} (San Diego, CA, USA) \emph{(\bibinfo{series}{DIS '19})}.
  \bibinfo{publisher}{Association for Computing Machinery},
  \bibinfo{address}{New York, NY, USA}, \bibinfo{pages}{421–433}.
\newblock
\showISBNx{9781450358507}
\urldef\tempurl%
\url{https://doi.org/10.1145/3322276.3323697}
\showDOI{\tempurl}


\bibitem[\protect\citeauthoryear{Barkhuus}{Barkhuus}{2012}]%
        {barkhuus2012mismeasurement}
\bibfield{author}{\bibinfo{person}{Louise Barkhuus}.}
  \bibinfo{year}{2012}\natexlab{}.
\newblock \showarticletitle{The Mismeasurement of Privacy: Using Contextual
  Integrity to Reconsider Privacy in HCI}. In
  \bibinfo{booktitle}{\emph{Proceedings of the SIGCHI Conference on Human
  Factors in Computing Systems}} (Austin, Texas, USA)
  \emph{(\bibinfo{series}{CHI '12})}. \bibinfo{publisher}{Association for
  Computing Machinery}, \bibinfo{address}{New York, NY, USA},
  \bibinfo{pages}{367–376}.
\newblock
\showISBNx{9781450310154}
\urldef\tempurl%
\url{https://doi.org/10.1145/2207676.2207727}
\showDOI{\tempurl}


\bibitem[\protect\citeauthoryear{Barry}{Barry}{2017}]%
        {barry2017not}
\bibfield{author}{\bibinfo{person}{Dwight Barry}.}
  \bibinfo{year}{2017}\natexlab{}.
\newblock \showarticletitle{Do not use averages with Likert scale data}.
\newblock \bibinfo{journal}{\emph{Enterp. Anal}}  \bibinfo{volume}{24}
  (\bibinfo{year}{2017}).
\newblock


\bibitem[\protect\citeauthoryear{Batson, Ahmad, and Tsang}{Batson
  et~al\mbox{.}}{2002}]%
        {batson2002four}
\bibfield{author}{\bibinfo{person}{C~Daniel Batson}, \bibinfo{person}{Nadia
  Ahmad}, {and} \bibinfo{person}{Jo-Ann Tsang}.}
  \bibinfo{year}{2002}\natexlab{}.
\newblock \showarticletitle{Four motives for community involvement}.
\newblock \bibinfo{journal}{\emph{Journal of social issues}}
  \bibinfo{volume}{58}, \bibinfo{number}{3} (\bibinfo{year}{2002}),
  \bibinfo{pages}{429--445}.
\newblock


\bibitem[\protect\citeauthoryear{Becker}{Becker}{1970}]%
        {becker1970field}
\bibfield{author}{\bibinfo{person}{Howard~Saul Becker}.}
  \bibinfo{year}{1970}\natexlab{}.
\newblock \bibinfo{booktitle}{\emph{Field work evidence}}.
\newblock \bibinfo{publisher}{Transaction publishers}. 39--62 pages.
\newblock


\bibitem[\protect\citeauthoryear{Berger, Totzauer, Lefeuvre, Storz, Kurze, and
  Bischof}{Berger et~al\mbox{.}}{2017}]%
        {berger2017wicked}
\bibfield{author}{\bibinfo{person}{Arne Berger}, \bibinfo{person}{S{\"o}ren
  Totzauer}, \bibinfo{person}{Kevin Lefeuvre}, \bibinfo{person}{Michael Storz},
  \bibinfo{person}{Albrecht Kurze}, {and} \bibinfo{person}{Andreas Bischof}.}
  \bibinfo{year}{2017}\natexlab{}.
\newblock \showarticletitle{Wicked, Open, Collaborative: Why Research through
  Design Matters for HCI Research}.
\newblock \bibinfo{journal}{\emph{i-com}} \bibinfo{volume}{16},
  \bibinfo{number}{2} (\bibinfo{year}{2017}), \bibinfo{pages}{131--142}.
\newblock


\bibitem[\protect\citeauthoryear{Berger}{Berger}{2015}]%
        {berger2015now}
\bibfield{author}{\bibinfo{person}{Roni Berger}.}
  \bibinfo{year}{2015}\natexlab{}.
\newblock \showarticletitle{Now I see it, now I don’t: Researcher’s
  position and reflexivity in qualitative research}.
\newblock \bibinfo{journal}{\emph{Qualitative research}} \bibinfo{volume}{15},
  \bibinfo{number}{2} (\bibinfo{year}{2015}), \bibinfo{pages}{219--234}.
\newblock


\bibitem[\protect\citeauthoryear{Bigham, Jayant, Ji, Little, Miller, Miller,
  Miller, Tatarowicz, White, White, and Yeh}{Bigham et~al\mbox{.}}{2010}]%
        {bigham2010vizwiz}
\bibfield{author}{\bibinfo{person}{Jeffrey~P. Bigham},
  \bibinfo{person}{Chandrika Jayant}, \bibinfo{person}{Hanjie Ji},
  \bibinfo{person}{Greg Little}, \bibinfo{person}{Andrew Miller},
  \bibinfo{person}{Robert~C. Miller}, \bibinfo{person}{Robin Miller},
  \bibinfo{person}{Aubrey Tatarowicz}, \bibinfo{person}{Brandyn White},
  \bibinfo{person}{Samual White}, {and} \bibinfo{person}{Tom Yeh}.}
  \bibinfo{year}{2010}\natexlab{}.
\newblock \showarticletitle{VizWiz: Nearly Real-Time Answers to Visual
  Questions}. In \bibinfo{booktitle}{\emph{Proceedings of the 23nd Annual ACM
  Symposium on User Interface Software and Technology}} (New York, New York,
  USA) \emph{(\bibinfo{series}{UIST '10})}. \bibinfo{publisher}{Association for
  Computing Machinery}, \bibinfo{address}{New York, NY, USA},
  \bibinfo{pages}{333–342}.
\newblock
\showISBNx{9781450302715}
\urldef\tempurl%
\url{https://doi.org/10.1145/1866029.1866080}
\showDOI{\tempurl}


\bibitem[\protect\citeauthoryear{Blaser and Ladner}{Blaser and Ladner}{2020}]%
        {blaser2020why}
\bibfield{author}{\bibinfo{person}{Brianna Blaser} {and}
  \bibinfo{person}{Richard~E. Ladner}.} \bibinfo{year}{2020}\natexlab{}.
\newblock \showarticletitle{Why is Data on Disability so Hard to Collect and
  Understand?}. In \bibinfo{booktitle}{\emph{Proceedings of the 5th
  International Conference on Research in Equity and Sustained Participation in
  Engineering, Computing, and Technology (RESPECT)}}.
\newblock


\bibitem[\protect\citeauthoryear{Bogina, Hartman, Kuflik, and
  Shulner-Tal}{Bogina et~al\mbox{.}}{2022}]%
        {bogina2022educating}
\bibfield{author}{\bibinfo{person}{Veronika Bogina}, \bibinfo{person}{Alan
  Hartman}, \bibinfo{person}{Tsvi Kuflik}, {and} \bibinfo{person}{Avital
  Shulner-Tal}.} \bibinfo{year}{2022}\natexlab{}.
\newblock \showarticletitle{Educating software and AI stakeholders about
  algorithmic fairness, accountability, transparency and ethics}.
\newblock \bibinfo{journal}{\emph{International Journal of Artificial
  Intelligence in Education}} \bibinfo{volume}{32}, \bibinfo{number}{3}
  (\bibinfo{year}{2022}), \bibinfo{pages}{808--833}.
\newblock


\bibitem[\protect\citeauthoryear{Bolukbasi, Chang, Zou, Saligrama, and
  Kalai}{Bolukbasi et~al\mbox{.}}{2016}]%
        {bolukbasi2016man}
\bibfield{author}{\bibinfo{person}{Tolga Bolukbasi}, \bibinfo{person}{Kai-Wei
  Chang}, \bibinfo{person}{James~Y Zou}, \bibinfo{person}{Venkatesh Saligrama},
  {and} \bibinfo{person}{Adam~T Kalai}.} \bibinfo{year}{2016}\natexlab{}.
\newblock \showarticletitle{Man is to computer programmer as woman is to
  homemaker? debiasing word embeddings}.
\newblock \bibinfo{journal}{\emph{Advances in neural information processing
  systems}}  \bibinfo{volume}{29} (\bibinfo{year}{2016}),
  \bibinfo{pages}{4349--4357}.
\newblock


\bibitem[\protect\citeauthoryear{Bot, Suver, Neto, Kellen, Klein, Bare, Doerr,
  Pratap, Wilbanks, Dorsey, Friend, and Trister}{Bot et~al\mbox{.}}{2016}]%
        {bot2016mpower}
\bibfield{author}{\bibinfo{person}{Brian~M. Bot}, \bibinfo{person}{Christine
  Suver}, \bibinfo{person}{Elias~Chaibub Neto}, \bibinfo{person}{Michael
  Kellen}, \bibinfo{person}{Arno Klein}, \bibinfo{person}{Christopher Bare},
  \bibinfo{person}{Megan Doerr}, \bibinfo{person}{Abhishek Pratap},
  \bibinfo{person}{John Wilbanks}, \bibinfo{person}{E.~Ray Dorsey},
  \bibinfo{person}{Stephen~H. Friend}, {and} \bibinfo{person}{Andrew~D
  Trister}.} \bibinfo{year}{2016}\natexlab{}.
\newblock \showarticletitle{The {mPower} study, {Parkinson} disease mobile data
  collected using {ResearchKit}}.
\newblock \bibinfo{journal}{\emph{Scientific Data}}  \bibinfo{volume}{3}
  (\bibinfo{date}{March} \bibinfo{year}{2016}), \bibinfo{pages}{160011}.
\newblock
\urldef\tempurl%
\url{https://doi.org/10.1038/sdata.2016.11}
\showURL{%
\tempurl}


\bibitem[\protect\citeauthoryear{Brady, Morris, Zhong, White, and Bigham}{Brady
  et~al\mbox{.}}{2013}]%
        {brady2013visual}
\bibfield{author}{\bibinfo{person}{Erin Brady},
  \bibinfo{person}{Meredith~Ringel Morris}, \bibinfo{person}{Yu Zhong},
  \bibinfo{person}{Samuel White}, {and} \bibinfo{person}{Jeffrey~P Bigham}.}
  \bibinfo{year}{2013}\natexlab{}.
\newblock \showarticletitle{Visual challenges in the everyday lives of blind
  people}. In \bibinfo{booktitle}{\emph{Proceedings of the SIGCHI conference on
  human factors in computing systems}}. \bibinfo{pages}{2117--2126}.
\newblock


\bibitem[\protect\citeauthoryear{Bragg, Glasser, Minakov, Caselli, and
  Thies}{Bragg et~al\mbox{.}}{2022}]%
        {bragg2022exploring}
\bibfield{author}{\bibinfo{person}{Danielle Bragg}, \bibinfo{person}{Abraham
  Glasser}, \bibinfo{person}{Fyodor Minakov}, \bibinfo{person}{Naomi Caselli},
  {and} \bibinfo{person}{William Thies}.} \bibinfo{year}{2022}\natexlab{}.
\newblock \showarticletitle{Exploring Collection of Sign Language Videos
  through Crowdsourcing}.
\newblock \bibinfo{journal}{\emph{Proceedings of the ACM on Human-Computer
  Interaction}} \bibinfo{volume}{6}, \bibinfo{number}{CSCW2}
  (\bibinfo{year}{2022}), \bibinfo{pages}{1--24}.
\newblock


\bibitem[\protect\citeauthoryear{Bragg, Koller, Bellard, Berke, Boudreault,
  Braffort, Caselli, Huenerfauth, Kacorri, Verhoef, Vogler, and
  Ringel~Morris}{Bragg et~al\mbox{.}}{2019}]%
        {bragg2019sign}
\bibfield{author}{\bibinfo{person}{Danielle Bragg}, \bibinfo{person}{Oscar
  Koller}, \bibinfo{person}{Mary Bellard}, \bibinfo{person}{Larwan Berke},
  \bibinfo{person}{Patrick Boudreault}, \bibinfo{person}{Annelies Braffort},
  \bibinfo{person}{Naomi Caselli}, \bibinfo{person}{Matt Huenerfauth},
  \bibinfo{person}{Hernisa Kacorri}, \bibinfo{person}{Tessa Verhoef},
  \bibinfo{person}{Christian Vogler}, {and} \bibinfo{person}{Meredith
  Ringel~Morris}.} \bibinfo{year}{2019}\natexlab{}.
\newblock \showarticletitle{Sign Language Recognition, Generation, and
  Translation: An Interdisciplinary Perspective}. In
  \bibinfo{booktitle}{\emph{The 21st International ACM SIGACCESS Conference on
  Computers and Accessibility}} (Pittsburgh, PA, USA)
  \emph{(\bibinfo{series}{ASSETS '19})}. \bibinfo{publisher}{Association for
  Computing Machinery}, \bibinfo{address}{New York, NY, USA},
  \bibinfo{pages}{16–31}.
\newblock
\showISBNx{9781450366762}
\urldef\tempurl%
\url{https://doi.org/10.1145/3308561.3353774}
\showDOI{\tempurl}


\bibitem[\protect\citeauthoryear{Bragg, Koller, Caselli, and Thies}{Bragg
  et~al\mbox{.}}{2020}]%
        {bragg2020exploring}
\bibfield{author}{\bibinfo{person}{Danielle Bragg}, \bibinfo{person}{Oscar
  Koller}, \bibinfo{person}{Naomi Caselli}, {and} \bibinfo{person}{William
  Thies}.} \bibinfo{year}{2020}\natexlab{}.
\newblock \showarticletitle{Exploring Collection of Sign Language Datasets:
  Privacy, Participation, and Model Performance}. In
  \bibinfo{booktitle}{\emph{The 22nd International ACM SIGACCESS Conference on
  Computers and Accessibility}} (Virtual Event, Greece)
  \emph{(\bibinfo{series}{ASSETS '20})}. \bibinfo{publisher}{Association for
  Computing Machinery}, \bibinfo{address}{New York, NY, USA}, Article
  \bibinfo{articleno}{33}, \bibinfo{numpages}{14}~pages.
\newblock
\showISBNx{9781450371032}
\urldef\tempurl%
\url{https://doi.org/10.1145/3373625.3417024}
\showDOI{\tempurl}


\bibitem[\protect\citeauthoryear{Braun and Clarke}{Braun and Clarke}{2006}]%
        {braun2006using}
\bibfield{author}{\bibinfo{person}{Virginia Braun} {and}
  \bibinfo{person}{Victoria Clarke}.} \bibinfo{year}{2006}\natexlab{}.
\newblock \showarticletitle{Using thematic analysis in psychology}.
\newblock \bibinfo{journal}{\emph{Qualitative research in psychology}}
  \bibinfo{volume}{3}, \bibinfo{number}{2} (\bibinfo{year}{2006}),
  \bibinfo{pages}{77--101}.
\newblock


\bibitem[\protect\citeauthoryear{Braun and Clarke}{Braun and Clarke}{2019}]%
        {braun2019reflecting}
\bibfield{author}{\bibinfo{person}{Virginia Braun} {and}
  \bibinfo{person}{Victoria Clarke}.} \bibinfo{year}{2019}\natexlab{}.
\newblock \showarticletitle{Reflecting on reflexive thematic analysis}.
\newblock \bibinfo{journal}{\emph{Qualitative research in sport, exercise and
  health}} \bibinfo{volume}{11}, \bibinfo{number}{4} (\bibinfo{year}{2019}),
  \bibinfo{pages}{589--597}.
\newblock


\bibitem[\protect\citeauthoryear{Braun and Clarke}{Braun and Clarke}{2021a}]%
        {braun2021can}
\bibfield{author}{\bibinfo{person}{Virginia Braun} {and}
  \bibinfo{person}{Victoria Clarke}.} \bibinfo{year}{2021}\natexlab{a}.
\newblock \showarticletitle{Can I use TA? Should I use TA? Should I not use TA?
  Comparing reflexive thematic analysis and other pattern-based qualitative
  analytic approaches}.
\newblock \bibinfo{journal}{\emph{Counselling and Psychotherapy Research}}
  \bibinfo{volume}{21}, \bibinfo{number}{1} (\bibinfo{year}{2021}),
  \bibinfo{pages}{37--47}.
\newblock


\bibitem[\protect\citeauthoryear{Braun and Clarke}{Braun and Clarke}{2021b}]%
        {braun2021one}
\bibfield{author}{\bibinfo{person}{Virginia Braun} {and}
  \bibinfo{person}{Victoria Clarke}.} \bibinfo{year}{2021}\natexlab{b}.
\newblock \showarticletitle{One size fits all? What counts as quality practice
  in (reflexive) thematic analysis?}
\newblock \bibinfo{journal}{\emph{Qualitative research in psychology}}
  \bibinfo{volume}{18}, \bibinfo{number}{3} (\bibinfo{year}{2021}),
  \bibinfo{pages}{328--352}.
\newblock


\bibitem[\protect\citeauthoryear{Brickley, Burgess, and Noy}{Brickley
  et~al\mbox{.}}{2019}]%
        {brickley2019google}
\bibfield{author}{\bibinfo{person}{Dan Brickley}, \bibinfo{person}{Matthew
  Burgess}, {and} \bibinfo{person}{Natasha Noy}.}
  \bibinfo{year}{2019}\natexlab{}.
\newblock \showarticletitle{Google Dataset Search: Building a search engine for
  datasets in an open Web ecosystem}. In \bibinfo{booktitle}{\emph{The World
  Wide Web Conference}}. \bibinfo{pages}{1365--1375}.
\newblock


\bibitem[\protect\citeauthoryear{Brown}{Brown}{2001}]%
        {brown2001studying}
\bibfield{author}{\bibinfo{person}{Barry Brown}.}
  \bibinfo{year}{2001}\natexlab{}.
\newblock \showarticletitle{Studying the internet experience}.
\newblock \bibinfo{journal}{\emph{HP laboratories technical report HPL}}
  \bibinfo{volume}{49} (\bibinfo{year}{2001}).
\newblock


\bibitem[\protect\citeauthoryear{Buolamwini and Gebru}{Buolamwini and
  Gebru}{2018}]%
        {buolamwini2018gender}
\bibfield{author}{\bibinfo{person}{Joy Buolamwini} {and}
  \bibinfo{person}{Timnit Gebru}.} \bibinfo{year}{2018}\natexlab{}.
\newblock \showarticletitle{Gender shades: Intersectional accuracy disparities
  in commercial gender classification}. In \bibinfo{booktitle}{\emph{Conference
  on fairness, accountability and transparency}}. PMLR,
  \bibinfo{pages}{77--91}.
\newblock


\bibitem[\protect\citeauthoryear{Caine}{Caine}{2016}]%
        {caine2016local}
\bibfield{author}{\bibinfo{person}{Kelly Caine}.}
  \bibinfo{year}{2016}\natexlab{}.
\newblock \showarticletitle{Local Standards for Sample Size at CHI}. In
  \bibinfo{booktitle}{\emph{Proceedings of the 2016 CHI Conference on Human
  Factors in Computing Systems}} (San Jose, California, USA)
  \emph{(\bibinfo{series}{CHI '16})}. \bibinfo{publisher}{Association for
  Computing Machinery}, \bibinfo{address}{New York, NY, USA},
  \bibinfo{pages}{981–992}.
\newblock
\showISBNx{9781450333627}
\urldef\tempurl%
\url{https://doi.org/10.1145/2858036.2858498}
\showDOI{\tempurl}


\bibitem[\protect\citeauthoryear{Calzolari, Soria, Del~Gratta, Goggi, Quochi,
  Russo, Choukri, Mariani, and Piperidis}{Calzolari et~al\mbox{.}}{2010}]%
        {calzolari2010lrec}
\bibfield{author}{\bibinfo{person}{Nicoletta Calzolari},
  \bibinfo{person}{Claudia Soria}, \bibinfo{person}{Riccardo Del~Gratta},
  \bibinfo{person}{Sara Goggi}, \bibinfo{person}{Valeria Quochi},
  \bibinfo{person}{Irene Russo}, \bibinfo{person}{Khalid Choukri},
  \bibinfo{person}{Joseph Mariani}, {and} \bibinfo{person}{Stelios Piperidis}.}
  \bibinfo{year}{2010}\natexlab{}.
\newblock \showarticletitle{The lrec map of language resources and
  technologies}. In \bibinfo{booktitle}{\emph{Proceedings of the Seventh
  International Conference on Language Resources and Evaluation (LREC'10)}}.
\newblock


\bibitem[\protect\citeauthoryear{Cao, Seelman, Lee, and Daum{\'e}~III}{Cao
  et~al\mbox{.}}{2022}]%
        {cao2022whats}
\bibfield{author}{\bibinfo{person}{Yang~Trista Cao}, \bibinfo{person}{Kyle
  Seelman}, \bibinfo{person}{Kyungjun Lee}, {and} \bibinfo{person}{Hal
  Daum{\'e}~III}.} \bibinfo{year}{2022}\natexlab{}.
\newblock \showarticletitle{What{'}s Different between Visual Question
  Answering for Machine {``}Understanding{''} Versus for Accessibility?}. In
  \bibinfo{booktitle}{\emph{Proceedings of the 2nd Conference of the
  Asia-Pacific Chapter of the Association for Computational Linguistics and the
  12th International Joint Conference on Natural Language Processing (Volume 1:
  Long Papers)}}. \bibinfo{publisher}{Association for Computational
  Linguistics}, \bibinfo{address}{Online only}, \bibinfo{pages}{1025--1034}.
\newblock
\urldef\tempurl%
\url{https://aclanthology.org/2022.aacl-main.75}
\showURL{%
\tempurl}


\bibitem[\protect\citeauthoryear{Carney, Webster, Alvarado, Phillips, Howell,
  Griffith, Jongejan, Pitaru, and Chen}{Carney et~al\mbox{.}}{2020}]%
        {carney2020teachable}
\bibfield{author}{\bibinfo{person}{Michelle Carney}, \bibinfo{person}{Barron
  Webster}, \bibinfo{person}{Irene Alvarado}, \bibinfo{person}{Kyle Phillips},
  \bibinfo{person}{Noura Howell}, \bibinfo{person}{Jordan Griffith},
  \bibinfo{person}{Jonas Jongejan}, \bibinfo{person}{Amit Pitaru}, {and}
  \bibinfo{person}{Alexander Chen}.} \bibinfo{year}{2020}\natexlab{}.
\newblock \showarticletitle{Teachable Machine: Approachable Web-Based Tool for
  Exploring Machine Learning Classification}. In
  \bibinfo{booktitle}{\emph{Extended Abstracts of the 2020 CHI Conference on
  Human Factors in Computing Systems}} (Honolulu, HI, USA)
  \emph{(\bibinfo{series}{CHI EA '20})}. \bibinfo{publisher}{Association for
  Computing Machinery}, \bibinfo{address}{New York, NY, USA},
  \bibinfo{pages}{1–8}.
\newblock
\showISBNx{9781450368193}
\urldef\tempurl%
\url{https://doi.org/10.1145/3334480.3382839}
\showDOI{\tempurl}


\bibitem[\protect\citeauthoryear{Chan, Daum{\'e}~III, Dickerson, Kacorri, and
  Shneiderman}{Chan et~al\mbox{.}}{2021}]%
        {chan2021supporting}
\bibfield{author}{\bibinfo{person}{Joel Chan}, \bibinfo{person}{Hal
  Daum{\'e}~III}, \bibinfo{person}{John~P Dickerson}, \bibinfo{person}{Hernisa
  Kacorri}, {and} \bibinfo{person}{Ben Shneiderman}.}
  \bibinfo{year}{2021}\natexlab{}.
\newblock \showarticletitle{Supporting human flourishing by ensuring human
  involvement in AI-infused systems}.
\newblock \bibinfo{journal}{\emph{NeurIPS 2021 Workshop on Human Centered AI}}
  (\bibinfo{year}{2021}).
\newblock


\bibitem[\protect\citeauthoryear{Deng, Wuyts, Scandariato, Preneel, and
  Joosen}{Deng et~al\mbox{.}}{2011}]%
        {deng2011privacy}
\bibfield{author}{\bibinfo{person}{Mina Deng}, \bibinfo{person}{Kim Wuyts},
  \bibinfo{person}{Riccardo Scandariato}, \bibinfo{person}{Bart Preneel}, {and}
  \bibinfo{person}{Wouter Joosen}.} \bibinfo{year}{2011}\natexlab{}.
\newblock \showarticletitle{A privacy threat analysis framework: supporting the
  elicitation and fulfillment of privacy requirements}.
\newblock \bibinfo{journal}{\emph{Requirements Engineering}}
  \bibinfo{volume}{16}, \bibinfo{number}{1} (\bibinfo{year}{2011}),
  \bibinfo{pages}{3--32}.
\newblock


\bibitem[\protect\citeauthoryear{Duerden and Witt}{Duerden and Witt}{2010}]%
        {duerden2010impact}
\bibfield{author}{\bibinfo{person}{Mat~D Duerden} {and}
  \bibinfo{person}{Peter~A Witt}.} \bibinfo{year}{2010}\natexlab{}.
\newblock \showarticletitle{The impact of direct and indirect experiences on
  the development of environmental knowledge, attitudes, and behavior}.
\newblock \bibinfo{journal}{\emph{Journal of environmental psychology}}
  \bibinfo{volume}{30}, \bibinfo{number}{4} (\bibinfo{year}{2010}),
  \bibinfo{pages}{379--392}.
\newblock


\bibitem[\protect\citeauthoryear{Dunne and Raby}{Dunne and Raby}{2013}]%
        {dunne2013speculative}
\bibfield{author}{\bibinfo{person}{Anthony Dunne} {and} \bibinfo{person}{Fiona
  Raby}.} \bibinfo{year}{2013}\natexlab{}.
\newblock \bibinfo{booktitle}{\emph{Speculative everything: design, fiction,
  and social dreaming}}.
\newblock \bibinfo{publisher}{MIT press}.
\newblock


\bibitem[\protect\citeauthoryear{Dwivedi, Gandhi, Parikh, Coenraad, Bonsignore,
  and Kacorri}{Dwivedi et~al\mbox{.}}{2021}]%
        {dwivedi2021exploring}
\bibfield{author}{\bibinfo{person}{Utkarsh Dwivedi}, \bibinfo{person}{Jaina
  Gandhi}, \bibinfo{person}{Raj Parikh}, \bibinfo{person}{Merijke Coenraad},
  \bibinfo{person}{Elizabeth Bonsignore}, {and} \bibinfo{person}{Hernisa
  Kacorri}.} \bibinfo{year}{2021}\natexlab{}.
\newblock \showarticletitle{Exploring Machine Teaching with Children}. In
  \bibinfo{booktitle}{\emph{2021 IEEE Symposium on Visual Languages and
  Human-Centric Computing (VL/HCC)}}. \bibinfo{pages}{1--11}.
\newblock
\urldef\tempurl%
\url{https://doi.org/10.1109/VL/HCC51201.2021.9576171}
\showDOI{\tempurl}


\bibitem[\protect\citeauthoryear{Earp, Ant{\'o}n, Aiman-Smith, and
  Stufflebeam}{Earp et~al\mbox{.}}{2005}]%
        {earp2005examining}
\bibfield{author}{\bibinfo{person}{Julia~Brande Earp}, \bibinfo{person}{Annie~I
  Ant{\'o}n}, \bibinfo{person}{Lynda Aiman-Smith}, {and}
  \bibinfo{person}{William~H Stufflebeam}.} \bibinfo{year}{2005}\natexlab{}.
\newblock \showarticletitle{Examining Internet privacy policies within the
  context of user privacy values}.
\newblock \bibinfo{journal}{\emph{IEEE Transactions on Engineering Management}}
  \bibinfo{volume}{52}, \bibinfo{number}{2} (\bibinfo{year}{2005}),
  \bibinfo{pages}{227--237}.
\newblock


\bibitem[\protect\citeauthoryear{Engler}{Engler}{2019}]%
        {engler2019some}
\bibfield{author}{\bibinfo{person}{A Engler}.} \bibinfo{year}{2019}\natexlab{}.
\newblock \showarticletitle{For some employment algorithms, disability
  discrimination by default}.
\newblock \bibinfo{journal}{\emph{Brookings. Available at: https://www.
  brookings.
  edu/blog/techtank/2019/10/31/for-some-employment-algorithms-disability-discrimination-by-default/(accessed
  1 June 2020)}} (\bibinfo{year}{2019}).
\newblock


\bibitem[\protect\citeauthoryear{{European Organization For Nuclear Research}
  and {OpenAIRE}}{{European Organization For Nuclear Research} and
  {OpenAIRE}}{2013}]%
        {zenodo}
\bibfield{author}{\bibinfo{person}{{European Organization For Nuclear
  Research}} {and} \bibinfo{person}{{OpenAIRE}}.}
  \bibinfo{year}{2013}\natexlab{}.
\newblock \bibinfo{title}{Zenodo}.
\newblock
\newblock
\urldef\tempurl%
\url{https://doi.org/10.25495/7GXK-RD71}
\showDOI{\tempurl}


\bibitem[\protect\citeauthoryear{Fathi, Ren, and Rehg}{Fathi
  et~al\mbox{.}}{2011}]%
        {fathi2011learning}
\bibfield{author}{\bibinfo{person}{Alireza Fathi}, \bibinfo{person}{Xiaofeng
  Ren}, {and} \bibinfo{person}{James~M Rehg}.} \bibinfo{year}{2011}\natexlab{}.
\newblock \showarticletitle{Learning to recognize objects in egocentric
  activities}. In \bibinfo{booktitle}{\emph{CVPR 2011}}. IEEE,
  \bibinfo{pages}{3281--3288}.
\newblock


\bibitem[\protect\citeauthoryear{Fiesler}{Fiesler}{2019}]%
        {fiesler2019ethical}
\bibfield{author}{\bibinfo{person}{Casey Fiesler}.}
  \bibinfo{year}{2019}\natexlab{}.
\newblock \showarticletitle{Ethical considerations for research involving
  (speculative) public data}.
\newblock \bibinfo{journal}{\emph{Proceedings of the ACM on Human-Computer
  Interaction}} \bibinfo{volume}{3}, \bibinfo{number}{GROUP}
  (\bibinfo{year}{2019}), \bibinfo{pages}{1--13}.
\newblock


\bibitem[\protect\citeauthoryear{Findler, Vilchinsky, and Werner}{Findler
  et~al\mbox{.}}{2007}]%
        {findler2007multidimensional}
\bibfield{author}{\bibinfo{person}{Liora Findler}, \bibinfo{person}{Noa
  Vilchinsky}, {and} \bibinfo{person}{Shirli Werner}.}
  \bibinfo{year}{2007}\natexlab{}.
\newblock \showarticletitle{The multidimensional attitudes scale toward persons
  with disabilities (MAS) construction and validation}.
\newblock \bibinfo{journal}{\emph{Rehabilitation Counseling Bulletin}}
  \bibinfo{volume}{50}, \bibinfo{number}{3} (\bibinfo{year}{2007}),
  \bibinfo{pages}{166--176}.
\newblock


\bibitem[\protect\citeauthoryear{for Information Policy~Leadership}{for
  Information Policy~Leadership}{2016}]%
        {centre2016risk}
\bibfield{author}{\bibinfo{person}{Centre for Information Policy~Leadership}.}
  \bibinfo{year}{2016}\natexlab{}.
\newblock \showarticletitle{Risk, High Risk, Risk Assessments and Data
  Protection Impact Assessments under the GDPR}.
\newblock
  \bibinfo{howpublished}{\url{https://www.informationpolicycentre.com/uploads/5/7/1/0/57104281/cipl_gdpr_project_risk_white_paper_21_december_2016.pdf}}.
\newblock  (\bibinfo{year}{2016}).
\newblock


\bibitem[\protect\citeauthoryear{Friedman, Kahn, and Borning}{Friedman
  et~al\mbox{.}}{2002}]%
        {friedman2002value}
\bibfield{author}{\bibinfo{person}{Batya Friedman}, \bibinfo{person}{Peter
  Kahn}, {and} \bibinfo{person}{Alan Borning}.}
  \bibinfo{year}{2002}\natexlab{}.
\newblock \showarticletitle{Value sensitive design: Theory and methods}.
\newblock \bibinfo{journal}{\emph{University of Washington technical report}}
  \bibinfo{volume}{2} (\bibinfo{year}{2002}), \bibinfo{pages}{12}.
\newblock


\bibitem[\protect\citeauthoryear{Garibay, Winslow, Andolina, Antona,
  Bodenschatz, Coursaris, Falco, Fiore, Garibay, Grieman, Havens, Jirotka,
  Kacorri, Karwowski, Kider, Konstan, Koon, Lopez-Gonzalez, Maifeld-Carucci,
  McGregor, Salvendy, Shneiderman, Stephanidis, Strobel, Holter, and
  Xu}{Garibay et~al\mbox{.}}{2023}]%
        {garibay2023six}
\bibfield{author}{\bibinfo{person}{Ozlem~Ozmen Garibay}, \bibinfo{person}{Brent
  Winslow}, \bibinfo{person}{Salvatore Andolina}, \bibinfo{person}{Margherita
  Antona}, \bibinfo{person}{Anja Bodenschatz}, \bibinfo{person}{Constantinos
  Coursaris}, \bibinfo{person}{Gregory Falco}, \bibinfo{person}{Stephen~M.
  Fiore}, \bibinfo{person}{Ivan Garibay}, \bibinfo{person}{Keri Grieman},
  \bibinfo{person}{John~C. Havens}, \bibinfo{person}{Marina Jirotka},
  \bibinfo{person}{Hernisa Kacorri}, \bibinfo{person}{Waldemar Karwowski},
  \bibinfo{person}{Joe Kider}, \bibinfo{person}{Joseph Konstan},
  \bibinfo{person}{Sean Koon}, \bibinfo{person}{Monica Lopez-Gonzalez},
  \bibinfo{person}{Iliana Maifeld-Carucci}, \bibinfo{person}{Sean McGregor},
  \bibinfo{person}{Gavriel Salvendy}, \bibinfo{person}{Ben Shneiderman},
  \bibinfo{person}{Constantine Stephanidis}, \bibinfo{person}{Christina
  Strobel}, \bibinfo{person}{Carolyn~Ten Holter}, {and} \bibinfo{person}{Wei
  Xu}.} \bibinfo{year}{2023}\natexlab{}.
\newblock \showarticletitle{Six Human-Centered Artificial Intelligence Grand
  Challenges}.
\newblock \bibinfo{journal}{\emph{International Journal of Human–Computer
  Interaction}} \bibinfo{volume}{39}, \bibinfo{number}{3}
  (\bibinfo{year}{2023}), \bibinfo{pages}{391--437}.
\newblock
\urldef\tempurl%
\url{https://doi.org/10.1080/10447318.2022.2153320}
\showDOI{\tempurl}
\showeprint{https://doi.org/10.1080/10447318.2022.2153320}


\bibitem[\protect\citeauthoryear{Gebru, Morgenstern, Vecchione, Vaughan,
  Wallach, III, and Crawford}{Gebru et~al\mbox{.}}{2018}]%
        {gebru2018datasheets}
\bibfield{author}{\bibinfo{person}{Timnit Gebru}, \bibinfo{person}{Jamie
  Morgenstern}, \bibinfo{person}{Briana Vecchione},
  \bibinfo{person}{Jennifer~Wortman Vaughan}, \bibinfo{person}{Hanna~M.
  Wallach}, \bibinfo{person}{Hal~Daum{\'{e}} III}, {and} \bibinfo{person}{Kate
  Crawford}.} \bibinfo{year}{2018}\natexlab{}.
\newblock \showarticletitle{Datasheets for Datasets}.
\newblock \bibinfo{journal}{\emph{CoRR}}  \bibinfo{volume}{abs/1803.09010}
  (\bibinfo{year}{2018}).
\newblock
\showeprint[arxiv]{1803.09010}
\urldef\tempurl%
\url{http://arxiv.org/abs/1803.09010}
\showURL{%
\tempurl}


\bibitem[\protect\citeauthoryear{Gilbert, Vitak, and Shilton}{Gilbert
  et~al\mbox{.}}{2021}]%
        {gilbert2021measuring}
\bibfield{author}{\bibinfo{person}{Sarah Gilbert}, \bibinfo{person}{Jessica
  Vitak}, {and} \bibinfo{person}{Katie Shilton}.}
  \bibinfo{year}{2021}\natexlab{}.
\newblock \showarticletitle{Measuring Americans’ comfort with research uses
  of their social media data}.
\newblock \bibinfo{journal}{\emph{Social Media+ Society}} \bibinfo{volume}{7},
  \bibinfo{number}{3} (\bibinfo{year}{2021}),
  \bibinfo{pages}{20563051211033824}.
\newblock


\bibitem[\protect\citeauthoryear{Google}{Google}{2019}]%
        {google2019lookout}
\bibfield{author}{\bibinfo{person}{Google}.} \bibinfo{year}{2019}\natexlab{}.
\newblock \bibinfo{title}{Lookout - Assisted Vision}.
\newblock
\newblock
\urldef\tempurl%
\url{https://play.google.com/store/apps/details?id=com.google.android.apps.accessibility.reveal&hl=en_US&gl=US}
\showURL{%
\tempurl}


\bibitem[\protect\citeauthoryear{Guo, Kamar, Vaughan, Wallach, and Morris}{Guo
  et~al\mbox{.}}{2019}]%
        {guo2019toward}
\bibfield{author}{\bibinfo{person}{Anhong Guo}, \bibinfo{person}{Ece Kamar},
  \bibinfo{person}{Jennifer~Wortman Vaughan}, \bibinfo{person}{Hanna Wallach},
  {and} \bibinfo{person}{Meredith~Ringel Morris}.}
  \bibinfo{year}{2019}\natexlab{}.
\newblock \showarticletitle{Toward Fairness in AI for People with Disabilities:
  A Research Roadmap}.
\newblock \bibinfo{journal}{\emph{arXiv preprint arXiv:1907.02227}}.
\newblock


\bibitem[\protect\citeauthoryear{{Gurari}, {Li}, {Lin}, {Zhao}, {Guo},
  {Stangl}, and {Bigham}}{{Gurari} et~al\mbox{.}}{2019}]%
        {gurari2019vizwiz}
\bibfield{author}{\bibinfo{person}{D. {Gurari}}, \bibinfo{person}{Q. {Li}},
  \bibinfo{person}{C. {Lin}}, \bibinfo{person}{Y. {Zhao}}, \bibinfo{person}{A.
  {Guo}}, \bibinfo{person}{A. {Stangl}}, {and} \bibinfo{person}{J.~P.
  {Bigham}}.} \bibinfo{year}{2019}\natexlab{}.
\newblock \showarticletitle{VizWiz-Priv: A Dataset for Recognizing the Presence
  and Purpose of Private Visual Information in Images Taken by Blind People}.
  In \bibinfo{booktitle}{\emph{2019 IEEE/CVF Conference on Computer Vision and
  Pattern Recognition (CVPR)}}. \bibinfo{pages}{939--948}.
\newblock
\urldef\tempurl%
\url{https://doi.org/10.1109/CVPR.2019.00103}
\showDOI{\tempurl}


\bibitem[\protect\citeauthoryear{Gurari, Li, Stangl, Guo, Lin, Grauman, Luo,
  and Bigham}{Gurari et~al\mbox{.}}{2018}]%
        {gurari2018vizwiz}
\bibfield{author}{\bibinfo{person}{Danna Gurari}, \bibinfo{person}{Qing Li},
  \bibinfo{person}{Abigale~J. Stangl}, \bibinfo{person}{Anhong Guo},
  \bibinfo{person}{Chi Lin}, \bibinfo{person}{Kristen Grauman},
  \bibinfo{person}{Jiebo Luo}, {and} \bibinfo{person}{Jeffrey~P. Bigham}.}
  \bibinfo{year}{2018}\natexlab{}.
\newblock \showarticletitle{VizWiz Grand Challenge: Answering Visual Questions
  from Blind People}.
\newblock \bibinfo{journal}{\emph{2018 IEEE/CVF Conference on Computer Vision
  and Pattern Recognition}} (\bibinfo{date}{Jun} \bibinfo{year}{2018}).
\newblock
\showISBNx{9781538664209}
\urldef\tempurl%
\url{https://doi.org/10.1109/cvpr.2018.00380}
\showDOI{\tempurl}


\bibitem[\protect\citeauthoryear{Hamidi, Poneres, Massey, and Hurst}{Hamidi
  et~al\mbox{.}}{2018}]%
        {hamidi2018should}
\bibfield{author}{\bibinfo{person}{Foad Hamidi}, \bibinfo{person}{Kellie
  Poneres}, \bibinfo{person}{Aaron Massey}, {and} \bibinfo{person}{Amy Hurst}.}
  \bibinfo{year}{2018}\natexlab{}.
\newblock \showarticletitle{Who Should Have Access to My Pointing Data?:
  Privacy Tradeoffs of Adaptive Assistive Technologies}. In
  \bibinfo{booktitle}{\emph{Proceedings of the 20th International ACM SIGACCESS
  Conference on Computers and Accessibility}} (Galway, Ireland)
  \emph{(\bibinfo{series}{ASSETS '18})}. \bibinfo{publisher}{ACM},
  \bibinfo{address}{New York, NY, USA}, \bibinfo{pages}{203--216}.
\newblock
\showISBNx{978-1-4503-5650-3}
\urldef\tempurl%
\url{https://doi.org/10.1145/3234695.3239331}
\showDOI{\tempurl}


\bibitem[\protect\citeauthoryear{Hassan, Berke, Vahdani, Jing, Tian, and
  Huenerfauth}{Hassan et~al\mbox{.}}{2020}]%
        {hassan2020isolated}
\bibfield{author}{\bibinfo{person}{Saad Hassan}, \bibinfo{person}{Larwan
  Berke}, \bibinfo{person}{Elahe Vahdani}, \bibinfo{person}{Longlong Jing},
  \bibinfo{person}{Yingli Tian}, {and} \bibinfo{person}{Matt Huenerfauth}.}
  \bibinfo{year}{2020}\natexlab{}.
\newblock \showarticletitle{An Isolated-Signing {RGBD} Dataset of 100
  {A}merican {S}ign {L}anguage Signs Produced by Fluent {ASL} Signers}. In
  \bibinfo{booktitle}{\emph{Proceedings of the LREC2020 9th Workshop on the
  Representation and Processing of Sign Languages: Sign Language Resources in
  the Service of the Language Community, Technological Challenges and
  Application Perspectives}}. \bibinfo{publisher}{European Language Resources
  Association (ELRA)}, \bibinfo{address}{Marseille, France},
  \bibinfo{pages}{89--94}.
\newblock
\showISBNx{979-10-95546-54-2}
\urldef\tempurl%
\url{https://www.aclweb.org/anthology/2020.signlang-1.14}
\showURL{%
\tempurl}


\bibitem[\protect\citeauthoryear{Heumann, Cassak, Kang, and Twitchell}{Heumann
  et~al\mbox{.}}{2016}]%
        {heumann2016privacy}
\bibfield{author}{\bibinfo{person}{Milton Heumann}, \bibinfo{person}{Lance
  Cassak}, \bibinfo{person}{Esther Kang}, {and} \bibinfo{person}{Thomas
  Twitchell}.} \bibinfo{year}{2016}\natexlab{}.
\newblock \showarticletitle{Privacy and Surveillance: Public Attitudes on
  Cameras on the Street, in the Home, and in the Workplace}.
\newblock \bibinfo{journal}{\emph{Rutgers JL \& Pub. Pol'y}}
  \bibinfo{volume}{14} (\bibinfo{year}{2016}), \bibinfo{pages}{37}.
\newblock


\bibitem[\protect\citeauthoryear{Hill, Turner, Martin, and Donovan}{Hill
  et~al\mbox{.}}{2013}]%
        {hill2013let}
\bibfield{author}{\bibinfo{person}{Elizabeth~M Hill}, \bibinfo{person}{Emma~L
  Turner}, \bibinfo{person}{Richard~M Martin}, {and} \bibinfo{person}{Jenny~L
  Donovan}.} \bibinfo{year}{2013}\natexlab{}.
\newblock \showarticletitle{“Let’s get the best quality research we can”:
  public awareness and acceptance of consent to use existing data in health
  research: a systematic review and qualitative study}.
\newblock \bibinfo{journal}{\emph{BMC medical research methodology}}
  \bibinfo{volume}{13}, \bibinfo{number}{1} (\bibinfo{year}{2013}),
  \bibinfo{pages}{1--10}.
\newblock


\bibitem[\protect\citeauthoryear{Hitron, Orlev, Wald, Shamir, Erel, and
  Zuckerman}{Hitron et~al\mbox{.}}{2019}]%
        {hitron2019can}
\bibfield{author}{\bibinfo{person}{Tom Hitron}, \bibinfo{person}{Yoav Orlev},
  \bibinfo{person}{Iddo Wald}, \bibinfo{person}{Ariel Shamir},
  \bibinfo{person}{Hadas Erel}, {and} \bibinfo{person}{Oren Zuckerman}.}
  \bibinfo{year}{2019}\natexlab{}.
\newblock \showarticletitle{Can Children Understand Machine Learning Concepts?
  The Effect of Uncovering Black Boxes}. In
  \bibinfo{booktitle}{\emph{Proceedings of the 2019 CHI Conference on Human
  Factors in Computing Systems}} (Glasgow, Scotland Uk)
  \emph{(\bibinfo{series}{CHI '19})}. \bibinfo{publisher}{Association for
  Computing Machinery}, \bibinfo{address}{New York, NY, USA},
  \bibinfo{pages}{1–11}.
\newblock
\showISBNx{9781450359702}
\urldef\tempurl%
\url{https://doi.org/10.1145/3290605.3300645}
\showDOI{\tempurl}


\bibitem[\protect\citeauthoryear{Hong, Gandhi, Mensah, Zeraati, Jarjue, Lee,
  and Kacorri}{Hong et~al\mbox{.}}{2022}]%
        {hong2022blind}
\bibfield{author}{\bibinfo{person}{Jonggi Hong}, \bibinfo{person}{Jaina
  Gandhi}, \bibinfo{person}{Ernest~Essuah Mensah},
  \bibinfo{person}{Farnaz~Zamiri Zeraati}, \bibinfo{person}{Ebrima Jarjue},
  \bibinfo{person}{Kyungjun Lee}, {and} \bibinfo{person}{Hernisa Kacorri}.}
  \bibinfo{year}{2022}\natexlab{}.
\newblock \showarticletitle{Blind Users Accessing Their Training Images in
  Teachable Object Recognizers}. In \bibinfo{booktitle}{\emph{Proceedings of
  the 24th International ACM SIGACCESS Conference on Computers and
  Accessibility}} (Athens, Greece) \emph{(\bibinfo{series}{ASSETS '22})}.
  \bibinfo{publisher}{Association for Computing Machinery},
  \bibinfo{address}{New York, NY, USA}, Article \bibinfo{articleno}{14},
  \bibinfo{numpages}{18}~pages.
\newblock
\showISBNx{9781450392587}
\urldef\tempurl%
\url{https://doi.org/10.1145/3517428.3544824}
\showDOI{\tempurl}


\bibitem[\protect\citeauthoryear{Hong, Lee, Xu, and Kacorri}{Hong
  et~al\mbox{.}}{2020}]%
        {hong2020crowdsourcing}
\bibfield{author}{\bibinfo{person}{Jonggi Hong}, \bibinfo{person}{Kyungjun
  Lee}, \bibinfo{person}{June Xu}, {and} \bibinfo{person}{Hernisa Kacorri}.}
  \bibinfo{year}{2020}\natexlab{}.
\newblock \bibinfo{booktitle}{\emph{Crowdsourcing the Perception of Machine
  Teaching}}.
\newblock \bibinfo{publisher}{Association for Computing Machinery},
  \bibinfo{address}{New York, NY, USA}, \bibinfo{pages}{1–14}.
\newblock
\showISBNx{9781450367080}
\urldef\tempurl%
\url{https://doi.org/10.1145/3313831.3376428}
\showURL{%
\tempurl}


\bibitem[\protect\citeauthoryear{Huenerfauth and Kacorri}{Huenerfauth and
  Kacorri}{2014}]%
        {huenerfauth2014release}
\bibfield{author}{\bibinfo{person}{Matt Huenerfauth} {and}
  \bibinfo{person}{Hernisa Kacorri}.} \bibinfo{year}{2014}\natexlab{}.
\newblock \showarticletitle{Release of experimental stimuli and questions for
  evaluating facial expressions in animations of American Sign Language}. In
  \bibinfo{booktitle}{\emph{Proceedings of the 6th Workshop on the
  Representation and Processing of Sign Languages: Beyond the Manual Channel,
  The 9th International Conference on Language Resources and Evaluation (LREC
  2014), Reykjavik, Iceland}}.
\newblock


\bibitem[\protect\citeauthoryear{Ienca, Scheibner, Ferretti, Gille, Amann,
  Sleigh, Blasimme, and Vayena}{Ienca et~al\mbox{.}}{2019}]%
        {ienca2019general}
\bibfield{author}{\bibinfo{person}{Marcello Ienca}, \bibinfo{person}{James
  Scheibner}, \bibinfo{person}{Agata Ferretti}, \bibinfo{person}{Felix Gille},
  \bibinfo{person}{Julia Amann}, \bibinfo{person}{Joanna Sleigh},
  \bibinfo{person}{Alessandro Blasimme}, {and} \bibinfo{person}{Effy Vayena}.}
  \bibinfo{year}{2019}\natexlab{}.
\newblock \bibinfo{booktitle}{\emph{How the general data protection regulation
  changes the rules for scientific research: Study}}.
\newblock \bibinfo{type}{{T}echnical {R}eport}. \bibinfo{institution}{ETH
  Zurich}.
\newblock


\bibitem[\protect\citeauthoryear{Jackson, Harrison, Swinburn, and
  Lawrence}{Jackson et~al\mbox{.}}{2015}]%
        {jackson2015using}
\bibfield{author}{\bibinfo{person}{Michaela Jackson}, \bibinfo{person}{Paul
  Harrison}, \bibinfo{person}{Boyd Swinburn}, {and} \bibinfo{person}{Mark
  Lawrence}.} \bibinfo{year}{2015}\natexlab{}.
\newblock \showarticletitle{Using a qualitative vignette to explore a complex
  public health issue}.
\newblock \bibinfo{journal}{\emph{Qualitative health research}}
  \bibinfo{volume}{25}, \bibinfo{number}{10} (\bibinfo{year}{2015}),
  \bibinfo{pages}{1395--1409}.
\newblock


\bibitem[\protect\citeauthoryear{Jamieson}{Jamieson}{2004}]%
        {jamieson2004likert}
\bibfield{author}{\bibinfo{person}{Susan Jamieson}.}
  \bibinfo{year}{2004}\natexlab{}.
\newblock \showarticletitle{Likert scales: How to (ab) use them?}
\newblock \bibinfo{journal}{\emph{Medical education}} \bibinfo{volume}{38},
  \bibinfo{number}{12} (\bibinfo{year}{2004}), \bibinfo{pages}{1217--1218}.
\newblock


\bibitem[\protect\citeauthoryear{Jean}{Jean}{2012}]%
        {jean2012information}
\bibfield{author}{\bibinfo{person}{Beth Lenore~St Jean}.}
  \bibinfo{year}{2012}\natexlab{}.
\newblock \emph{\bibinfo{title}{Information behavior of people diagnosed with a
  chronic serious health condition: A longitudinal study}}.
\newblock \bibinfo{thesistype}{Ph.D. Dissertation}. \bibinfo{school}{University
  of Michigan}.
\newblock


\bibitem[\protect\citeauthoryear{Jo and Gebru}{Jo and Gebru}{2020}]%
        {jo2020lessons}
\bibfield{author}{\bibinfo{person}{Eun~Seo Jo} {and} \bibinfo{person}{Timnit
  Gebru}.} \bibinfo{year}{2020}\natexlab{}.
\newblock \showarticletitle{Lessons from Archives: Strategies for Collecting
  Sociocultural Data in Machine Learning}. In
  \bibinfo{booktitle}{\emph{Proceedings of the 2020 Conference on Fairness,
  Accountability, and Transparency}} (Barcelona, Spain)
  \emph{(\bibinfo{series}{FAT* '20})}. \bibinfo{publisher}{Association for
  Computing Machinery}, \bibinfo{address}{New York, NY, USA},
  \bibinfo{pages}{306–316}.
\newblock
\showISBNx{9781450369367}
\urldef\tempurl%
\url{https://doi.org/10.1145/3351095.3372829}
\showDOI{\tempurl}


\bibitem[\protect\citeauthoryear{Kacorri}{Kacorri}{2016}]%
        {kacorri2016data}
\bibfield{author}{\bibinfo{person}{Hernisa Kacorri}.}
  \bibinfo{year}{2016}\natexlab{}.
\newblock \emph{\bibinfo{title}{Data-Driven Synthesis and Evaluation of
  Syntactic Facial Expressions in American Sign Language Animation}}.
\newblock \bibinfo{thesistype}{Ph.D. Dissertation}. \bibinfo{school}{CUNY
  Academic Works}.
\newblock
\urldef\tempurl%
\url{https://academicworks.cuny.edu/gc_etds/1375}
\showURL{%
\tempurl}


\bibitem[\protect\citeauthoryear{Kacorri}{Kacorri}{2017}]%
        {kacorri2017teachable}
\bibfield{author}{\bibinfo{person}{Hernisa Kacorri}.}
  \bibinfo{year}{2017}\natexlab{}.
\newblock \showarticletitle{Teachable Machines for Accessibility}.
\newblock \bibinfo{journal}{\emph{SIGACCESS Access. Comput.}}
  \bibinfo{number}{119}, \bibinfo{pages}{10–18}.
\newblock
\showISSN{1558-2337}
\urldef\tempurl%
\url{https://doi.org/10.1145/3167902.3167904}
\showDOI{\tempurl}


\bibitem[\protect\citeauthoryear{Kacorri, Dwivedi, Amancherla, Jha, and
  Chanduka}{Kacorri et~al\mbox{.}}{2020a}]%
        {kacorri2020incluset}
\bibfield{author}{\bibinfo{person}{Hernisa Kacorri}, \bibinfo{person}{Utkarsh
  Dwivedi}, \bibinfo{person}{Sravya Amancherla}, \bibinfo{person}{Mayanka Jha},
  {and} \bibinfo{person}{Riya Chanduka}.} \bibinfo{year}{2020}\natexlab{a}.
\newblock \showarticletitle{IncluSet: A Data Surfacing Repository for
  Accessibility Datasets}. In \bibinfo{booktitle}{\emph{The 22nd International
  ACM SIGACCESS Conference on Computers and Accessibility}}
  \emph{(\bibinfo{series}{ASSETS '20})}. \bibinfo{publisher}{Association for
  Computing Machinery}, \bibinfo{address}{New York, NY, USA}, Article
  \bibinfo{articleno}{72}, \bibinfo{numpages}{4}~pages.
\newblock
\showISBNx{9781450371032}
\urldef\tempurl%
\url{https://doi.org/10.1145/3373625.3418026}
\showDOI{\tempurl}


\bibitem[\protect\citeauthoryear{Kacorri, Dwivedi, and Kamikubo}{Kacorri
  et~al\mbox{.}}{2020b}]%
        {kacorri2020data}
\bibfield{author}{\bibinfo{person}{Hernisa Kacorri}, \bibinfo{person}{Utkarsh
  Dwivedi}, {and} \bibinfo{person}{Rie Kamikubo}.}
  \bibinfo{year}{2020}\natexlab{b}.
\newblock \showarticletitle{Data Sharing in Wellness, Accessibility, and
  Aging}.
\newblock \bibinfo{journal}{\emph{NeurIPS 2020 Workshop on Dataset Curation and
  Security}} (\bibinfo{year}{2020}).
\newblock


\bibitem[\protect\citeauthoryear{Kacorri, Kitani, Bigham, and Asakawa}{Kacorri
  et~al\mbox{.}}{2017}]%
        {kacorri2017people}
\bibfield{author}{\bibinfo{person}{Hernisa Kacorri}, \bibinfo{person}{Kris~M.
  Kitani}, \bibinfo{person}{Jeffrey~P. Bigham}, {and} \bibinfo{person}{Chieko
  Asakawa}.} \bibinfo{year}{2017}\natexlab{}.
\newblock \showarticletitle{People with Visual Impairment Training Personal
  Object Recognizers: Feasibility and Challenges}. In
  \bibinfo{booktitle}{\emph{Proceedings of the 2017 CHI Conference on Human
  Factors in Computing Systems}} (Denver, Colorado, USA)
  \emph{(\bibinfo{series}{CHI '17})}. \bibinfo{publisher}{Association for
  Computing Machinery}, \bibinfo{address}{New York, NY, USA},
  \bibinfo{pages}{5839–5849}.
\newblock
\showISBNx{9781450346559}
\urldef\tempurl%
\url{https://doi.org/10.1145/3025453.3025899}
\showDOI{\tempurl}


\bibitem[\protect\citeauthoryear{Kacorri, Mascetti, Gerino, Ahmetovic, Takagi,
  and Asakawa}{Kacorri et~al\mbox{.}}{2016}]%
        {kacorri2016supporting}
\bibfield{author}{\bibinfo{person}{Hernisa Kacorri}, \bibinfo{person}{Sergio
  Mascetti}, \bibinfo{person}{Andrea Gerino}, \bibinfo{person}{Dragan
  Ahmetovic}, \bibinfo{person}{Hironobu Takagi}, {and} \bibinfo{person}{Chieko
  Asakawa}.} \bibinfo{year}{2016}\natexlab{}.
\newblock \showarticletitle{Supporting Orientation of People with Visual
  Impairment: Analysis of Large Scale Usage Data}. In
  \bibinfo{booktitle}{\emph{Proceedings of the 18th International ACM SIGACCESS
  Conference on Computers and Accessibility}} (Reno, Nevada, USA)
  \emph{(\bibinfo{series}{ASSETS '16})}. \bibinfo{publisher}{ACM},
  \bibinfo{address}{New York, NY, USA}, \bibinfo{pages}{151--159}.
\newblock
\showISBNx{978-1-4503-4124-0}
\urldef\tempurl%
\url{https://doi.org/10.1145/2982142.2982178}
\showDOI{\tempurl}


\bibitem[\protect\citeauthoryear{Kaggle}{Kaggle}{2021}]%
        {kaggle}
\bibfield{author}{\bibinfo{person}{Kaggle}.} \bibinfo{year}{2021}\natexlab{}.
\newblock \bibinfo{title}{Kaggle: Your Machine Learning and Data Science
  Community}.
\newblock \bibinfo{howpublished}{\url{https://www.kaggle.com}}.
\newblock


\bibitem[\protect\citeauthoryear{Kamikubo, Dwivedi, and Kacorri}{Kamikubo
  et~al\mbox{.}}{2021}]%
        {kamikubo2021sharing}
\bibfield{author}{\bibinfo{person}{Rie Kamikubo}, \bibinfo{person}{Utkarsh
  Dwivedi}, {and} \bibinfo{person}{Hernisa Kacorri}.}
  \bibinfo{year}{2021}\natexlab{}.
\newblock \showarticletitle{Sharing Practices for Datasets Related to
  Accessibility and Aging}. In \bibinfo{booktitle}{\emph{The 23rd International
  ACM SIGACCESS Conference on Computers and Accessibility}} (Virtual Event,
  USA) \emph{(\bibinfo{series}{ASSETS '21})}. \bibinfo{publisher}{Association
  for Computing Machinery}, \bibinfo{address}{New York, NY, USA}, Article
  \bibinfo{articleno}{28}, \bibinfo{numpages}{16}~pages.
\newblock
\showISBNx{9781450383066}
\urldef\tempurl%
\url{https://doi.org/10.1145/3441852.3471208}
\showDOI{\tempurl}


\bibitem[\protect\citeauthoryear{Kamikubo, Wang, Marte, Mahmood, and
  Kacorri}{Kamikubo et~al\mbox{.}}{2022}]%
        {kamikubo2022data}
\bibfield{author}{\bibinfo{person}{Rie Kamikubo}, \bibinfo{person}{Lining
  Wang}, \bibinfo{person}{Crystal Marte}, \bibinfo{person}{Amnah Mahmood},
  {and} \bibinfo{person}{Hernisa Kacorri}.} \bibinfo{year}{2022}\natexlab{}.
\newblock \showarticletitle{Data Representativeness in Accessibility Datasets:
  A Meta-Analysis}. In \bibinfo{booktitle}{\emph{Proceedings of the 24th
  International ACM SIGACCESS Conference on Computers and Accessibility}}
  (Athens, Greece) \emph{(\bibinfo{series}{ASSETS '22})}.
  \bibinfo{publisher}{Association for Computing Machinery},
  \bibinfo{address}{New York, NY, USA}, Article \bibinfo{articleno}{8},
  \bibinfo{numpages}{15}~pages.
\newblock
\showISBNx{9781450392587}
\urldef\tempurl%
\url{https://doi.org/10.1145/3517428.3544826}
\showDOI{\tempurl}


\bibitem[\protect\citeauthoryear{Kim, Jeon, Lee, Choe, and Seo}{Kim
  et~al\mbox{.}}{2017}]%
        {kim2017omnitrack}
\bibfield{author}{\bibinfo{person}{Young-Ho Kim}, \bibinfo{person}{Jae~Ho
  Jeon}, \bibinfo{person}{Bongshin Lee}, \bibinfo{person}{Eun~Kyoung Choe},
  {and} \bibinfo{person}{Jinwook Seo}.} \bibinfo{year}{2017}\natexlab{}.
\newblock \showarticletitle{OmniTrack: A flexible self-tracking approach
  leveraging semi-automated tracking}.
\newblock \bibinfo{journal}{\emph{Proceedings of the ACM on interactive,
  mobile, wearable and ubiquitous technologies}} \bibinfo{volume}{1},
  \bibinfo{number}{3} (\bibinfo{year}{2017}), \bibinfo{pages}{1--28}.
\newblock


\bibitem[\protect\citeauthoryear{Kokolakis}{Kokolakis}{2017}]%
        {kokolakis2017privacy}
\bibfield{author}{\bibinfo{person}{Spyros Kokolakis}.}
  \bibinfo{year}{2017}\natexlab{}.
\newblock \showarticletitle{Privacy attitudes and privacy behaviour: A review
  of current research on the privacy paradox phenomenon}.
\newblock \bibinfo{journal}{\emph{Computers \& security}}  \bibinfo{volume}{64}
  (\bibinfo{year}{2017}), \bibinfo{pages}{122--134}.
\newblock


\bibitem[\protect\citeauthoryear{Kop}{Kop}{2021}]%
        {kop2021machine}
\bibfield{author}{\bibinfo{person}{Mauritz Kop}.}
  \bibinfo{year}{2021}\natexlab{}.
\newblock \showarticletitle{Machine learning and EU data-sharing practices:
  Legal aspects of machine learning training datasets for AI systems}.
\newblock In \bibinfo{booktitle}{\emph{Research Handbook on Big Data Law}}.
  \bibinfo{publisher}{Edward Elgar Publishing}, \bibinfo{pages}{432--453}.
\newblock


\bibitem[\protect\citeauthoryear{Kostkova, Brewer, De~Lusignan, Fottrell,
  Goldacre, Hart, Koczan, Knight, Marsolier, McKendry, et~al\mbox{.}}{Kostkova
  et~al\mbox{.}}{2016}]%
        {kostkova2016owns}
\bibfield{author}{\bibinfo{person}{Patty Kostkova}, \bibinfo{person}{Helen
  Brewer}, \bibinfo{person}{Simon De~Lusignan}, \bibinfo{person}{Edward
  Fottrell}, \bibinfo{person}{Ben Goldacre}, \bibinfo{person}{Graham Hart},
  \bibinfo{person}{Phil Koczan}, \bibinfo{person}{Peter Knight},
  \bibinfo{person}{Corinne Marsolier}, \bibinfo{person}{Rachel~A McKendry},
  {et~al\mbox{.}}} \bibinfo{year}{2016}\natexlab{}.
\newblock \showarticletitle{Who owns the data? Open data for healthcare}.
\newblock \bibinfo{journal}{\emph{Frontiers in public health}}
  \bibinfo{volume}{4} (\bibinfo{year}{2016}), \bibinfo{pages}{7}.
\newblock


\bibitem[\protect\citeauthoryear{Kr{\"o}ger, Miceli, and M{\"u}ller}{Kr{\"o}ger
  et~al\mbox{.}}{2021}]%
        {kroger2021data}
\bibfield{author}{\bibinfo{person}{Jacob~Leon Kr{\"o}ger},
  \bibinfo{person}{Milagros Miceli}, {and} \bibinfo{person}{Florian
  M{\"u}ller}.} \bibinfo{year}{2021}\natexlab{}.
\newblock \showarticletitle{How Data Can Be Used Against People: A
  Classification of Personal Data Misuses}.
\newblock \bibinfo{journal}{\emph{Available at SSRN 3887097}}
  (\bibinfo{year}{2021}).
\newblock


\bibitem[\protect\citeauthoryear{Krutzinna and Floridi}{Krutzinna and
  Floridi}{2019}]%
        {krutzinna2019ethical}
\bibfield{author}{\bibinfo{person}{Jenny Krutzinna} {and}
  \bibinfo{person}{Luciano Floridi}.} \bibinfo{year}{2019}\natexlab{}.
\newblock \showarticletitle{Ethical medical data donation: a pressing issue}.
\newblock \bibinfo{journal}{\emph{The Ethics of Medical Data Donation}}
  (\bibinfo{year}{2019}), \bibinfo{pages}{1--6}.
\newblock


\bibitem[\protect\citeauthoryear{Kulynych and Korn}{Kulynych and Korn}{2003}]%
        {kulynych2003new}
\bibfield{author}{\bibinfo{person}{Jennifer Kulynych} {and}
  \bibinfo{person}{David Korn}.} \bibinfo{year}{2003}\natexlab{}.
\newblock \showarticletitle{The new HIPAA (Health Insurance Portability and
  Accountability Act of 1996) Medical Privacy Rule: help or hindrance for
  clinical research?}
\newblock \bibinfo{journal}{\emph{Circulation}} \bibinfo{volume}{108},
  \bibinfo{number}{8} (\bibinfo{year}{2003}), \bibinfo{pages}{912--914}.
\newblock


\bibitem[\protect\citeauthoryear{Lee and Kacorri}{Lee and Kacorri}{2019}]%
        {lee2019hands}
\bibfield{author}{\bibinfo{person}{Kyungjun Lee} {and} \bibinfo{person}{Hernisa
  Kacorri}.} \bibinfo{year}{2019}\natexlab{}.
\newblock \showarticletitle{Hands Holding Clues for Object Recognition in
  Teachable Machines}. In \bibinfo{booktitle}{\emph{Proceedings of the 2019 CHI
  Conference on Human Factors in Computing Systems}} (Glasgow, Scotland Uk)
  \emph{(\bibinfo{series}{CHI '19})}. \bibinfo{publisher}{Association for
  Computing Machinery}, \bibinfo{address}{New York, NY, USA},
  \bibinfo{pages}{1–12}.
\newblock
\showISBNx{9781450359702}
\urldef\tempurl%
\url{https://doi.org/10.1145/3290605.3300566}
\showDOI{\tempurl}


\bibitem[\protect\citeauthoryear{Lee, Sato, Asakawa, Kacorri, and Asakawa}{Lee
  et~al\mbox{.}}{2020}]%
        {lee2020pedestrian}
\bibfield{author}{\bibinfo{person}{Kyungjun Lee}, \bibinfo{person}{Daisuke
  Sato}, \bibinfo{person}{Saki Asakawa}, \bibinfo{person}{Hernisa Kacorri},
  {and} \bibinfo{person}{Chieko Asakawa}.} \bibinfo{year}{2020}\natexlab{}.
\newblock \bibinfo{booktitle}{\emph{Pedestrian Detection with Wearable Cameras
  for the Blind: A Two-Way Perspective}}.
\newblock \bibinfo{publisher}{Association for Computing Machinery},
  \bibinfo{address}{New York, NY, USA}, \bibinfo{pages}{1–12}.
\newblock
\showISBNx{9781450367080}
\urldef\tempurl%
\url{https://doi.org/10.1145/3313831.3376398}
\showURL{%
\tempurl}


\bibitem[\protect\citeauthoryear{Linden, Khandelwal, Harkous, and Fawaz}{Linden
  et~al\mbox{.}}{2018}]%
        {linden2018privacy}
\bibfield{author}{\bibinfo{person}{Thomas Linden}, \bibinfo{person}{Rishabh
  Khandelwal}, \bibinfo{person}{Hamza Harkous}, {and} \bibinfo{person}{Kassem
  Fawaz}.} \bibinfo{year}{2018}\natexlab{}.
\newblock \showarticletitle{The privacy policy landscape after the GDPR}.
\newblock \bibinfo{journal}{\emph{arXiv preprint arXiv:1809.08396}}
  (\bibinfo{year}{2018}).
\newblock


\bibitem[\protect\citeauthoryear{Lotfian, Ingensand, and Brovelli}{Lotfian
  et~al\mbox{.}}{2020}]%
        {lotfian2020framework}
\bibfield{author}{\bibinfo{person}{Maryam Lotfian}, \bibinfo{person}{Jens
  Ingensand}, {and} \bibinfo{person}{Maria~Antonia Brovelli}.}
  \bibinfo{year}{2020}\natexlab{}.
\newblock \showarticletitle{A framework for classifying participant motivation
  that considers the typology of citizen science projects}.
\newblock \bibinfo{journal}{\emph{ISPRS International Journal of
  Geo-Information}} \bibinfo{volume}{9}, \bibinfo{number}{12}
  (\bibinfo{year}{2020}), \bibinfo{pages}{704}.
\newblock


\bibitem[\protect\citeauthoryear{Luo, Liu, and Choe}{Luo et~al\mbox{.}}{2019}]%
        {luo2019co}
\bibfield{author}{\bibinfo{person}{Yuhan Luo}, \bibinfo{person}{Peiyi Liu},
  {and} \bibinfo{person}{Eun~Kyoung Choe}.} \bibinfo{year}{2019}\natexlab{}.
\newblock \showarticletitle{Co-Designing food trackers with dietitians:
  Identifying design opportunities for food tracker customization}. In
  \bibinfo{booktitle}{\emph{Proceedings of the 2019 CHI Conference on Human
  Factors in Computing Systems}}. \bibinfo{pages}{1--13}.
\newblock


\bibitem[\protect\citeauthoryear{Luo, Oh, St~Jean, Choe, et~al\mbox{.}}{Luo
  et~al\mbox{.}}{2020}]%
        {luo2020interrelationships}
\bibfield{author}{\bibinfo{person}{Yuhan Luo}, \bibinfo{person}{Chi~Young Oh},
  \bibinfo{person}{Beth St~Jean}, \bibinfo{person}{Eun~Kyoung Choe},
  {et~al\mbox{.}}} \bibinfo{year}{2020}\natexlab{}.
\newblock \showarticletitle{Interrelationships Between Patients’ Data
  Tracking Practices, Data Sharing Practices, and Health Literacy: Onsite
  Survey Study}.
\newblock \bibinfo{journal}{\emph{Journal of medical Internet research}}
  \bibinfo{volume}{22}, \bibinfo{number}{12} (\bibinfo{year}{2020}),
  \bibinfo{pages}{e18937}.
\newblock


\bibitem[\protect\citeauthoryear{Lyle}{Lyle}{2014}]%
        {lyle2014openicpsr}
\bibfield{author}{\bibinfo{person}{Jared Lyle}.}
  \bibinfo{year}{2014}\natexlab{}.
\newblock \showarticletitle{OpenICPSR}.
\newblock \bibinfo{journal}{\emph{Bulletin of the Association for Information
  Science and Technology}} \bibinfo{volume}{40}, \bibinfo{number}{5}
  (\bibinfo{year}{2014}), \bibinfo{pages}{55--56}.
\newblock


\bibitem[\protect\citeauthoryear{Machirori and Patel}{Machirori and
  Patel}{2021}]%
        {machirori2021turning}
\bibfield{author}{\bibinfo{person}{Mavis Machirori} {and}
  \bibinfo{person}{Reema Patel}.} \bibinfo{year}{2021}\natexlab{}.
\newblock \showarticletitle{Turning distrust in data sharing into “engage,
  deliberate, decide”}.
\newblock  (\bibinfo{year}{2021}).
\newblock
\urldef\tempurl%
\url{https://www.adalovelaceinstitute.org/blog/distrust-data-sharing-engage-deliberate-decide/}
\showURL{%
\tempurl}


\bibitem[\protect\citeauthoryear{Madaio, Stark, Wortman~Vaughan, and
  Wallach}{Madaio et~al\mbox{.}}{2020}]%
        {madaio2020co}
\bibfield{author}{\bibinfo{person}{Michael~A Madaio}, \bibinfo{person}{Luke
  Stark}, \bibinfo{person}{Jennifer Wortman~Vaughan}, {and}
  \bibinfo{person}{Hanna Wallach}.} \bibinfo{year}{2020}\natexlab{}.
\newblock \showarticletitle{Co-designing checklists to understand
  organizational challenges and opportunities around fairness in AI}. In
  \bibinfo{booktitle}{\emph{Proceedings of the 2020 CHI Conference on Human
  Factors in Computing Systems}}. \bibinfo{pages}{1--14}.
\newblock


\bibitem[\protect\citeauthoryear{Massiceti, Zintgraf, Bronskill, Theodorou,
  Harris, Cutrell, Morrison, Hofmann, and Stumpf}{Massiceti
  et~al\mbox{.}}{2021}]%
        {massiceti2021orbit}
\bibfield{author}{\bibinfo{person}{Daniela Massiceti}, \bibinfo{person}{Luisa
  Zintgraf}, \bibinfo{person}{John Bronskill}, \bibinfo{person}{Lida
  Theodorou}, \bibinfo{person}{Matthew~Tobias Harris}, \bibinfo{person}{Edward
  Cutrell}, \bibinfo{person}{Cecily Morrison}, \bibinfo{person}{Katja Hofmann},
  {and} \bibinfo{person}{Simone Stumpf}.} \bibinfo{year}{2021}\natexlab{}.
\newblock \showarticletitle{Orbit: A real-world few-shot dataset for teachable
  object recognition}. In \bibinfo{booktitle}{\emph{Proceedings of the IEEE/CVF
  International Conference on Computer Vision}}. \bibinfo{pages}{10818--10828}.
\newblock


\bibitem[\protect\citeauthoryear{Maxwell}{Maxwell}{2010}]%
        {maxwell2010using}
\bibfield{author}{\bibinfo{person}{Joseph~A Maxwell}.}
  \bibinfo{year}{2010}\natexlab{}.
\newblock \showarticletitle{Using numbers in qualitative research}.
\newblock \bibinfo{journal}{\emph{Qualitative inquiry}} \bibinfo{volume}{16},
  \bibinfo{number}{6} (\bibinfo{year}{2010}), \bibinfo{pages}{475--482}.
\newblock


\bibitem[\protect\citeauthoryear{McDonald and Cranor}{McDonald and
  Cranor}{2008}]%
        {mcdonald2008cost}
\bibfield{author}{\bibinfo{person}{Aleecia~M McDonald} {and}
  \bibinfo{person}{Lorrie~Faith Cranor}.} \bibinfo{year}{2008}\natexlab{}.
\newblock \showarticletitle{The cost of reading privacy policies}.
\newblock \bibinfo{journal}{\emph{Isjlp}}  \bibinfo{volume}{4}
  (\bibinfo{year}{2008}), \bibinfo{pages}{543}.
\newblock


\bibitem[\protect\citeauthoryear{McNaney, Morgan, Kulkarni, Vega,
  Heidarivincheh, McConville, Whone, Kim, Kirkham, and Craddock}{McNaney
  et~al\mbox{.}}{2022}]%
        {mcnaney2022exploring}
\bibfield{author}{\bibinfo{person}{Roisin McNaney}, \bibinfo{person}{Catherine
  Morgan}, \bibinfo{person}{Pranav Kulkarni}, \bibinfo{person}{Julio Vega},
  \bibinfo{person}{Farnoosh Heidarivincheh}, \bibinfo{person}{Ryan McConville},
  \bibinfo{person}{Alan Whone}, \bibinfo{person}{Mickey Kim},
  \bibinfo{person}{Reuben Kirkham}, {and} \bibinfo{person}{Ian Craddock}.}
  \bibinfo{year}{2022}\natexlab{}.
\newblock \showarticletitle{Exploring Perceptions of Cross-Sectoral Data
  Sharing with People with Parkinson’s}. In \bibinfo{booktitle}{\emph{CHI
  Conference on Human Factors in Computing Systems}} (New Orleans, LA, USA)
  \emph{(\bibinfo{series}{CHI '22})}. \bibinfo{publisher}{Association for
  Computing Machinery}, \bibinfo{address}{New York, NY, USA}, Article
  \bibinfo{articleno}{280}, \bibinfo{numpages}{14}~pages.
\newblock
\showISBNx{9781450391573}
\urldef\tempurl%
\url{https://doi.org/10.1145/3491102.3501984}
\showDOI{\tempurl}


\bibitem[\protect\citeauthoryear{McNaney, Tsekleves, and Synnott}{McNaney
  et~al\mbox{.}}{2020}]%
        {mcnaney2020future}
\bibfield{author}{\bibinfo{person}{Roisin McNaney}, \bibinfo{person}{Emmanuel
  Tsekleves}, {and} \bibinfo{person}{Jonathan Synnott}.}
  \bibinfo{year}{2020}\natexlab{}.
\newblock \bibinfo{booktitle}{\emph{Future Opportunities for IoT to Support
  People with Parkinson's}}.
\newblock \bibinfo{publisher}{Association for Computing Machinery},
  \bibinfo{address}{New York, NY, USA}, \bibinfo{pages}{1–15}.
\newblock
\showISBNx{9781450367080}
\urldef\tempurl%
\url{https://doi.org/10.1145/3313831.3376871}
\showURL{%
\tempurl}


\bibitem[\protect\citeauthoryear{Mentis, Komlodi, Schrader, Phipps,
  Gruber-Baldini, Yarbrough, and Shulman}{Mentis et~al\mbox{.}}{2017}]%
        {mentis2017crafting}
\bibfield{author}{\bibinfo{person}{Helena~M Mentis}, \bibinfo{person}{Anita
  Komlodi}, \bibinfo{person}{Katrina Schrader}, \bibinfo{person}{Michael
  Phipps}, \bibinfo{person}{Ann Gruber-Baldini}, \bibinfo{person}{Karen
  Yarbrough}, {and} \bibinfo{person}{Lisa Shulman}.}
  \bibinfo{year}{2017}\natexlab{}.
\newblock \showarticletitle{Crafting a view of self-tracking data in the
  clinical visit}. In \bibinfo{booktitle}{\emph{Proceedings of the 2017 CHI
  Conference on Human Factors in Computing Systems}}.
  \bibinfo{pages}{5800--5812}.
\newblock


\bibitem[\protect\citeauthoryear{Microsoft}{Microsoft}{2017}]%
        {microsoft2017seeing}
\bibfield{author}{\bibinfo{person}{Microsoft}.}
  \bibinfo{year}{2017}\natexlab{}.
\newblock \bibinfo{title}{Seeing AI: An app for visually impaired people that
  narrates the world around you}.
\newblock
\newblock
\urldef\tempurl%
\url{https://www.microsoft.com/en-us/garage/wall-of-fame/seeing-ai/}
\showURL{%
\tempurl}


\bibitem[\protect\citeauthoryear{Mishra, Miller, Haldar, Khelifi, Eschler,
  Elera, Pollack, and Pratt}{Mishra et~al\mbox{.}}{2018}]%
        {mishra2018supporting}
\bibfield{author}{\bibinfo{person}{Sonali~R Mishra}, \bibinfo{person}{Andrew~D
  Miller}, \bibinfo{person}{Shefali Haldar}, \bibinfo{person}{Maher Khelifi},
  \bibinfo{person}{Jordan Eschler}, \bibinfo{person}{Rashmi~G Elera},
  \bibinfo{person}{Ari~H Pollack}, {and} \bibinfo{person}{Wanda Pratt}.}
  \bibinfo{year}{2018}\natexlab{}.
\newblock \showarticletitle{Supporting collaborative health tracking in the
  hospital: patients' perspectives}. In \bibinfo{booktitle}{\emph{Proceedings
  of the 2018 CHI Conference on Human Factors in Computing Systems}}.
  \bibinfo{pages}{1--14}.
\newblock


\bibitem[\protect\citeauthoryear{Morris}{Morris}{2020}]%
        {morris2020ai}
\bibfield{author}{\bibinfo{person}{Meredith~Ringel Morris}.}
  \bibinfo{year}{2020}\natexlab{}.
\newblock \showarticletitle{AI and Accessibility}.
\newblock \bibinfo{journal}{\emph{Commun. ACM}} \bibinfo{volume}{63},
  \bibinfo{number}{6}, \bibinfo{pages}{35–37}.
\newblock
\showISSN{0001-0782}
\urldef\tempurl%
\url{https://doi.org/10.1145/3356727}
\showDOI{\tempurl}


\bibitem[\protect\citeauthoryear{Mozersky, Parsons, Walsh, Baldwin, McIntosh,
  and DuBois}{Mozersky et~al\mbox{.}}{2020}]%
        {mozersky2020research}
\bibfield{author}{\bibinfo{person}{Jessica Mozersky}, \bibinfo{person}{Meredith
  Parsons}, \bibinfo{person}{Heidi Walsh}, \bibinfo{person}{Kari Baldwin},
  \bibinfo{person}{Tristan McIntosh}, {and} \bibinfo{person}{James~M DuBois}.}
  \bibinfo{year}{2020}\natexlab{}.
\newblock \showarticletitle{Research participant views regarding qualitative
  data sharing}.
\newblock \bibinfo{journal}{\emph{Ethics \& human research}}
  \bibinfo{volume}{42}, \bibinfo{number}{2} (\bibinfo{year}{2020}),
  \bibinfo{pages}{13--27}.
\newblock


\bibitem[\protect\citeauthoryear{Nakamura}{Nakamura}{2019}]%
        {nakamura2019my}
\bibfield{author}{\bibinfo{person}{Karen Nakamura}.}
  \bibinfo{year}{2019}\natexlab{}.
\newblock \showarticletitle{My Algorithms Have Determined You're Not Human:
  AI-ML, Reverse Turing-Tests, and the Disability Experience}. In
  \bibinfo{booktitle}{\emph{The 21st International ACM SIGACCESS Conference on
  Computers and Accessibility}} (Pittsburgh, PA, USA)
  \emph{(\bibinfo{series}{ASSETS '19})}. \bibinfo{publisher}{Association for
  Computing Machinery}, \bibinfo{address}{New York, NY, USA},
  \bibinfo{pages}{1–2}.
\newblock
\showISBNx{9781450366762}
\urldef\tempurl%
\url{https://doi.org/10.1145/3308561.3353812}
\showDOI{\tempurl}


\bibitem[\protect\citeauthoryear{Nicholas, Shilton, Schueller, Gray, Kwasny,
  Mohr, et~al\mbox{.}}{Nicholas et~al\mbox{.}}{2019}]%
        {nicholas2019role}
\bibfield{author}{\bibinfo{person}{Jennifer Nicholas}, \bibinfo{person}{Katie
  Shilton}, \bibinfo{person}{Stephen~M Schueller}, \bibinfo{person}{Elizabeth~L
  Gray}, \bibinfo{person}{Mary~J Kwasny}, \bibinfo{person}{David~C Mohr},
  {et~al\mbox{.}}} \bibinfo{year}{2019}\natexlab{}.
\newblock \showarticletitle{The role of data type and recipient in
  individuals’ perspectives on sharing passively collected smartphone data
  for mental health: Cross-sectional questionnaire study}.
\newblock \bibinfo{journal}{\emph{JMIR mHealth and uHealth}}
  \bibinfo{volume}{7}, \bibinfo{number}{4} (\bibinfo{year}{2019}),
  \bibinfo{pages}{e12578}.
\newblock


\bibitem[\protect\citeauthoryear{Nissenbaum}{Nissenbaum}{2004}]%
        {nissenbaum2004privacy}
\bibfield{author}{\bibinfo{person}{Helen Nissenbaum}.}
  \bibinfo{year}{2004}\natexlab{}.
\newblock \showarticletitle{Privacy as contextual integrity}.
\newblock \bibinfo{journal}{\emph{Washington Law Review}}  \bibinfo{volume}{79}
  (\bibinfo{year}{2004}), \bibinfo{pages}{119}.
\newblock


\bibitem[\protect\citeauthoryear{Nusbaum, Douglas, Damus, Paasche-Orlow, and
  Estrella-Luna}{Nusbaum et~al\mbox{.}}{2017}]%
        {nusbaum2017communicating}
\bibfield{author}{\bibinfo{person}{Lika Nusbaum}, \bibinfo{person}{Brenda
  Douglas}, \bibinfo{person}{Karla Damus}, \bibinfo{person}{Michael
  Paasche-Orlow}, {and} \bibinfo{person}{Neenah Estrella-Luna}.}
  \bibinfo{year}{2017}\natexlab{}.
\newblock \showarticletitle{Communicating risks and benefits in informed
  consent for research: a qualitative study}.
\newblock \bibinfo{journal}{\emph{Global Qualitative Nursing Research}}
  \bibinfo{volume}{4} (\bibinfo{year}{2017}),
  \bibinfo{pages}{2333393617732017}.
\newblock


\bibitem[\protect\citeauthoryear{Park, Bragg, Kamar, and Morris}{Park
  et~al\mbox{.}}{2021}]%
        {park2021designing}
\bibfield{author}{\bibinfo{person}{Joon~Sung Park}, \bibinfo{person}{Danielle
  Bragg}, \bibinfo{person}{Ece Kamar}, {and} \bibinfo{person}{Meredith~Ringel
  Morris}.} \bibinfo{year}{2021}\natexlab{}.
\newblock \showarticletitle{Designing an online infrastructure for collecting
  AI data from people with disabilities}. In
  \bibinfo{booktitle}{\emph{Proceedings of the 2021 ACM Conference on Fairness,
  Accountability, and Transparency}}. \bibinfo{pages}{52--63}.
\newblock


\bibitem[\protect\citeauthoryear{Queiroz, Sampaio, Lima, and Lima}{Queiroz
  et~al\mbox{.}}{2020}]%
        {queiroz2020ai}
\bibfield{author}{\bibinfo{person}{Rubens~Lacerda Queiroz},
  \bibinfo{person}{Fábio~Ferrentini Sampaio}, \bibinfo{person}{Cabral Lima},
  {and} \bibinfo{person}{Priscila Machado~Vieira Lima}.}
  \bibinfo{year}{2020}\natexlab{}.
\newblock \bibinfo{title}{AI from concrete to abstract: demystifying artificial
  intelligence to the general public}.
\newblock
\newblock
\urldef\tempurl%
\url{https://doi.org/10.1007/s00146-021-01151-x}
\showDOI{\tempurl}
\showeprint[arxiv]{2006.04013}~[cs.CY]


\bibitem[\protect\citeauthoryear{Rake, Van~Gelder, Grim, Heeren, Engelen, and
  Van De~Belt}{Rake et~al\mbox{.}}{2017}]%
        {rake2017personalized}
\bibfield{author}{\bibinfo{person}{Ester~A Rake}, \bibinfo{person}{Marleen~MHJ
  Van~Gelder}, \bibinfo{person}{David~C Grim}, \bibinfo{person}{Barend Heeren},
  \bibinfo{person}{Lucien~JLPG Engelen}, {and} \bibinfo{person}{Tom~H Van
  De~Belt}.} \bibinfo{year}{2017}\natexlab{}.
\newblock \showarticletitle{Personalized consent flow in contemporary data
  sharing for medical research: A viewpoint}.
\newblock \bibinfo{journal}{\emph{BioMed Research International}}
  \bibinfo{volume}{2017} (\bibinfo{year}{2017}).
\newblock


\bibitem[\protect\citeauthoryear{Redden, Brand, and Terzieva}{Redden
  et~al\mbox{.}}{2020}]%
        {redden2020data}
\bibfield{author}{\bibinfo{person}{Joanna Redden}, \bibinfo{person}{Jessica
  Brand}, {and} \bibinfo{person}{Vanesa Terzieva}.}
  \bibinfo{year}{2020}\natexlab{}.
\newblock \showarticletitle{Data harm record}.
\newblock  (\bibinfo{year}{2020}).
\newblock


\bibitem[\protect\citeauthoryear{Ren and Gu}{Ren and Gu}{2010}]%
        {ren2010figure}
\bibfield{author}{\bibinfo{person}{Xiaofeng Ren} {and} \bibinfo{person}{Chunhui
  Gu}.} \bibinfo{year}{2010}\natexlab{}.
\newblock \showarticletitle{Figure-ground segmentation improves handled object
  recognition in egocentric video}. In \bibinfo{booktitle}{\emph{2010 IEEE
  Computer Society Conference on Computer Vision and Pattern Recognition}}.
  IEEE, \bibinfo{pages}{3137--3144}.
\newblock


\bibitem[\protect\citeauthoryear{Research}{Research}{2018}]%
        {microsoft2018microsoft}
\bibfield{author}{\bibinfo{person}{Microsoft Research}.}
  \bibinfo{year}{2018}\natexlab{}.
\newblock \bibinfo{booktitle}{\emph{Microsoft Research Open Data}}.
\newblock
\urldef\tempurl%
\url{https://msropendata.com/}
\showURL{%
\tempurl}


\bibitem[\protect\citeauthoryear{Rotman, Preece, Hammock, Procita, Hansen,
  Parr, Lewis, and Jacobs}{Rotman et~al\mbox{.}}{2012}]%
        {rotman2012dynamic}
\bibfield{author}{\bibinfo{person}{Dana Rotman}, \bibinfo{person}{Jenny
  Preece}, \bibinfo{person}{Jen Hammock}, \bibinfo{person}{Kezee Procita},
  \bibinfo{person}{Derek Hansen}, \bibinfo{person}{Cynthia Parr},
  \bibinfo{person}{Darcy Lewis}, {and} \bibinfo{person}{David Jacobs}.}
  \bibinfo{year}{2012}\natexlab{}.
\newblock \showarticletitle{Dynamic Changes in Motivation in Collaborative
  Citizen-Science Projects}. In \bibinfo{booktitle}{\emph{Proceedings of the
  ACM 2012 Conference on Computer Supported Cooperative Work}} (Seattle,
  Washington, USA) \emph{(\bibinfo{series}{CSCW '12})}.
  \bibinfo{publisher}{Association for Computing Machinery},
  \bibinfo{address}{New York, NY, USA}, \bibinfo{pages}{217–226}.
\newblock
\showISBNx{9781450310864}
\urldef\tempurl%
\url{https://doi.org/10.1145/2145204.2145238}
\showDOI{\tempurl}


\bibitem[\protect\citeauthoryear{Russakovsky, Deng, Su, Krause, Satheesh, Ma,
  Huang, Karpathy, Khosla, Bernstein, et~al\mbox{.}}{Russakovsky
  et~al\mbox{.}}{2015}]%
        {russakovsky2015imagenet}
\bibfield{author}{\bibinfo{person}{Olga Russakovsky}, \bibinfo{person}{Jia
  Deng}, \bibinfo{person}{Hao Su}, \bibinfo{person}{Jonathan Krause},
  \bibinfo{person}{Sanjeev Satheesh}, \bibinfo{person}{Sean Ma},
  \bibinfo{person}{Zhiheng Huang}, \bibinfo{person}{Andrej Karpathy},
  \bibinfo{person}{Aditya Khosla}, \bibinfo{person}{Michael Bernstein},
  {et~al\mbox{.}}} \bibinfo{year}{2015}\natexlab{}.
\newblock \showarticletitle{Imagenet large scale visual recognition challenge}.
\newblock \bibinfo{journal}{\emph{International journal of computer vision}}
  \bibinfo{volume}{115}, \bibinfo{number}{3} (\bibinfo{year}{2015}),
  \bibinfo{pages}{211--252}.
\newblock


\bibitem[\protect\citeauthoryear{Sandelowski}{Sandelowski}{2001}]%
        {sandelowski2001real}
\bibfield{author}{\bibinfo{person}{Margarete Sandelowski}.}
  \bibinfo{year}{2001}\natexlab{}.
\newblock \showarticletitle{Real qualitative researchers do not count: The use
  of numbers in qualitative research}.
\newblock \bibinfo{journal}{\emph{Research in nursing \& health}}
  \bibinfo{volume}{24}, \bibinfo{number}{3} (\bibinfo{year}{2001}),
  \bibinfo{pages}{230--240}.
\newblock


\bibitem[\protect\citeauthoryear{Sandelowski, Voils, and Knafl}{Sandelowski
  et~al\mbox{.}}{2009}]%
        {sandelowski2009quantitizing}
\bibfield{author}{\bibinfo{person}{Margarete Sandelowski},
  \bibinfo{person}{Corrine~I Voils}, {and} \bibinfo{person}{George Knafl}.}
  \bibinfo{year}{2009}\natexlab{}.
\newblock \showarticletitle{On quantitizing}.
\newblock \bibinfo{journal}{\emph{Journal of mixed methods research}}
  \bibinfo{volume}{3}, \bibinfo{number}{3} (\bibinfo{year}{2009}),
  \bibinfo{pages}{208--222}.
\newblock


\bibitem[\protect\citeauthoryear{Schoenberg and Ravdal}{Schoenberg and
  Ravdal}{2000}]%
        {schoenberg2000using}
\bibfield{author}{\bibinfo{person}{Nancy~E Schoenberg} {and}
  \bibinfo{person}{Hege Ravdal}.} \bibinfo{year}{2000}\natexlab{}.
\newblock \showarticletitle{Using vignettes in awareness and attitudinal
  research}.
\newblock \bibinfo{journal}{\emph{International journal of social research
  methodology}} \bibinfo{volume}{3}, \bibinfo{number}{1}
  (\bibinfo{year}{2000}), \bibinfo{pages}{63--74}.
\newblock


\bibitem[\protect\citeauthoryear{Sears and Hanson}{Sears and Hanson}{2012}]%
        {sears2011representing}
\bibfield{author}{\bibinfo{person}{Andrew Sears} {and}
  \bibinfo{person}{Vicki~L. Hanson}.} \bibinfo{year}{2012}\natexlab{}.
\newblock \showarticletitle{Representing Users in Accessibility Research}.
\newblock \bibinfo{journal}{\emph{ACM Trans. Access. Comput.}}
  \bibinfo{volume}{4}, \bibinfo{number}{2}, Article \bibinfo{articleno}{7},
  \bibinfo{numpages}{6}~pages.
\newblock
\showISSN{1936-7228}
\urldef\tempurl%
\url{https://doi.org/10.1145/2141943.2141945}
\showDOI{\tempurl}


\bibitem[\protect\citeauthoryear{Shah, Coathup, Teare, Forgie, Giordano,
  Hansen, Groeneveld, Hudson, Pearson, Ruetten, et~al\mbox{.}}{Shah
  et~al\mbox{.}}{2019}]%
        {shah2019motivations}
\bibfield{author}{\bibinfo{person}{Nisha Shah}, \bibinfo{person}{Victoria
  Coathup}, \bibinfo{person}{Harriet Teare}, \bibinfo{person}{Ian Forgie},
  \bibinfo{person}{Giuseppe~Nicola Giordano}, \bibinfo{person}{Tue~Haldor
  Hansen}, \bibinfo{person}{Lenka Groeneveld}, \bibinfo{person}{Michelle
  Hudson}, \bibinfo{person}{Ewan Pearson}, \bibinfo{person}{Hartmut Ruetten},
  {et~al\mbox{.}}} \bibinfo{year}{2019}\natexlab{}.
\newblock \showarticletitle{Motivations for data sharing—views of research
  participants from four European countries: a DIRECT study}.
\newblock \bibinfo{journal}{\emph{European Journal of Human Genetics}}
  \bibinfo{volume}{27}, \bibinfo{number}{5} (\bibinfo{year}{2019}),
  \bibinfo{pages}{721--729}.
\newblock


\bibitem[\protect\citeauthoryear{Sharif, McCall, and Bolante}{Sharif
  et~al\mbox{.}}{2022}]%
        {sharif2022should}
\bibfield{author}{\bibinfo{person}{Ather Sharif}, \bibinfo{person}{Aedan~Liam
  McCall}, {and} \bibinfo{person}{Kianna~Roces Bolante}.}
  \bibinfo{year}{2022}\natexlab{}.
\newblock \showarticletitle{Should I Say “Disabled People” or “People
  with Disabilities”? Language Preferences of Disabled People Between
  Identity- and Person-First Language}. In
  \bibinfo{booktitle}{\emph{Proceedings of the 24th International ACM SIGACCESS
  Conference on Computers and Accessibility}} (Athens, Greece)
  \emph{(\bibinfo{series}{ASSETS '22})}. \bibinfo{publisher}{Association for
  Computing Machinery}, \bibinfo{address}{New York, NY, USA}, Article
  \bibinfo{articleno}{10}, \bibinfo{numpages}{18}~pages.
\newblock
\showISBNx{9781450392587}
\urldef\tempurl%
\url{https://doi.org/10.1145/3517428.3544813}
\showDOI{\tempurl}


\bibitem[\protect\citeauthoryear{Shilton}{Shilton}{2009}]%
        {shilton2009four}
\bibfield{author}{\bibinfo{person}{Katie Shilton}.}
  \bibinfo{year}{2009}\natexlab{}.
\newblock \showarticletitle{Four billion little brothers? Privacy, mobile
  phones, and ubiquitous data collection}.
\newblock \bibinfo{journal}{\emph{Commun. ACM}} \bibinfo{volume}{52},
  \bibinfo{number}{11} (\bibinfo{year}{2009}), \bibinfo{pages}{48--53}.
\newblock


\bibitem[\protect\citeauthoryear{Shirazi, Clawson, Hassanpour, Tourian,
  Schmidt, Chi, Borazio, and Van~Laerhoven}{Shirazi et~al\mbox{.}}{2013}]%
        {shirazi2013already}
\bibfield{author}{\bibinfo{person}{Alireza~Sahami Shirazi},
  \bibinfo{person}{James Clawson}, \bibinfo{person}{Yashar Hassanpour},
  \bibinfo{person}{Mohammad~J Tourian}, \bibinfo{person}{Albrecht Schmidt},
  \bibinfo{person}{Ed~H Chi}, \bibinfo{person}{Marko Borazio}, {and}
  \bibinfo{person}{Kristof Van~Laerhoven}.} \bibinfo{year}{2013}\natexlab{}.
\newblock \showarticletitle{Already up? using mobile phones to track \& share
  sleep behavior}.
\newblock \bibinfo{journal}{\emph{International Journal of Human-Computer
  Studies}} \bibinfo{volume}{71}, \bibinfo{number}{9} (\bibinfo{year}{2013}),
  \bibinfo{pages}{878--888}.
\newblock


\bibitem[\protect\citeauthoryear{Shneiderman}{Shneiderman}{2020a}]%
        {shneiderman2020bridging}
\bibfield{author}{\bibinfo{person}{Ben Shneiderman}.}
  \bibinfo{year}{2020}\natexlab{a}.
\newblock \showarticletitle{Bridging the gap between ethics and practice:
  guidelines for reliable, safe, and trustworthy human-centered AI systems}.
\newblock \bibinfo{journal}{\emph{ACM Transactions on Interactive Intelligent
  Systems (TiiS)}} \bibinfo{volume}{10}, \bibinfo{number}{4}
  (\bibinfo{year}{2020}), \bibinfo{pages}{1--31}.
\newblock


\bibitem[\protect\citeauthoryear{Shneiderman}{Shneiderman}{2020b}]%
        {shneiderman2020human}
\bibfield{author}{\bibinfo{person}{Ben Shneiderman}.}
  \bibinfo{year}{2020}\natexlab{b}.
\newblock \showarticletitle{Human-centered artificial intelligence: Reliable,
  safe \& trustworthy}.
\newblock \bibinfo{journal}{\emph{International Journal of Human--Computer
  Interaction}} \bibinfo{volume}{36}, \bibinfo{number}{6}
  (\bibinfo{year}{2020}), \bibinfo{pages}{495--504}.
\newblock


\bibitem[\protect\citeauthoryear{Sivan-Sevilla}{Sivan-Sevilla}{2022}]%
        {sivan2022varieties}
\bibfield{author}{\bibinfo{person}{Ido Sivan-Sevilla}.}
  \bibinfo{year}{2022}\natexlab{}.
\newblock \showarticletitle{Varieties of enforcement strategies post-GDPR: a
  fuzzy-set qualitative comparative analysis (fsQCA) across data protection
  authorities}.
\newblock \bibinfo{journal}{\emph{Journal of European Public Policy}}
  (\bibinfo{year}{2022}), \bibinfo{pages}{1--34}.
\newblock


\bibitem[\protect\citeauthoryear{Soden, Skirpan, Fiesler, Ashktorab, Baumer,
  Blythe, and Jones}{Soden et~al\mbox{.}}{2019}]%
        {soden2019chi}
\bibfield{author}{\bibinfo{person}{Robert Soden}, \bibinfo{person}{Michael
  Skirpan}, \bibinfo{person}{Casey Fiesler}, \bibinfo{person}{Zahra Ashktorab},
  \bibinfo{person}{Eric P.~S. Baumer}, \bibinfo{person}{Mark Blythe}, {and}
  \bibinfo{person}{Jasmine Jones}.} \bibinfo{year}{2019}\natexlab{}.
\newblock \showarticletitle{CHI4EVIL: Creative Speculation on the Negative
  Impacts of HCI Research}. In \bibinfo{booktitle}{\emph{Extended Abstracts of
  the 2019 CHI Conference on Human Factors in Computing Systems}} (Glasgow,
  Scotland Uk) \emph{(\bibinfo{series}{CHI EA '19})}.
  \bibinfo{publisher}{Association for Computing Machinery},
  \bibinfo{address}{New York, NY, USA}, \bibinfo{pages}{1–8}.
\newblock
\showISBNx{9781450359719}
\urldef\tempurl%
\url{https://doi.org/10.1145/3290607.3299033}
\showDOI{\tempurl}


\bibitem[\protect\citeauthoryear{Solove}{Solove}{2014}]%
        {solove2014reasons}
\bibfield{author}{\bibinfo{person}{Daniel Solove}.}
  \bibinfo{year}{2014}\natexlab{}.
\newblock \bibinfo{title}{10 Reasons Why Privacy Matters}.
\newblock
  \bibinfo{howpublished}{\url{https://teachprivacy.com/10-reasons-privacy-matters/}}.
\newblock


\bibitem[\protect\citeauthoryear{Solove}{Solove}{2008}]%
        {solove2008understanding}
\bibfield{author}{\bibinfo{person}{Daniel~J Solove}.}
  \bibinfo{year}{2008}\natexlab{}.
\newblock \showarticletitle{Understanding privacy}.
\newblock  (\bibinfo{year}{2008}).
\newblock


\bibitem[\protect\citeauthoryear{Sosa-Garc{\'\i}a and Odone}{Sosa-Garc{\'\i}a
  and Odone}{2017}]%
        {sosa2017hands}
\bibfield{author}{\bibinfo{person}{Joan Sosa-Garc{\'\i}a} {and}
  \bibinfo{person}{Francesca Odone}.} \bibinfo{year}{2017}\natexlab{}.
\newblock \showarticletitle{“Hands on” visual recognition for visually
  impaired users}.
\newblock \bibinfo{journal}{\emph{ACM Transactions on Accessible Computing
  (TACCESS)}} \bibinfo{volume}{10}, \bibinfo{number}{3} (\bibinfo{year}{2017}),
  \bibinfo{pages}{1--30}.
\newblock


\bibitem[\protect\citeauthoryear{Stangl, Shiroma, Davis, Xie, Fleischmann,
  Findlater, and Gurari}{Stangl et~al\mbox{.}}{2022}]%
        {stangl2022privacy}
\bibfield{author}{\bibinfo{person}{Abigale Stangl}, \bibinfo{person}{Kristina
  Shiroma}, \bibinfo{person}{Nathan Davis}, \bibinfo{person}{Bo Xie},
  \bibinfo{person}{Kenneth~R Fleischmann}, \bibinfo{person}{Leah Findlater},
  {and} \bibinfo{person}{Danna Gurari}.} \bibinfo{year}{2022}\natexlab{}.
\newblock \showarticletitle{Privacy Concerns for Visual Assistance
  Technologies}.
\newblock \bibinfo{journal}{\emph{ACM Transactions on Accessible Computing
  (TACCESS)}} \bibinfo{volume}{15}, \bibinfo{number}{2} (\bibinfo{year}{2022}),
  \bibinfo{pages}{1--43}.
\newblock


\bibitem[\protect\citeauthoryear{Sullivan and Artino~Jr}{Sullivan and
  Artino~Jr}{2013}]%
        {sullivan2013analyzing}
\bibfield{author}{\bibinfo{person}{Gail~M Sullivan} {and}
  \bibinfo{person}{Anthony~R Artino~Jr}.} \bibinfo{year}{2013}\natexlab{}.
\newblock \showarticletitle{Analyzing and interpreting data from Likert-type
  scales}.
\newblock \bibinfo{journal}{\emph{Journal of graduate medical education}}
  \bibinfo{volume}{5}, \bibinfo{number}{4} (\bibinfo{year}{2013}),
  \bibinfo{pages}{541--542}.
\newblock


\bibitem[\protect\citeauthoryear{Taichman, Sahni, Pinborg, Peiperl, Laine,
  James, Hong, Haileamlak, Gollogly, Godlee, Frizelle, Florenzano, Drazen,
  Bauchner, Baethge, and Backus}{Taichman et~al\mbox{.}}{2017}]%
        {taichman2017data}
\bibfield{author}{\bibinfo{person}{Darren~B Taichman}, \bibinfo{person}{Peush
  Sahni}, \bibinfo{person}{Anja Pinborg}, \bibinfo{person}{Larry Peiperl},
  \bibinfo{person}{Christine Laine}, \bibinfo{person}{Astrid James},
  \bibinfo{person}{Sung-Tae Hong}, \bibinfo{person}{Abraham Haileamlak},
  \bibinfo{person}{Laragh Gollogly}, \bibinfo{person}{Fiona Godlee},
  \bibinfo{person}{Frank~A Frizelle}, \bibinfo{person}{Fernando Florenzano},
  \bibinfo{person}{Jeffrey~M Drazen}, \bibinfo{person}{Howard Bauchner},
  \bibinfo{person}{Christopher Baethge}, {and} \bibinfo{person}{Joyce Backus}.}
  \bibinfo{year}{2017}\natexlab{}.
\newblock \showarticletitle{Data sharing statements for clinical trials: a
  requirement of the {International} {Committee} of {Medical} {Journal}
  {Editors}}.
\newblock \bibinfo{journal}{\emph{The Lancet}} \bibinfo{volume}{389},
  \bibinfo{number}{10086} (\bibinfo{date}{June} \bibinfo{year}{2017}),
  \bibinfo{pages}{e12--e14}.
\newblock
\showISSN{0140-6736}
\urldef\tempurl%
\url{https://doi.org/10.1016/S0140-6736(17)31282-5}
\showDOI{\tempurl}


\bibitem[\protect\citeauthoryear{Tangari, Ikram, Ijaz, Kaafar, and
  Berkovsky}{Tangari et~al\mbox{.}}{2021}]%
        {tangari2021mobile}
\bibfield{author}{\bibinfo{person}{Gioacchino Tangari},
  \bibinfo{person}{Muhammad Ikram}, \bibinfo{person}{Kiran Ijaz},
  \bibinfo{person}{Mohamed~Ali Kaafar}, {and} \bibinfo{person}{Shlomo
  Berkovsky}.} \bibinfo{year}{2021}\natexlab{}.
\newblock \showarticletitle{Mobile health and privacy: cross sectional study}.
\newblock \bibinfo{journal}{\emph{bmj}}  \bibinfo{volume}{373}
  (\bibinfo{year}{2021}).
\newblock


\bibitem[\protect\citeauthoryear{Tatman}{Tatman}{2017}]%
        {tatman2017gender}
\bibfield{author}{\bibinfo{person}{Rachael Tatman}.}
  \bibinfo{year}{2017}\natexlab{}.
\newblock \showarticletitle{Gender and Dialect Bias in {Y}ou{T}ube{'}s
  Automatic Captions}. In \bibinfo{booktitle}{\emph{Proceedings of the First
  {ACL} Workshop on Ethics in Natural Language Processing}}.
  \bibinfo{publisher}{Association for Computational Linguistics},
  \bibinfo{address}{Valencia, Spain}, \bibinfo{pages}{53--59}.
\newblock
\urldef\tempurl%
\url{https://doi.org/10.18653/v1/W17-1606}
\showDOI{\tempurl}


\bibitem[\protect\citeauthoryear{Theodorou, Massiceti, Zintgraf, Stumpf,
  Morrison, Cutrell, Harris, and Hofmann}{Theodorou et~al\mbox{.}}{2021}]%
        {theodorou2021disability}
\bibfield{author}{\bibinfo{person}{Lida Theodorou}, \bibinfo{person}{Daniela
  Massiceti}, \bibinfo{person}{Luisa Zintgraf}, \bibinfo{person}{Simone
  Stumpf}, \bibinfo{person}{Cecily Morrison}, \bibinfo{person}{Edward Cutrell},
  \bibinfo{person}{Matthew~Tobias Harris}, {and} \bibinfo{person}{Katja
  Hofmann}.} \bibinfo{year}{2021}\natexlab{}.
\newblock \showarticletitle{Disability-first Dataset Creation: Lessons from
  Constructing a Dataset for Teachable Object Recognition with Blind and Low
  Vision Data Collectors}. In \bibinfo{booktitle}{\emph{The 23rd International
  ACM SIGACCESS Conference on Computers and Accessibility}}.
  \bibinfo{pages}{1--12}.
\newblock


\bibitem[\protect\citeauthoryear{Thomas and Leiponen}{Thomas and
  Leiponen}{2016}]%
        {thomas2016big}
\bibfield{author}{\bibinfo{person}{Llewellyn~DW Thomas} {and}
  \bibinfo{person}{Aija Leiponen}.} \bibinfo{year}{2016}\natexlab{}.
\newblock \showarticletitle{Big data commercialization}.
\newblock \bibinfo{journal}{\emph{IEEE Engineering Management Review}}
  \bibinfo{volume}{44}, \bibinfo{number}{2} (\bibinfo{year}{2016}),
  \bibinfo{pages}{74--90}.
\newblock


\bibitem[\protect\citeauthoryear{Treviranus}{Treviranus}{2019}]%
        {treviranus2019value}
\bibfield{author}{\bibinfo{person}{Jutta Treviranus}.}
  \bibinfo{year}{2019}\natexlab{}.
\newblock \showarticletitle{The Value of Being Different}. In
  \bibinfo{booktitle}{\emph{Proceedings of the 16th Web For All 2019
  Personalization - Personalizing the Web}} (San Francisco, CA, USA)
  \emph{(\bibinfo{series}{W4A '19})}. \bibinfo{publisher}{ACM},
  \bibinfo{address}{New York, NY, USA}, Article \bibinfo{articleno}{1},
  \bibinfo{numpages}{7}~pages.
\newblock
\showISBNx{978-1-4503-6716-5}
\urldef\tempurl%
\url{https://doi.org/10.1145/3315002.3332429}
\showDOI{\tempurl}


\bibitem[\protect\citeauthoryear{Trewin}{Trewin}{2018}]%
        {trewin2018ai}
\bibfield{author}{\bibinfo{person}{Shari Trewin}.}
  \bibinfo{year}{2018}\natexlab{}.
\newblock \showarticletitle{AI fairness for people with disabilities: Point of
  view}.
\newblock \bibinfo{journal}{\emph{arXiv preprint arXiv:1811.10670}}
  (\bibinfo{year}{2018}).
\newblock


\bibitem[\protect\citeauthoryear{Tseng, Bell, and Gurari}{Tseng
  et~al\mbox{.}}{2022}]%
        {tseng2022vizwiz}
\bibfield{author}{\bibinfo{person}{Yu-Yun Tseng}, \bibinfo{person}{Alexander
  Bell}, {and} \bibinfo{person}{Danna Gurari}.}
  \bibinfo{year}{2022}\natexlab{}.
\newblock \showarticletitle{VizWiz-FewShot: Locating Objects in Images Taken
  by People with Visual Impairments}. In \bibinfo{booktitle}{\emph{Computer
  Vision -- ECCV 2022}}, \bibfield{editor}{\bibinfo{person}{Shai Avidan},
  \bibinfo{person}{Gabriel Brostow}, \bibinfo{person}{Moustapha Ciss{\'e}},
  \bibinfo{person}{Giovanni~Maria Farinella}, {and} \bibinfo{person}{Tal
  Hassner}} (Eds.). \bibinfo{publisher}{Springer Nature Switzerland},
  \bibinfo{address}{Cham}, \bibinfo{pages}{575--591}.
\newblock
\showISBNx{978-3-031-20074-8}


\bibitem[\protect\citeauthoryear{Verheggen, Nieman, and Jonkers}{Verheggen
  et~al\mbox{.}}{1998}]%
        {verheggen1998determinants}
\bibfield{author}{\bibinfo{person}{Frank~WSM Verheggen}, \bibinfo{person}{Fred
  Nieman}, {and} \bibinfo{person}{Ruud Jonkers}.}
  \bibinfo{year}{1998}\natexlab{}.
\newblock \showarticletitle{Determinants of patient participation in clinical
  studies requiring informed consent: why patients enter a clinical trial}.
\newblock \bibinfo{journal}{\emph{Patient education and counseling}}
  \bibinfo{volume}{35}, \bibinfo{number}{2} (\bibinfo{year}{1998}),
  \bibinfo{pages}{111--125}.
\newblock


\bibitem[\protect\citeauthoryear{Verhulst, Noveck, Caplan, Brown, and
  Paz}{Verhulst et~al\mbox{.}}{2014}]%
        {verhulst2014open}
\bibfield{author}{\bibinfo{person}{Stefaan Verhulst},
  \bibinfo{person}{Beth~Simone Noveck}, \bibinfo{person}{Robyn Caplan},
  \bibinfo{person}{Kristy Brown}, {and} \bibinfo{person}{Claudia Paz}.}
  \bibinfo{year}{2014}\natexlab{}.
\newblock \showarticletitle{The open data era in health and social care}.
\newblock \bibinfo{journal}{\emph{NYLS Legal Studies Research Paper}}
  \bibinfo{number}{2563788} (\bibinfo{year}{2014}).
\newblock


\bibitem[\protect\citeauthoryear{Viscusi and Zeckhauser}{Viscusi and
  Zeckhauser}{2015}]%
        {viscusi2015relative}
\bibfield{author}{\bibinfo{person}{W~Kip Viscusi} {and}
  \bibinfo{person}{Richard~J Zeckhauser}.} \bibinfo{year}{2015}\natexlab{}.
\newblock \showarticletitle{The relative weights of direct and indirect
  experiences in the formation of environmental risk beliefs}.
\newblock \bibinfo{journal}{\emph{Risk Analysis}} \bibinfo{volume}{35},
  \bibinfo{number}{2} (\bibinfo{year}{2015}), \bibinfo{pages}{318--331}.
\newblock


\bibitem[\protect\citeauthoryear{Vitak and Shilton}{Vitak and Shilton}{2020}]%
        {vitak2020trust}
\bibfield{author}{\bibinfo{person}{Jessica Vitak} {and} \bibinfo{person}{Katie
  Shilton}.} \bibinfo{year}{2020}\natexlab{}.
\newblock \showarticletitle{Trust, privacy and security, and accessibility
  considerations when conducting mobile technologies research with older
  adults}. In \bibinfo{booktitle}{\emph{Mobile Technology for Adaptive Aging:
  Proceedings of a Workshop}}. National Academies Press,
  \bibinfo{pages}{1--20}.
\newblock


\bibitem[\protect\citeauthoryear{Wacharamanotham, Eisenring, Haroz, and
  Echtler}{Wacharamanotham et~al\mbox{.}}{2020}]%
        {wacharamanotham2020transparency}
\bibfield{author}{\bibinfo{person}{Chat Wacharamanotham},
  \bibinfo{person}{Lukas Eisenring}, \bibinfo{person}{Steve Haroz}, {and}
  \bibinfo{person}{Florian Echtler}.} \bibinfo{year}{2020}\natexlab{}.
\newblock \showarticletitle{Transparency of CHI Research Artifacts: Results of
  a Self-Reported Survey}. In \bibinfo{booktitle}{\emph{Proceedings of the 2020
  CHI Conference on Human Factors in Computing Systems}} (Honolulu, HI, USA)
  \emph{(\bibinfo{series}{CHI '20})}. \bibinfo{publisher}{Association for
  Computing Machinery}, \bibinfo{address}{New York, NY, USA},
  \bibinfo{pages}{1–14}.
\newblock
\showISBNx{9781450367080}
\urldef\tempurl%
\url{https://doi.org/10.1145/3313831.3376448}
\showDOI{\tempurl}


\bibitem[\protect\citeauthoryear{Walport and Brest}{Walport and Brest}{2011}]%
        {walport2011sharing}
\bibfield{author}{\bibinfo{person}{Mark Walport} {and} \bibinfo{person}{Paul
  Brest}.} \bibinfo{year}{2011}\natexlab{}.
\newblock \showarticletitle{Sharing research data to improve public health}.
\newblock \bibinfo{journal}{\emph{The Lancet}} \bibinfo{volume}{377},
  \bibinfo{number}{9765} (\bibinfo{date}{2019/09/15} \bibinfo{year}{2011}),
  \bibinfo{pages}{537--539}.
\newblock
\showISBNx{0140-6736}
\urldef\tempurl%
\url{https://doi.org/10.1016/S0140-6736(10)62234-9}
\showDOI{\tempurl}


\bibitem[\protect\citeauthoryear{White, Doraiswamy, and Horvitz}{White
  et~al\mbox{.}}{2018}]%
        {white2018detecting}
\bibfield{author}{\bibinfo{person}{Ryen~W White}, \bibinfo{person}{P~Murali
  Doraiswamy}, {and} \bibinfo{person}{Eric Horvitz}.}
  \bibinfo{year}{2018}\natexlab{}.
\newblock \showarticletitle{Detecting neurodegenerative disorders from web
  search signals}.
\newblock \bibinfo{journal}{\emph{NPJ digital medicine}} \bibinfo{volume}{1},
  \bibinfo{number}{1} (\bibinfo{year}{2018}), \bibinfo{pages}{1--4}.
\newblock


\bibitem[\protect\citeauthoryear{Whittaker, Alper, Bennett, Hendren, Kaziunas,
  Mills, Morris, Rankin, Rogers, Salas, et~al\mbox{.}}{Whittaker
  et~al\mbox{.}}{2019}]%
        {whittaker2019disability}
\bibfield{author}{\bibinfo{person}{Meredith Whittaker}, \bibinfo{person}{Meryl
  Alper}, \bibinfo{person}{Cynthia~L Bennett}, \bibinfo{person}{Sara Hendren},
  \bibinfo{person}{Liz Kaziunas}, \bibinfo{person}{Mara Mills},
  \bibinfo{person}{Meredith~Ringel Morris}, \bibinfo{person}{Joy Rankin},
  \bibinfo{person}{Emily Rogers}, \bibinfo{person}{Marcel Salas},
  {et~al\mbox{.}}} \bibinfo{year}{2019}\natexlab{}.
\newblock \showarticletitle{Disability, Bias, and AI}.
\newblock \bibinfo{journal}{\emph{AI Now Institute, November}}
  (\bibinfo{year}{2019}).
\newblock


\bibitem[\protect\citeauthoryear{Wilczek}{Wilczek}{2021}]%
        {wilczek2021archival}
\bibfield{author}{\bibinfo{person}{Eliot Wilczek}.}
  \bibinfo{year}{2021}\natexlab{}.
\newblock \showarticletitle{Archival Engagements with Wicked Problems}.
\newblock \bibinfo{journal}{\emph{The American Archivist}}
  \bibinfo{volume}{84}, \bibinfo{number}{2} (\bibinfo{year}{2021}),
  \bibinfo{pages}{468--501}.
\newblock


\bibitem[\protect\citeauthoryear{Xu, Luo, Carroll, and Rosson}{Xu
  et~al\mbox{.}}{2011}]%
        {xu2011personalization}
\bibfield{author}{\bibinfo{person}{Heng Xu}, \bibinfo{person}{Xin~Robert Luo},
  \bibinfo{person}{John~M Carroll}, {and} \bibinfo{person}{Mary~Beth Rosson}.}
  \bibinfo{year}{2011}\natexlab{}.
\newblock \showarticletitle{The personalization privacy paradox: An exploratory
  study of decision making process for location-aware marketing}.
\newblock \bibinfo{journal}{\emph{Decision support systems}}
  \bibinfo{volume}{51}, \bibinfo{number}{1} (\bibinfo{year}{2011}),
  \bibinfo{pages}{42--52}.
\newblock


\bibitem[\protect\citeauthoryear{Zhong, Garrigues, and Bigham}{Zhong
  et~al\mbox{.}}{2013}]%
        {zhong2013real}
\bibfield{author}{\bibinfo{person}{Yu Zhong}, \bibinfo{person}{Pierre~J.
  Garrigues}, {and} \bibinfo{person}{Jeffrey~P. Bigham}.}
  \bibinfo{year}{2013}\natexlab{}.
\newblock \showarticletitle{Real Time Object Scanning Using a Mobile Phone and
  Cloud-Based Visual Search Engine}. In \bibinfo{booktitle}{\emph{Proceedings
  of the 15th International ACM SIGACCESS Conference on Computers and
  Accessibility}} (Bellevue, Washington) \emph{(\bibinfo{series}{ASSETS '13})}.
  \bibinfo{publisher}{Association for Computing Machinery},
  \bibinfo{address}{New York, NY, USA}, Article \bibinfo{articleno}{20},
  \bibinfo{numpages}{8}~pages.
\newblock
\showISBNx{9781450324052}
\urldef\tempurl%
\url{https://doi.org/10.1145/2513383.2513443}
\showDOI{\tempurl}


\bibitem[\protect\citeauthoryear{Zuiderwijk, Janssen, Van De~Kaa, and
  Poulis}{Zuiderwijk et~al\mbox{.}}{2016}]%
        {zuiderwijk2016wicked}
\bibfield{author}{\bibinfo{person}{Anneke Zuiderwijk}, \bibinfo{person}{Marijn
  Janssen}, \bibinfo{person}{Geerten Van De~Kaa}, {and} \bibinfo{person}{Kostas
  Poulis}.} \bibinfo{year}{2016}\natexlab{}.
\newblock \showarticletitle{The wicked problem of commercial value creation in
  open data ecosystems: Policy guidelines for governments}.
\newblock \bibinfo{journal}{\emph{Information polity}} \bibinfo{volume}{21},
  \bibinfo{number}{3} (\bibinfo{year}{2016}), \bibinfo{pages}{223--236}.
\newblock


\end{thebibliography}
\end{document}